\documentclass[brazil,12pt,a4paper,openany]{book}
\usepackage[utf8x]{inputenc}
\usepackage{anysize}
\usepackage[ruled,chapter]{algorithm}
\usepackage{makeidx}
\usepackage{subfig}
\usepackage{amsmath}
\usepackage{amssymb}
\usepackage{amsthm}
\usepackage{mathrsfs}
\usepackage{graphicx}
\usepackage{fancyhdr}
\usepackage{array}
\usepackage{geometry}
\usepackage{simplewick}
\usepackage{latexsym}
\usepackage[all]{xy}
\usepackage{euscript}
\usepackage{enumerate}
\usepackage{dsfont}
\usepackage{slashed}
\usepackage[dvipsnames]{xcolor}
\usepackage{cite}

\usepackage{pdfpages}

\usepackage{sectsty}
\chapterfont{\color{NavyBlue}}  
\sectionfont{\color{NavyBlue}}  
\subsectionfont{\color{NavyBlue}}  
\subsubsectionfont{\color{NavyBlue}}  

\usepackage{csquotes} 

\usepackage{titlesec}
\titleformat{\chapter}[hang]
  {\normalfont\huge\bfseries\color{NavyBlue}}{\thechapter}
  {1em}{\Huge}

\def\mbf#1{\mathbf{#1}}
\def\bs#1{\boldsymbol{#1}}
\def\tbf#1{\textbf{#1}}

\def\be{\begin{equation}}
\def\ee{\end{equation}}
\def\bea{\begin{eqnarray}}
\def\eea{\end{eqnarray}}

\usepackage[plainpages=false,pagebackref=true,pdftex]{hyperref}

\makeindex

\marginsize{25mm}{25mm}{25mm}{25mm}

\hypersetup{colorlinks=true,breaklinks=true,linkcolor=NavyBlue,citecolor=cyan,
urlcolor=cyan}

\begin{document}

\setlength{\abovecaptionskip}{0.0cm}
\setlength{\belowcaptionskip}{0.0cm}
\setlength{\baselineskip}{24pt}

\pagestyle{fancy}
\lhead{}
\chead{}
\rhead{\thepage}
\lfoot{}
\cfoot{}
\rfoot{}

\fancypagestyle{plain}
{
	\fancyhf{}
	\lhead{}
	\chead{}
	\rhead{\thepage}
	\lfoot{}
	\cfoot{}
	\rfoot{}
}

\renewcommand{\headrulewidth}{0pt}


\frontmatter 

\thispagestyle{empty}

\begin{figure}[h]
	\includegraphics[scale=0.8]{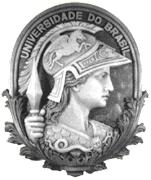}
\end{figure}

\vspace{15pt}

\begin{center}

\textbf{UNIVERSIDADE FEDERAL DO RIO DE JANEIRO}

\textbf{INSTITUTO DE F\'ISICA}

\vspace{30pt}

{\Large \bf Coherent Structures and Lattice-Boltzmann Hydrodynamics in Turbulent Pipe Flows}

\vspace{25pt}

{\large \bf Bruno Magacho da Silva}

\vspace{35pt}

\begin{flushright}
\parbox{10.3cm}{Tese de Doutorado apresentada ao Programa de Pós-Graduação em Física do Instituto de Física da Universidade Federal do Rio de Janeiro - UFRJ, como parte dos requisitos necessários à obtenção do título de Doutor em Ciências (Física).

\vspace{18pt}

{\large \bf Orientador: Luca Moriconi}}
\end{flushright}

\vspace{90pt}

\textbf{Rio de Janeiro}

\textbf{Maio de 2024}

\end{center}



\newpage

\thispagestyle{empty}

\noindent


\clearpage



\includepdf[pages={1}]{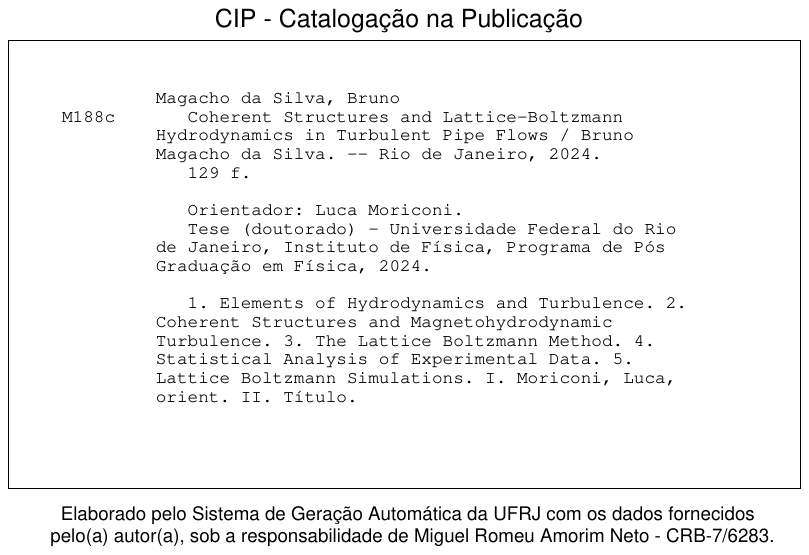}


\newpage

\begin{center}
    Coherent Structures and Lattice-Boltzmann Hydrodynamics in Turbulent Pipe Flows\\
    \textbf{Bruno Magacho da Silva}\\
    Orientador: Luca Moriconi
\end{center}

Tese de Doutorado submetida ao Programa de Pós-Graduação em Física, Instituto de Física, da Universidade Federal do Rio de Janeiro – UFRJ, como parte dos requisitos necessários à obtenção do título de Doutor em Ciências (Física).

Aprovada por:

\vspace{0.5 cm}

\begin{flushright}

............................................................................................\\
\vspace{-0.4 cm}
Prof. Dr. Luca Roberto Augusto Moriconi, IF-UFRJ   \hbox{ } \hbox{ }  \\
\vspace{-0.4 cm}
(Presidente e Orientador) \hbox{ } \hbox{ } \hbox{ } \hbox{ } \hbox{ } \hbox{ } \hbox{ }\hbox{ } \hbox{ } \hbox{ }

 \vspace{0.4 cm}

............................................................................................\\
\vspace{-0.4 cm}
 Prof. Dr. Felipe Arruda de Araujo Pinheiro, IF-UFRJ   \hbox{ } \\

\vspace{0.4 cm}

............................................................................................\\
\vspace{-0.4 cm}
 Prof. Dr. Alexei Mailybaev, IMPA  \hbox{ }  \hbox{ } \hbox{ }  \hbox{ } \hbox{ }  \hbox{ } \hbox{ } \hbox{ } \\

\vspace{0.4 cm}

............................................................................................\\
\vspace{-0.4 cm}
 Prof. Dr. Fábio Antonio Tavares Ramos, IM-UFRJ  \hbox{ }  \hbox{ }  \\

 \vspace{0.4 cm}

............................................................................................\\
\vspace{-0.4 cm}
Prof. Dr. Rodrigo Miranda Pereira, UFF  \hbox{ } \hbox{ }  \hbox{ } \hbox{ } \hbox{ } \hbox{ }  \\

\vspace{0.4 cm}

............................................................................................\\
\vspace{-0.4 cm}
Prof. Dr. Rodrigo Lage Sacramento, IF-UFRJ  \hbox{ } \hbox{ } \hbox{ } \hbox{ } \\	

\vspace{0.4 cm}

............................................................................................\\
\vspace{-0.4 cm}
 Prof. Dr. Leonardo de Sousa Grigorio, CEFET   \hbox{ }  \hbox{ } \hbox{ } \hbox{ } \\

\end{flushright}

\vspace{2.0 cm}

\begin{center}
    Rio de Janeiro, RJ - Brasil\\
    Maio de 2024
\end{center}


\newpage

\noindent

\vspace*{20pt}
\begin{center}
{\LARGE\bf Resumo}\\
\vspace{15pt}
{\Large\bf Coherent Structures and Lattice-Boltzmann Hydrodynamics in Turbulent Pipe Flows}\\
\vspace{6pt}
{\bf Bruno Magacho da Silva}\\
\vspace{12pt}
{\bf Orientador: Luca Moriconi}\\
\vspace{20pt}
\parbox{14cm}{Resumo da Tese de Doutorado apresentada ao Programa de Pós-Graduação em Física do Instituto de Física da Universidade Federal do Rio de Janeiro - UFRJ, como parte dos requisitos necessários à obtenção do título de Doutor em Ciências (Física).}
\end{center}
\vspace*{35pt}

Estruturas coerentes (EC) são conhecidas por serem parte dos fundamentos da dinâmica dos escoamentos turbulentos. Por muito tempo, acreditou-se que sua manifestação fosse caótica e desorganizada. No entanto, nas últimas duas décadas, tem sido demonstrado através de simulações numéricas e experimentos que um alto grau de organização das EC poderia ser atribuído à constituição de um estado turbulento. Compreender essas dinâmicas organizacionais promete trazer previsões teóricas e aplicadas valiosas, como a vida média das estruturas turbulentas, a compreensão do papel das EC no transporte de particulados e no desenvolvimento de reatores de fusão uma vez que uma redução na transferência de calor é esperada para escoamentos menos turbulentos.

Nesta tese de doutorado foi realizada uma análise estatística de um banco de dados experimental de escoamentos turbulentos em tubo --que tem como objetivo investigar a dinâmica das EC para uma ampla gama de intensidades turbulentas, realizado no Núcleo Interdisciplinar de Dinâmica de Fluidos - UFRJ. A identificação das EC foi alcançada selecionando-se o modo mais energético na direção do escoamento dentro de uma camada de referência especificada. Além disso, a dinâmica de transição entre as EC identificadas foi investigada como um processo estocástico, revelando um efeito não-Markoviano por meio de um decaimento algébrico da autocorrelação temporal das EC identificadas. Por fim, o comportamento não-Markoviano observado entre as transições das EC foi reproduzido por um modelo Markoviano de baixo nível, que leva em conta os efeitos de degenerescência na definição das EC identificadas.

A fim de se obter um algoritmo capaz de simular o regime quase-estático em escoamentos magnetohidrodinâmicos (MHD) --relevante na maioria das aplicações industriais-- um modelo de tempo de relaxamento múltiplo (TRM) e uma condição de contorno dependente da distância foram introduzidos para o método de Boltzmann de rede (MBR) associado à equação de indução para escoamentos MHD. Finalmente, foi realizada uma simulação de escoamento turbulento em tubo pelo MBR com um modelo TRM para distribuições hidrodinâmicas. A identificação das EC revelou um efeito de memória não trivial com relação à força que desencadeou o estado turbulento. A dinâmica de transição das EC revelou um comportamento Markoviano para os dados finamente resolvidos no tempo, indicando que o comportamento experimental poderia ser recuperado para maiores separações de tempo e, consequentemente, um conjunto de dados maior.

\vspace{15pt}

\textbf{Palavras-chave:} Estruturas Coerentes, Turbulência, Método de Boltzmann de Rede, Escoamentos em Tubos, Magnetohidrodinâmica.



\newpage

\noindent

\vspace*{20pt}
\begin{center}
{\LARGE\bf Abstract}\\
\vspace{15pt}
{\Large\bf Coherent Structures and Lattice-Boltzmann Hydrodynamics in Turbulent Pipe Flows}\\
\vspace{6pt}
{\bf Bruno Magacho da Silva}\\
\vspace{12pt}
{\bf Orientador: Luca Moriconi}\\
\vspace{20pt}
\parbox{14cm}{\emph{Abstract} da Tese de Doutorado apresentada ao Programa de Pós-Graduação em Física do Instituto de Física da Universidade Federal do Rio de Janeiro - UFRJ, como parte dos requisitos necessários à obtenção do título de Doutor em Ciências (Física).} 
\end{center}
\vspace*{35pt}

Coherent structures (CS) are known to be part of the foundations of turbulent flow dynamics. For a long time, their appearance was believed to be chaotic and unorganized. However, over the past two decades, it has been demonstrated through numerical simulations and experiments that a high degree of organization of CS could be attributed to the constitution of a turbulent state. Understanding these organizational dynamics promises to bring valuable theoretical and applied predictions, such as the average lifetime of turbulent structures, understanding the role of CS in particulate transport, and in the development of fusion reactors since a reduction of heat transfer is expected for less turbulent flows.

In this Ph.D. dissertation, a statistical analysis of an experimental database of a turbulent pipe flow --aimed to investigate the dynamics of CS for a wide range of turbulent intensities-- was carried out at the Interdisciplinary Center for Fluid Dynamics - UFRJ. The identification of CS was achieved by selecting the most energetic mode in the flow direction within a specified reference shell. Furthermore, the transition dynamics between the identified CS was investigated as a stochastic process, revealing a non-Markovian effect through an algebraic decay of the temporal self-correlation of the identified CS. Finally, the non-Markovian behavior observed between the transitions of CS was reproduced by a low-level Markovian model, which takes into account the degeneracy effects in the definition of the identified CS.

In order to obtain an algorithm capable of simulating the quasi-static regime in magnetohydrodynamic (MHD) flows --relevant in most industrial applications-- a multiple-relaxation-time (MRT) model and a distance-dependent boundary condition were introduced for the lattice Boltzmann method (LBM) associated with the induction equation for MHD flows. Finally, a turbulent pipe flow simulation was performed by the LBM with a MRT model for hydrodynamic distributions. The identification of CS revealed a non-trivial memory effect with respect to the force that triggered the turbulent state. The transition dynamics of CS revealed a Markovian behavior for finely resolved time data, indicating that experimental behavior could be recovered for larger time separations and, consequently, a larger dataset.

\vspace{15pt}




\newpage

\noindent

\vspace*{20pt}

\begin{center}

{\LARGE\bf Agradecimentos}

\end{center}

\vspace*{40pt}

Gostaria de aproveitar este momento para agradecer a todos que incentivaram a minha trajetória, seja na vida pessoal ou acadêmica e que desempenharam um papel fundamental até o término deste doutorado.

Em particular, agradeço aos meus pais, Mauro Saraiva e Mathilde Magacho que sempre confiaram em mim e me deram todo o suporte possível para que eu chegasse até aqui.

Agradeço a minha noiva Laís Marques, que está comigo desde o final da minha graduação e acompanhou de perto toda a minha trajetória durante a pós graduação. Fundamental para meu bem estar e companheira em todas as situações possíveis.

Agradeço ao meu orientador, Prof. Dr. Luca Moriconi, o qual acreditou no meu potencial e me deu a oportunidade de realizar o mestrado e doutorado na área que sempre despertou um interesse maior em minha pessoa. Agradeço pelos ensinamentos, incentivos e por sempre me motivar a encarar os desafios da vida acadêmica.

Agradeço aos meus companheiro do \emph{Journal Club} de turbulência e dinâmica de fluidos, que desempenharam um papel fundamental na complementariedade da minha formação através da discussão de inumeráveis artigos ao longo desses 6 anos de participação. Em particular agradeço aos que estiveram próximos de mim nesses últimos 4 anos de doutorado, Leonardo Grigório, Maiara Neumann, Giovanni Saisse e Rodrigo Pereira. Em particular ao Hugo Tavares pelas inúmeras discussões sobre o método de Boltzmann de rede e magnetohidrodinâmica.

Agradeço aos integrantes do Núcleo Interdisciplinar de Dinâmica dos Fluidos, por todo o suporte para que esse doutoramento fosse possível. Em particular a Prof. Dra. Juliana Loureiro, Augusto Correia e Bayode Owolabi. Um agradecimento especial para o Eduardo Ramos, pelas incontáveis ajudas para a utilização do cluster e para o Robert Jäckel por todas as discussões sobre turbulência e todo o seu esforço e motivação na realização do experimento de escoamento turbulento em dutos, o qual eu cuidei da análise estatística. 

Agradeço também os amigos que fiz durante toda a minha formação, que sempre contribuiram no meu desenvolvimento, seja pessoal ou profissional. Em particular gostaria de agradecer ao Guilherme Costa, Gabriel Apolinário e João Victor Bertolon. Gostaria de agradecer especialmente o Victor Valadão, o qual além de dividir sala, ir junto comigo para a faculdade em diversas ocasiões, também ajudou a fortificar meus conhecimentos sobre turbulência. O último, porém não menos importante é o Gabriel Picanço, o qual foi fundamental em toda a minha trajetória acadêmica, que desde o começo da minha trajetória na Física me acompanhou, incentivou e incentiva até hoje.

Por fim, agradeço ao CNPq e a COPPETEC, pelas bolsas que tornaram possível o desenvolvimento e conclusão dessa pesquisa de doutorado.



\newpage
\phantomsection
\addcontentsline{toc}{chapter}{Summary}
\tableofcontents

\newpage
\phantomsection
\addcontentsline{toc}{chapter}{List of Figures}
\listoffigures

\newpage
\phantomsection
\addcontentsline{toc}{chapter}{List of Tables}
\listoftables

\mainmatter
\begin{chapter}{Introduction}
\label{intro}

\hspace{0.5 cm} Turbulence is one of the most intriguing and unsolved phenomena in classical physics. The strict definition is related to ``violent" or unsteady movement of a given fluid. This definition agrees with our visual evidence of turbulent flows, e.g., a wave breaking on the beach or the chaotic movement that a plane feels with heavy accelerations when passing through a turbulent region.

From an academic point of view, however, the classification and understanding shall go way deeper, and it does since turbulent flows already have some key aspects to be matched when simulated through some numerical approach or to be validated when measured in an experiment. These aspects are related to the dynamical similarity, which describes the turbulent intensity by an adimensional and universal parameter --the Reynolds number, which will be defined later in this text-- the statistical properties of the flow for a fixed Reynolds number, such as the turbulent kinetic energy, statistical moments of velocity fluctuations, not to mention many others.

The current theoretical understanding of turbulence has advanced in many frontiers, such as in homogeneous and isotropic or wall-bounded turbulence. Nonetheless, many gaps persist in the literature, with open questions that have direct applications, such as the development of facilities related to the energy sector, e.g., the mitigation of scale formation in the oil industry or the development of liquid-metal-cooled fusion reactors, both related to wall-bounded turbulence.

A related area to the above-mentioned applications is the so-called magnetohydrodynamic (MHD) flow, the flow of electrically conducting fluid in the presence of an external magnetic field. Magnetic fields are known for their effects of turbulent suppression, although their dynamics and a more broad statistical description of this mechanism are still poorly understood. 

This thesis aims to give a direct description of hydrodynamics and turbulence, coherent structures, MHD flows, and the Lattice Boltzmann Method (LBM), a numerical tool to simulate a broad range of multiphysics phenomena. The main results are related to the study of coherent structure (CS) dynamics through a statistical analysis of an experimental dataset of a turbulent pipe flow, the development of an improved numerical approach within the LBM to simulate MHD flows, and the simulation of a turbulent pipe flow through the LBM.

Chapter \ref{cap2}, starts motivating the turbulence phenomenon with a few empirical evidences and a description of the fundamental observables. The equation that describes the dynamics of viscous flows is derived, followed by the definition of dynamical similarity. The chapter ends with the definition of statistical moments, which will give some of the main observables important to the validation of wall-bounded turbulence.

The description of CS and MHD turbulence is shown in Chap.~\ref{cap3}. It starts with the definition of some empirical observations of the so-called hairpin vortex, then, we move to CS identification, where some of the main techniques used in the literature are mentioned, such as proper orthogonal decomposition, dimensionality reduction, and the dynamical systems point of view. MHD turbulence is presented by describing the induction equation and how it is coupled with the hydrodynamic part. Lastly, the quasi-static (QS) approximation --relevant to most industrial applications of MHD flows-- is briefly discussed.

The core of the LBM is described in Chap. \ref{cap4}, which starts by defining the macroscopic length scale and the Boltzmann equation.  The most simple collision operator, the single relaxation time (SRT), is presented, followed by the definition of the LBM. A Dirichlet boundary condition is described by the bounce-back approach and an interpolated version of the bounce-back for non-cartesian boundaries. The hydrodynamic part of the LBM is concluded with the explanation of the advanced central-moments collision operator. The chapter ends with the description of a SRT collision model for the magnetic induction equation.

The first part of the results of this thesis lies in Chap. \ref{cap5}, where the statistical analysis of turbulent pipe flows is presented. First, the experimental setup is explained briefly, followed by the validations of turbulent data. Secondly, the identification of CS is made for all the investigated data sets. Finally, the transition between the identified CS is investigated as a stochastic process, revealing a non-Markovian memory effect. This behavior was then replicated by a low-level Markovian model here developed.

The last part of the work is associated with the LBM results and is presented in Chap. \ref{cap6}. It starts with the development of a multiple relaxation time (MRT) collision operator for the LBM for the induction equation and a new boundary condition for the magnetic distributions, which is shown in Appendix \ref{apendicea}. Then, the algorithm is validated in different scenarios for MHD flows, with and without boundaries, in the QS approximation. The chapter ends with the results of turbulent pipe flow, performed with a MRT collision model for the hydrodynamic part, which is shown in detail in Appendix \ref{apendiceb}. 

The turbulent pipe flow simulated with the LBM is firstly analyzed through its instantaneous observations, then, the turbulent statistics are analyzed and compared with previous results in the literature. After all validations, a CS identification was performed in the turbulent data set, revealing a non-trivial memory effect regarding the force used to trigger the turbulent state. The stochastic mode transition was also analyzed, showing a Markovian behavior for the finely time-resolved data. The chapter ends with the indication that the experimentally observed behavior (non-Markovian) would be recovered for coarser time separation and with the length of the identified CSs in radii units.

\end{chapter}

\begin{chapter}{Elements of Hydrodynamics and Turbulence}
\label{cap2}

\hspace{5 mm} In order to have a well-constructed description of turbulence and its statistical properties \cite{Frisch}, we shall begin describing some basic hydrodynamic features \cite{Acheson}.

\section{Prelude to the Navier-Stokes Equations}

\subsection{Empirical Evidences}

\hspace{5 mm} The effects of fluid dynamics and turbulence appear on a broad range of scales, as when we prepare a hot cup of coffee, where we can clearly see the turbulent dynamics of the smoke with a typical scale of centimeters, as shown in Fig.~\ref{coffe}, up to climate evidence of vortex shedding by the dynamics of the airflow through an island with a typical scale of kilometers, as illustrated in Fig. \ref{islandvonkarman}.

\begin{figure}
    \centering
    \includegraphics[width=0.8\textwidth]{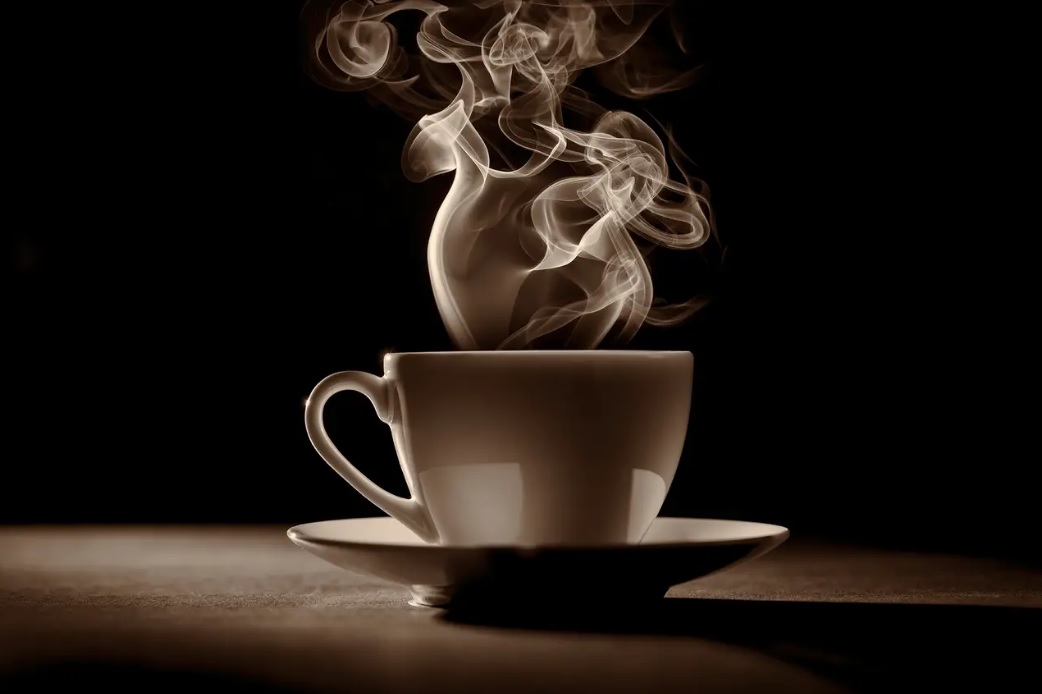}
    \caption{Turbulent vapor of a cup of coffe. Source: \href{https://expresso.pt/sociedade/lifestyle/2020-06-20-Ha-quanto-tempo-bebemos-cafe--Portugueses-experimentaram-no-no-seculo-XV-na-Etiopia}{Expresso}}
    \label{coffe}
\end{figure}

\begin{figure}
    \centering
        \includegraphics[width=0.8\textwidth]{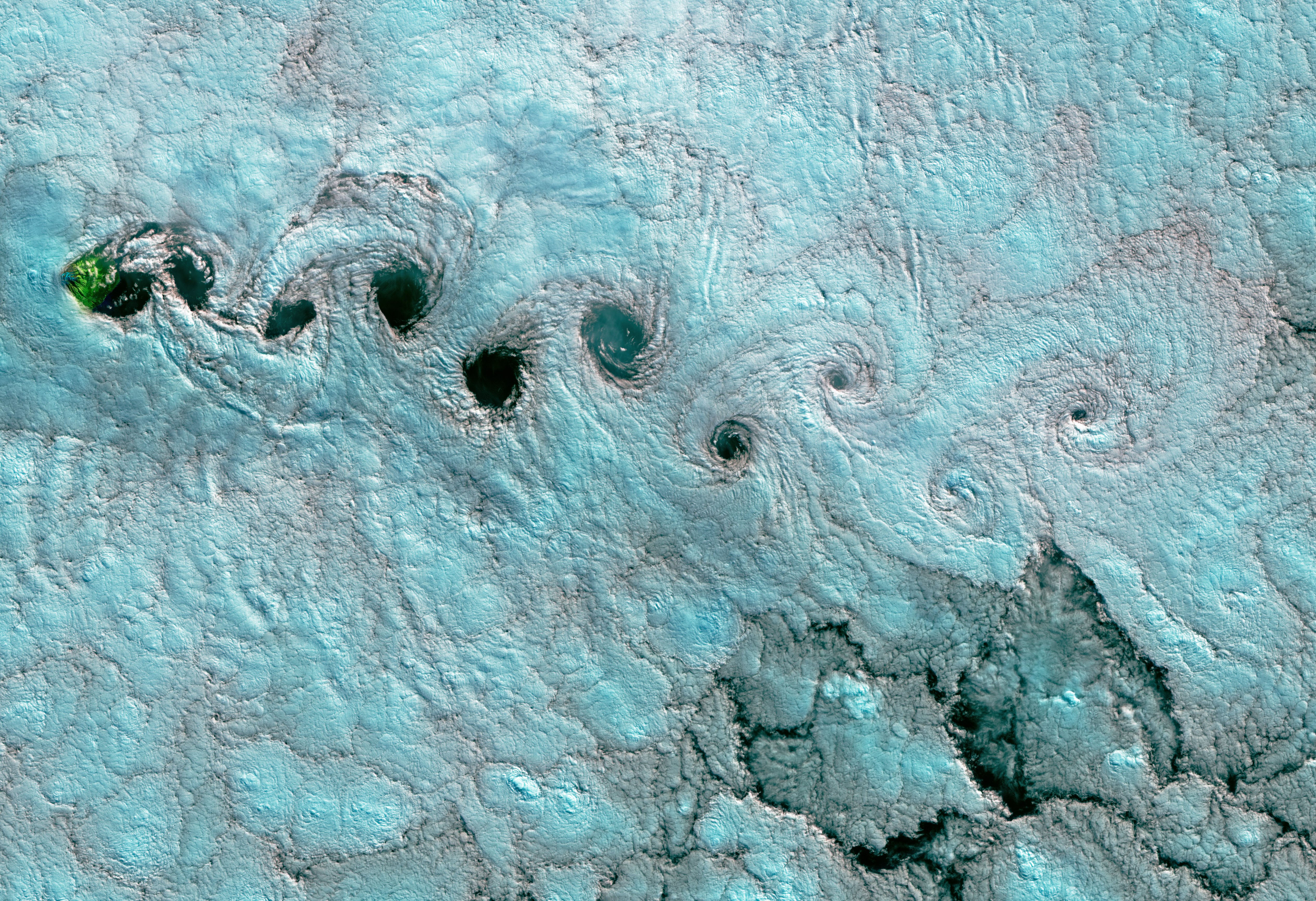}
    \caption{Von Kármán vortex shedding in Tristan da Cunha - South Atlantic. \\ Source: \href{https://earthobservatory.nasa.gov/images/90734/two-views-of-von-karman-vortices}{NASA earth observatory}}
    \label{islandvonkarman}
\end{figure}

The strict meaning of the word turbulence may not be so clear to the reader at this moment. To have a sharper definition, one can observe from experiments the two opposite regimes that the state of the flow may be associated with. Either the flow can be laminar, where its dynamics are usually well-behaved, or the flow can be turbulent, which in this case, the dynamics of the flow are more chaotic. Fig. \ref{laminarturbulent} illustrates those two opposite states and the transitional state in between them \cite{McGurk2020}. The precise parameter definition that determines in which regime the state of the flow may be, will be discussed shortly.

\begin{figure}
    \centering
        \includegraphics[width=1.0\textwidth]{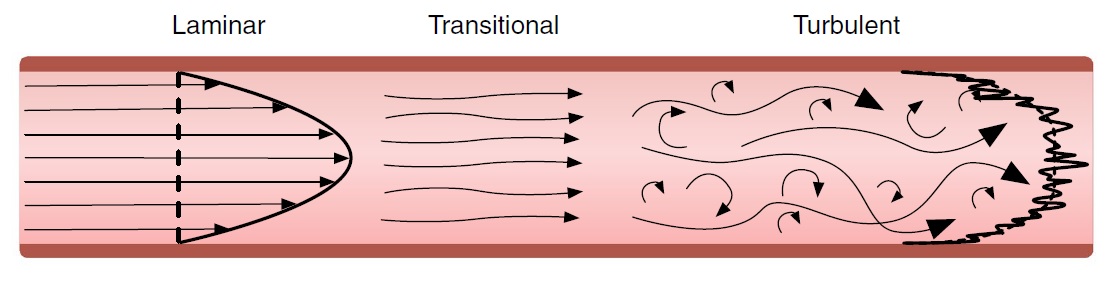}
    \caption{Laminar, transitional and turbulent states illustration [3].}
    \label{laminarturbulent}
\end{figure}

\subsection{Fundamental Observables and Operators}

\hspace{0.5 cm} To describe the dynamics of a flow, one could track the velocity field
\be
    \mbf{u} = \mbf{u}(\mbf{x},t) \ , \\
    \label{velocity}
\ee

\noindent which is a function of both the space $\mbf{x}$ and time $t$.

In general, we are interested in the dynamics of the velocity field, so it is interesting to look at the so-called material or convective derivative, given by
\be
    \frac{d\mbf{u}(\mbf{x},t)}{dt} = \frac{\partial \mbf{u}}{\partial t} + (\mbf{u} \cdot \nabla )\mbf{u} \ ,
    \label{materialderivative}
\ee

\noindent where the right-hand side (RHS) follows straightforwardly from the chain rule.
From now on, in order to make things simpler, we also would like to define a convenient notation for those operators, which will be given by
\be 
    \frac{d}{dt} \equiv d_t \ , \ \ \frac{\partial}{\partial t} \equiv \partial_t \ , \ \ \frac{\partial}{\partial x_i} \equiv \partial_i \ , \ \  \mbox{and} \ \  \mbf{a} \cdot \mbf{b} \equiv a_ib_i \ , 
\ee

\noindent where the last definition is simply the Einstein notation of summation over repeated indices. Using the above definitions, Eq.~(\ref{materialderivative}) now becomes
\be
    d_t u_i = \partial_t u_i + u_j \partial_j u_i \ .
\ee

A steady flow satisfies the condition 
\be 
    \partial_t \mbf{u} = 0 \ , 
\ee 

\noindent which means that the velocity field doesn't change explicitly on time. 

The circulation of the velocity field in a closed curve is defined as 
\be 
    \Gamma = \oint_C \mbf{u} \cdot d\mbf{l} \ ,
    \label{circulation}
\ee 

\noindent which measures how much the velocity field is parallel to the path of the curve $C$. If the closed curve $C$ can be spanned by a surface $S$, such that $C = \partial S$, then, one can use Stokes' theorem to write 
\be 
    \Gamma = \int_{S=\partial C} (\nabla \times \mbf{u}) \cdot d\mbf{S} \ ,
\ee 

\noindent in which we find the vorticity $\bs{\omega}$ 
\be 
    \bs{\omega} = \nabla \times \mbf{u} \ ,
\ee 

\noindent an obervable of great interest for turbulent flows which measures how much the flow is rotating at a given point. We expect to see coherent structures which will present those rotation patterns and may persist in the flow for some time. There are several applications where understanding the role played by the vortex dynamics is relevant. An example given in Fig. \ref{wing} shows a wing with its ``starting vortex" on its tail \cite{Acheson}.
\begin{figure}[h!]
    \centering
        \includegraphics[width=1.0\textwidth]{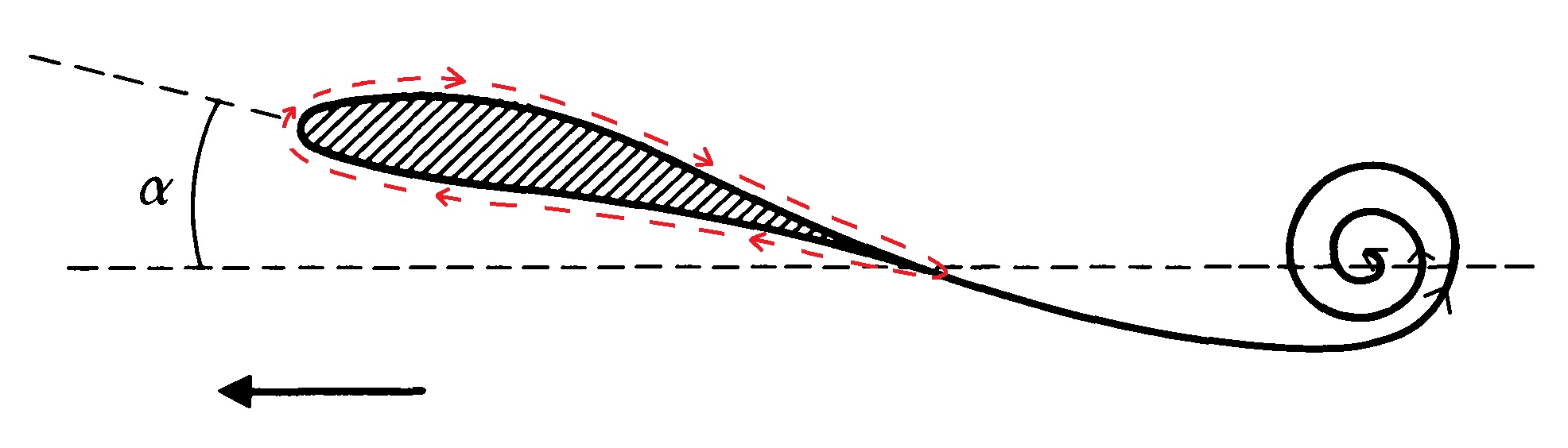}
    \caption{Starting vortex and angular momentum conservation. The flow comes from the right to the left. Adapted from [2].}
    \label{wing}
\end{figure}
Simple angular momentum conservation would imply a circulation in the opposite direction around the wing, which, by Bernoulli's principle, would generate a pressure difference with higher/lower pressure below/above the wing and, thus, generate the lift force. 

\section{The Navier-Stokes Equations}

\subsection{Elementary View on Viscous Flows}

\hspace{0.5 cm}Viscous flows experimentally display shear stress, which acts as a dissipative force on real fluids. Near to physical boundaries, it imposes a no-slip boundary condition leading to the creation of boundary layers which may change depending on the flow regime. Fig. \ref{boundary} illustrates the different boundary layers near a surface on a turbulent flow \cite{Robinson1990}. One can see that each layer has coherent structures of different shapes --with their creation as quasi-streamwise vortices at the buffer layer-- which will be mentioned later.

\begin{figure}[h!]
    \centering
        \includegraphics[width=0.6\textwidth]{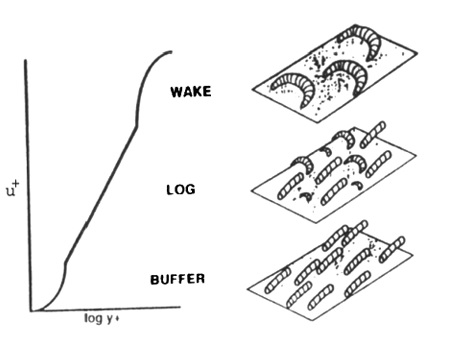}
        \caption{Idealized scheme of vortical structure populations in the different regions of the canonical boundary layer [4].}
    \label{boundary}
\end{figure}

For Newtonian fluids, the shear stress is proportional to the velocity gradient, such as
\be 
    \tau = \mu \frac{du}{dy} \ ,
    \label{shearforce}
\ee 

\noindent where $\mu$ is a fluid property, often referred to as dynamical viscosity. It measures the resistance of the fluid to deformation and its physical unit is Pascal-second (Pa.s) in the SI. To give an idea of real world values, honey can have values in the range $2.000-10.000$ while water $1$ mPa.s respectively. Therefore, under similar conditions, i.e., similar boundary conditions, characteristic length and external forcing, it's reasonable to expect that honey will move much slower than water.

\subsection{Stress Tensor and Cauchy's Equation of Motion}

\hspace{0.5 cm}To understand the dynamics of a viscous fluid element, one could consider the stress $t_i$ that this element has on its surface, which can be normal or tangential to the surface. In this case, it will be given by
\be 
    t_i = T_{ij}n_j \ , 
\ee 

\noindent where $T_{ij}$ is the stress tensor and $n_j$ is the normal vector to the surface of the fluid element. The stress tensor $T_{ij}$ in principle is defined by nine quantities
\be 
    T_{ij} = \begin{pmatrix}
                T_{11} & T_{12} & T_{13}\\
                T_{21} & T_{22} & T_{23}\\
                T_{31} & T_{32} & T_{33}
            \end{pmatrix} \ , 
\ee 

\noindent but due to Euler's \emph{principle of moment of momentum}, which says
\begin{displayquote}
    \emph{The total torque on a body about some fixed point is equal to the rate of change of the moment of momentum of the body about that same point.}
\end{displayquote}

\noindent we have that $T_{ij}=T_{ji}$, leading the stress tensor to six unknown quantities.

Taking this type of forcing into account, the Cauchy's equation of motion can easily be derived from Newton's second law together with the divergence theorem, and it will be given by
\be 
    \rho (\partial_t u_i + u_j\partial_j u_i) = \partial_j T_{ij} + \rho f_i\ , 
    \label{Cauchyequation}
\ee 

\noindent where $\rho$ is the fluid density and $f_i$ is an arbitrary body force. One can actually see that Eq.~(\ref{Cauchyequation}) is a simple equation of momentum conservation.

\subsection{Navier-Stokes Equations of Motion for an Incompressible Viscous Fluid}

\hspace{0.5 cm} Now, as a matter of interest, we describe incompressible fluids, which satisfies
\be 
    \nabla \cdot \mbf{u} = 0 \ ,
\ee 

\noindent known as the incompressibility condition. This condition also implies that the density inside a infinitesimal volume is constant.

The stress tensor can be decomposed in a diagonal contribution and a non-diagonal part. Stokes, assuming that the non-diagonal part should be linear with respect to velocity gradients, should vanish if the flow doesn't deform, and that the relation between its components are isotropic --since there is no a prior preferred direction, derived that
\be 
    T_{ij} = -p\delta_{ij} + \mu (\partial_i u_j + \partial_j u_i) \ ,
    \label{stresstensor}
\ee 

\noindent where the first term is associated with the perpendicular stress on the surface of the fluid element and $p = p(\mbf{x},t)$ is the so-called scalar pressure field. The non diagonal part of the stress tensor is related to the tangential shear stress, already mentioned in Eq.~(\ref{shearforce}).

Inserting Eq.~(\ref{stresstensor}) into Eq.~(\ref{Cauchyequation}) and making using of the incompressibility condition, is easy to show that
\bea 
    &&\partial_t u_i + u_j\partial_j u_i = -\partial_i p' + \nu \partial^2_j u_i +  f_i \ , \label{NS}\\
    && \partial_i u_i = 0 \ , \label{incompressibility}
\eea 

\noindent where $p' = p/\rho$ (since $\rho$ is constant, it can be passed inside the gradient) and $\nu = \mu/\rho$ is the kinematic viscosity. Eqs.~(\ref{NS}) and (\ref{incompressibility}) are the Navier-Stokes (NS) equations of motion for incompressible viscous flows. Currently, the Clay Mathematics Institute offers a 1 million dollar prize to anyone who could show that this equation has or not a unique and smooth solution for arbitrary smooth initial conditions and arbitrarily long times \cite{Fefferman2006}. This is one of the seven millenium prize problems in mathematics.

One can observe that Eq.~(\ref{NS}) is completely determined by the velocity field, since the pressure field satisfies
\be
    p'(\mbf{x},t) = -\nabla^{-2}\nabla \cdot [(\mbf{u}\cdot\nabla) \mbf{u}] \ ,
    \label{poissonP}
\ee

\noindent where $\nabla^{-2}$ is the inverse Laplace operator and it is assumed that external forces are divergence-free.

\subsection{Dynamical Similarity}

\hspace{0.5 cm}Predicting the dynamics of turbulent flows through the NS equations poses a challenging task because of its intricate nonlinear behavior. It was related to this unique problem that Osborne Reynolds, in his pioneering experimental work in 1883 (illustrated in Fig. \ref{Reynoldsdraw}) \cite{Reynolds1883}, clearly observed the existence of two well-defined flow states, as already mentioned, either the flow can be laminar or turbulent.

\begin{figure}[h!]
    \centering
        \includegraphics[width=0.7\textwidth]{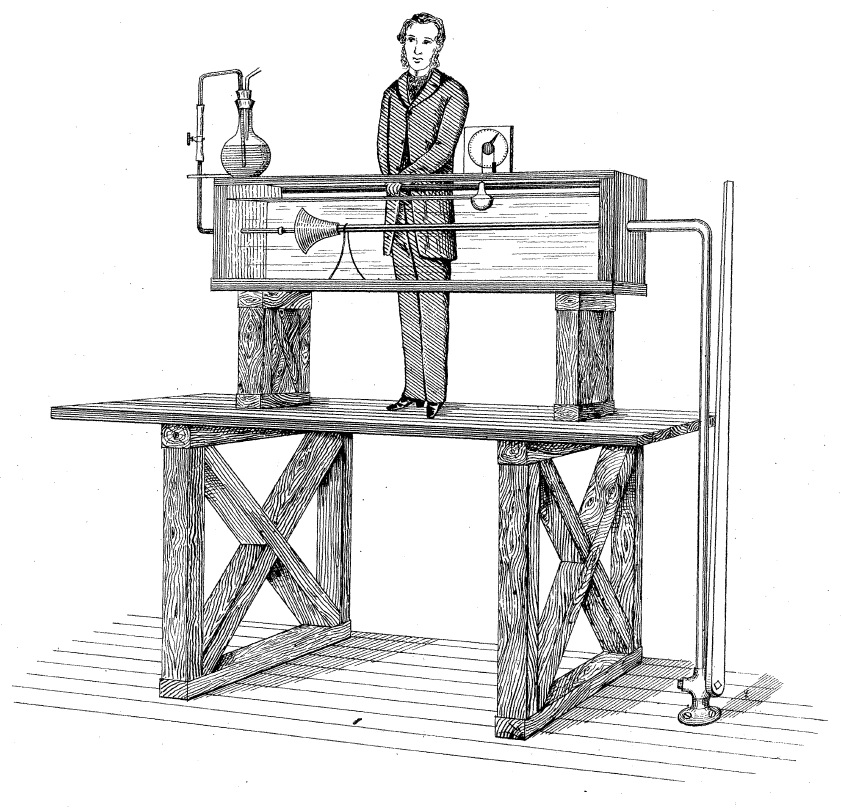}
        \caption{Illustration of Reynolds experimental apparatus [6].}
    \label{Reynoldsdraw}
\end{figure}

He also observed that the transition between those states always tended to happen for the same dimensionless combination of mean flow velocity $U$, characteristic length $L$, and kinematic viscosity $\nu$ as
\be 
    Re = \frac{UL}{\nu} \ ,
\ee 

\noindent which nowadays bears his name, the Reynolds number.

The fact that the dynamics of NS equations can be described by a single parameter is also known as dynamical similarity. It is easy to see from the NS that the Reynolds number appears explicitly in its dimensionless form. To do so, one can make the following transformations
\be
    \partial_t \to \frac{U}{L} \Tilde{\partial_t} \ , \ \ \partial_i \to \frac{1}{L} \Tilde{\partial_i} \ , \ \ u_i \to U \Tilde{u_i} \ , \ \mbox{and} \ p \to U^2\Tilde{p} \ ,
\ee

\noindent where all the tilde variables and operators are nondimensional.

Using the above transformations and assuming no external force, the NS equations in its nondimensional form reads
\bea 
    &&\Tilde{\partial_t} \Tilde{u_i} + \Tilde{u_j}\Tilde{\partial_j} \Tilde{u_i} = -\Tilde{\partial_i} \Tilde{p} +  \frac{\Tilde{\partial}^2_j \Tilde{u_i}}{Re} \ , \\
    && \Tilde{\partial_i} \Tilde{u_i} = 0 \ , 
\eea 

\noindent where we made $p' = p$ for simplicity.

\subsection{Statistical Moments}

\hspace{0.5 cm}An important quantity which will be analysed through this thesis is the velocity fluctuation
\be
    \delta \mbf{u}(\mbf{x},t) \equiv \mbf{u}(\mbf{x},t) - \langle \mbf{u}(\mbf{x},t) \rangle_t \ ,
\ee

\noindent where $\langle ... \rangle_t$ denotes a time average, which, applied to an arbitrary observable $\mathcal{O}(\mbf{x},t)$, should return a time independent function $\mathcal{F}$, such as
\be
    \langle \mathcal{O}(\mbf{x},t) \rangle_t = \mathcal{F}(\mbf{x}) = \frac{1}{T}\sum^T_{t=0}\mathcal{O}(\mbf{x},t) \ ,
    \label{timeaverage}
\ee

\noindent where $T$ is the size of the samples. The aforementioned average and the subsequent ones, only make sense in a statisticaly stationary regime. The motivation to look at velocity fluctuations is that they can provide important information about turbulent flows, such as turbulent kinetic energy, amount of anisotropy, information about coherent structures, etc. The root-mean-square (\emph{rms}) velocity is defined by
\be
    \mbf{u}_{rms}(\mbf{x}) \equiv \sqrt{\langle \delta\mbf{u}(\mbf{x},t)^2 \rangle_t} \ ,
    \label{urmsformula}
\ee

\begin{figure}[h!]
    \centering
    \includegraphics[width=.47\linewidth]{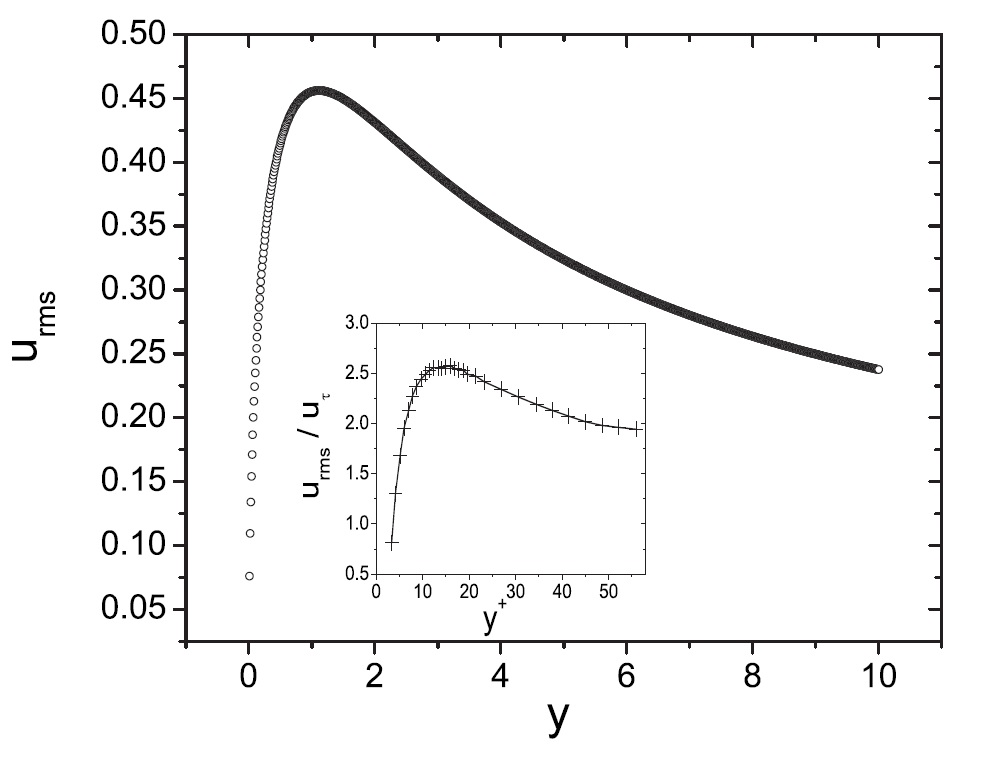}
    \label{urmsluca}
\end{figure}

\vspace{-1.0 cm}

\begin{figure}[h!]
\centering
\begin{minipage}{.5\textwidth}
  \centering
  \includegraphics[width=0.94\linewidth]{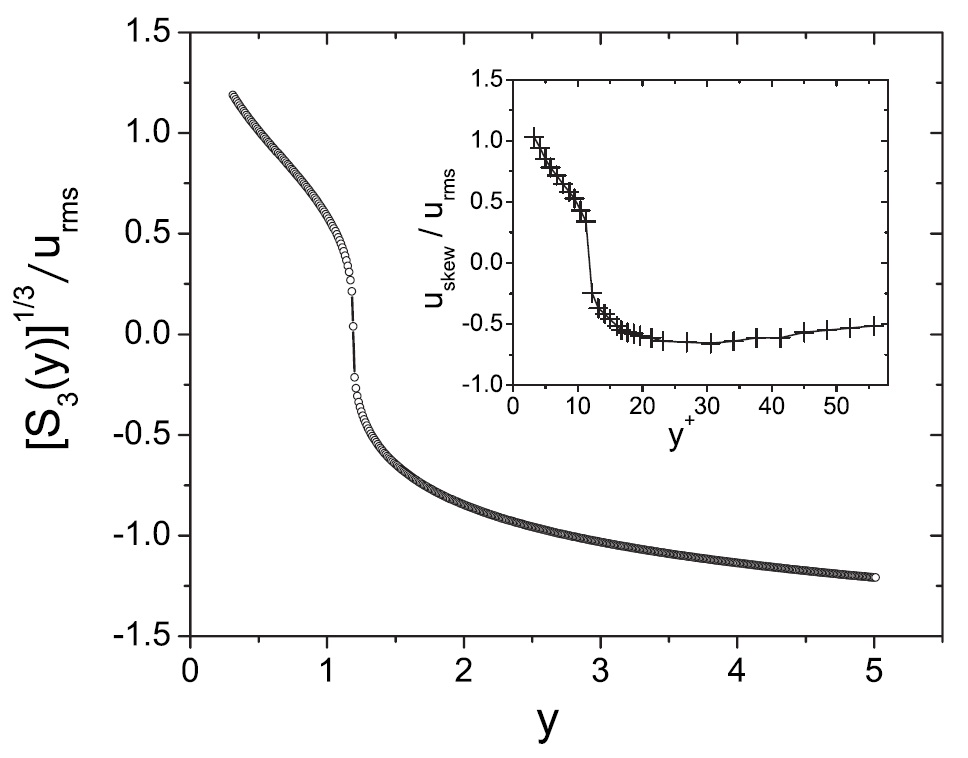}
  \label{skewnessluca}
\end{minipage}%
\begin{minipage}{.5\textwidth}
  \centering
  \includegraphics[width=0.94\linewidth]{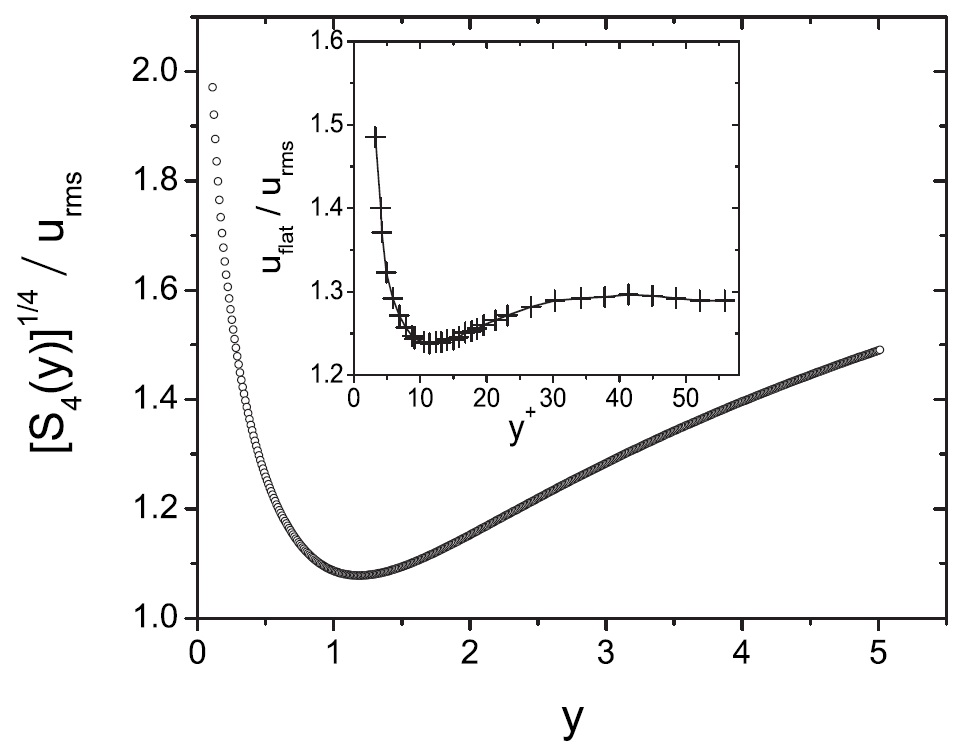}
  \label{flatnessluca}
\end{minipage}
\caption{Modelled structure functions in wall-bounded turbulence as a function of the wall distance in qualitative agreement with observed experimental data. Top: root-mean-square velocity, bottom left: skewness, and bottom right: flatness. Insets shows experimental data [9].}
\label{statisticalmoments}
\end{figure}

\noindent which can be sought as a first measure of turbulence intensity. The statistical moment of order $n$ is defined by 
\be
    S_n(\mbf{x}) \equiv \frac{\langle \delta \mbf{u}(\mbf{x},t)^n \rangle_t}{\mbf{u}_{rms}(\mbf{x})^n} \ , 
    \label{statisticalmomentsn}
\ee

\noindent the skewness $S_3$ measures the amount of velocity fluctuations events of same sign, while the flatness $S_4$ has the capability to capture isolated extreme events \cite{Loulou1997}. For Gaussian distributed velocity fields, $S_3(\mbf{x}) = 0$ and $S_4(\mbf{x}) = 3$.

In particular, all those observables are fundamental in wall-bounded (WB) turbulence, which is the main focus of this dissertation. The expected shape for those statistical moments were already predicted in a model with an ensemble of analytical and randomly distributed Lamb-Oseen vortices (see Fig. \ref{statisticalmoments}) \cite{Moriconi2009}, which agrees with experimental measurements \cite{Lorkowski1997}.

\end{chapter}

\begin{chapter}{Coherent Structures and Magnetohydrodynamic Turbulence}
\label{cap3}

\hspace{5 mm} In this chapter we make an overview regarding the literature on coherent structures and magnetohydrodynamic turbulence.

\section{Coherent Structures}

\hspace{0.5 cm}Coherent Structures have been a matter of study for a long time, both from a fundamental and an applied point of view. Those CSs, which can be defined and identified in several ways, can persist for a reasonable period of time and usually carry relevant information about the flow. As already mentioned, in WB turbulence, they can give a reasonable hint of the shape of statistical moments \cite{Moriconi2009}.

\subsection{Hairpin vortex}

\hspace{0.5 cm}In WB turbulence those CS can appear in many ways, usually depending on the boundary layer region, as illustrated in Fig. \ref{boundary}. They can be shaped as the so-called hairpin vortex, or as a quasi-streamwise vortex, among other types of structures. Their existence was verified experimentally with the use of the particle image velocimetry (PIV) technique \cite{Adrian2000}. In Fig. \ref{hairpin} we illustrate the structure of a typical hairpin vortex. One can note that the structure of a hairpin vortex also has quasi-streamwise vortices on its ``legs" and that the vortex topology induces a backward flow, i.e, contrary to the mean flow direction. These regions of backward flow between the legs are responsible for negative streamwise velocity fluctuations by lifting slow fluid parcels close to the wall. These regions are also known as low-speed streaks.

\begin{figure}[h!]
    \centering
        \includegraphics[width=0.8\textwidth]{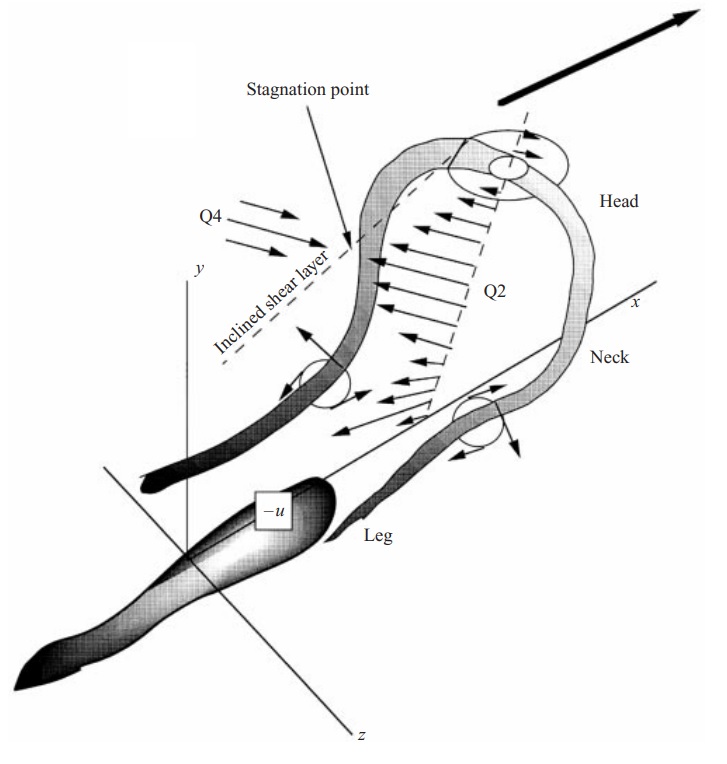}
        \caption{Schematic representation of a hairpin vortex [10].}
    \label{hairpin}
\end{figure}

\newpage

It is also known that hairpin vortices usually come in packets, which propagate together and have a self generation mechanism. They can start close to the wall, defining small regions of low-speed streaks, and evolve to a much larger structure as they detach from the wall and pass through the different regions of the boundary layer. Fig. \ref{hairpinpacket} show three different type of packets in different boundary layer regions. 

\begin{figure}[h!]
    \centering
        \includegraphics[width=0.8\textwidth]{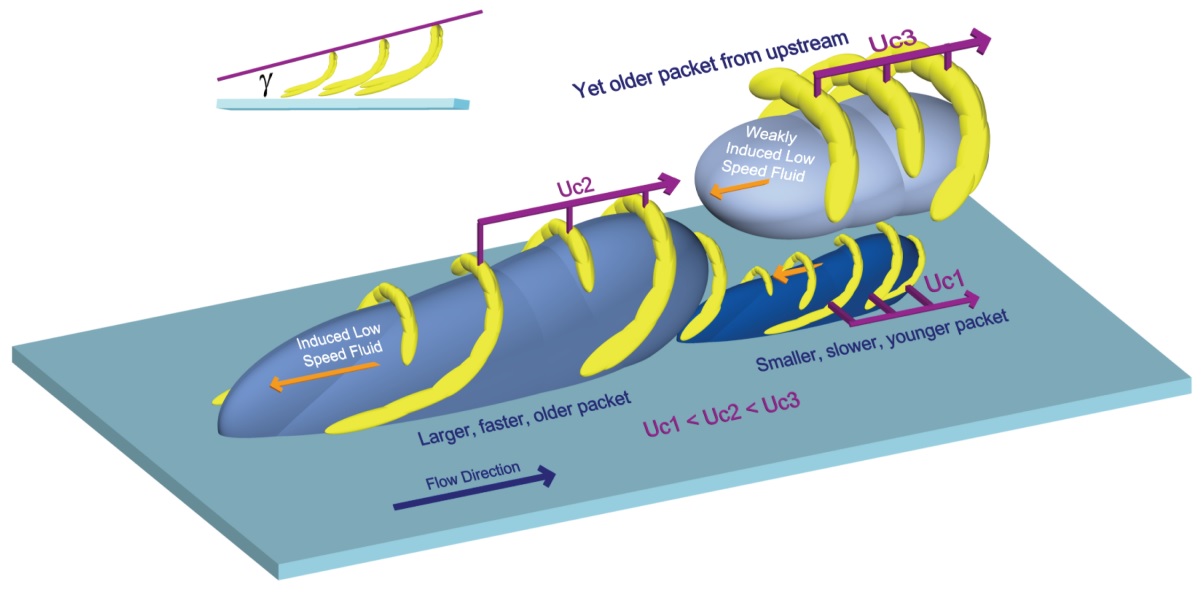}
        \caption{Illustration of different hairpin packets [10].}
    \label{hairpinpacket}
\end{figure}

\subsection{CS Identification}

\hspace{0.5 cm}There are several ways to identify a coherent structure, as by the vorticity, the $\lambda_{2}$-criterion -- defined by the imaginary part of the complex eigenvalue of the antisymmetric part velocity gradient tensor \cite{Jeong1995}, the $\lambda_{\omega}$-criterion -- defined from local properties of the vorticity field \cite{Elsas2017}, the proper orthogonal decomposition (POD) -- based on the average most energetic modes, and by dimensionality reduction -- based on the instantaneous most streamwise energetic mode -- among many others. Here we are going to make a brief review on these last two.

\subsubsection{Proper Orthogonal Decomposition}

\hspace{0.5 cm}The POD method \cite{Aubry1988,Berkooz1993} can be used to identify coherent structures which carry the most energetic contributions to the turbulent kinetic energy. In a nutshell, the method consists in solving the eigenvalue equation
\be 
    \int_{x'} R_{ij}(\mbf{x},\mbf{x}')\phi_j(\mbf{x}')d^3x' = \lambda\phi_i(\mbf{x}) \ ,
\ee 

\noindent where $R_{ij}$ is the time average of the velocity selfcorrelation matrix and $\phi_i$ is the eigenfunction which maximizes the projection of the velocity field. Using the aforementioned eigenfunctions, one can expand the velocity field as
\be 
    u_i(\mbf{x}) = \sum_{n=1}^{\infty} a^{(n)}\phi_i^{(n)}(\mbf{x}) \ .
\ee 

The eigenfunctions are normalized and the time average of the coefficients are the eigenvalues
\be 
    \int_{x}\bs{\phi}^{(n)}(\mbf{x})\bs{\phi}^{*(m)}(\mbf{x})d^3x = \delta_{nm} \ \mbox{ and } \ \langle a^{(n)}a^{(m)} \rangle_t = \delta_{nm}\lambda^{(n)} \ .
\ee 

Expanding the time average of the selfcorrelation matrix, it is trivial to show that
\be 
    R_{ij}(\mbf{x},\mbf{x}') = \langle u_i(\mbf{x})u^*_j(\mbf{x}') \rangle_t = \sum_{n=1}^{\infty} \lambda^{(n)} \phi_i^{(n)}(\mbf{x}) \phi_j^{*(n)}(\mbf{x}') \ .
\ee 

The turbulent kinect energy is obtained simply as the sum of the eigenvalues, as
\be 
    E = \int_{x} \langle u_i(\mbf{x}) u_i(\mbf{x}) \rangle_t d^3x = \sum_{n=1}^{\infty} \lambda^{(n)}  \ .
\ee 

Using the POD method, the reconstructed instantaneous vector fields have shown two pairs of counter-rotating vortices in WB turbulence with Heisenberg parameter $\alpha = 2$ \cite{Aubry1988}\footnote{The Heinsenberg parameter appears in an energy transfer model using POD to filter the unresolved small scale modes.}, as it can be seen in Fig. \ref{vecfieldaubry}.

\begin{figure}[h!]
    \centering
        \includegraphics[width=0.8\textwidth]{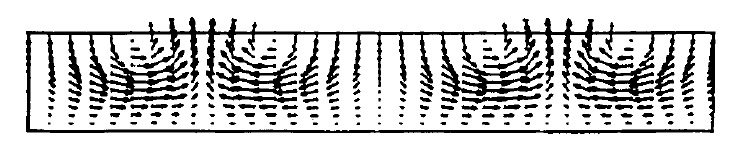}
        \caption{Reconstructed cross sectional vector field with POD and Heisenberg parameter $\alpha = 2$  [13].}
    \label{vecfieldaubry}
\end{figure}

Regarding turbulent pipe flow simulations, one has periodicity on the azimuthal direction and is usually common to assume streamwise periodicity as well. In this case, the velocity field can be further decomposed into azimuthal $(m)$ and streamwise mode numbers $(k)$. 
\begin{figure}[h!]
    \centering
        \includegraphics[width=0.8\textwidth]{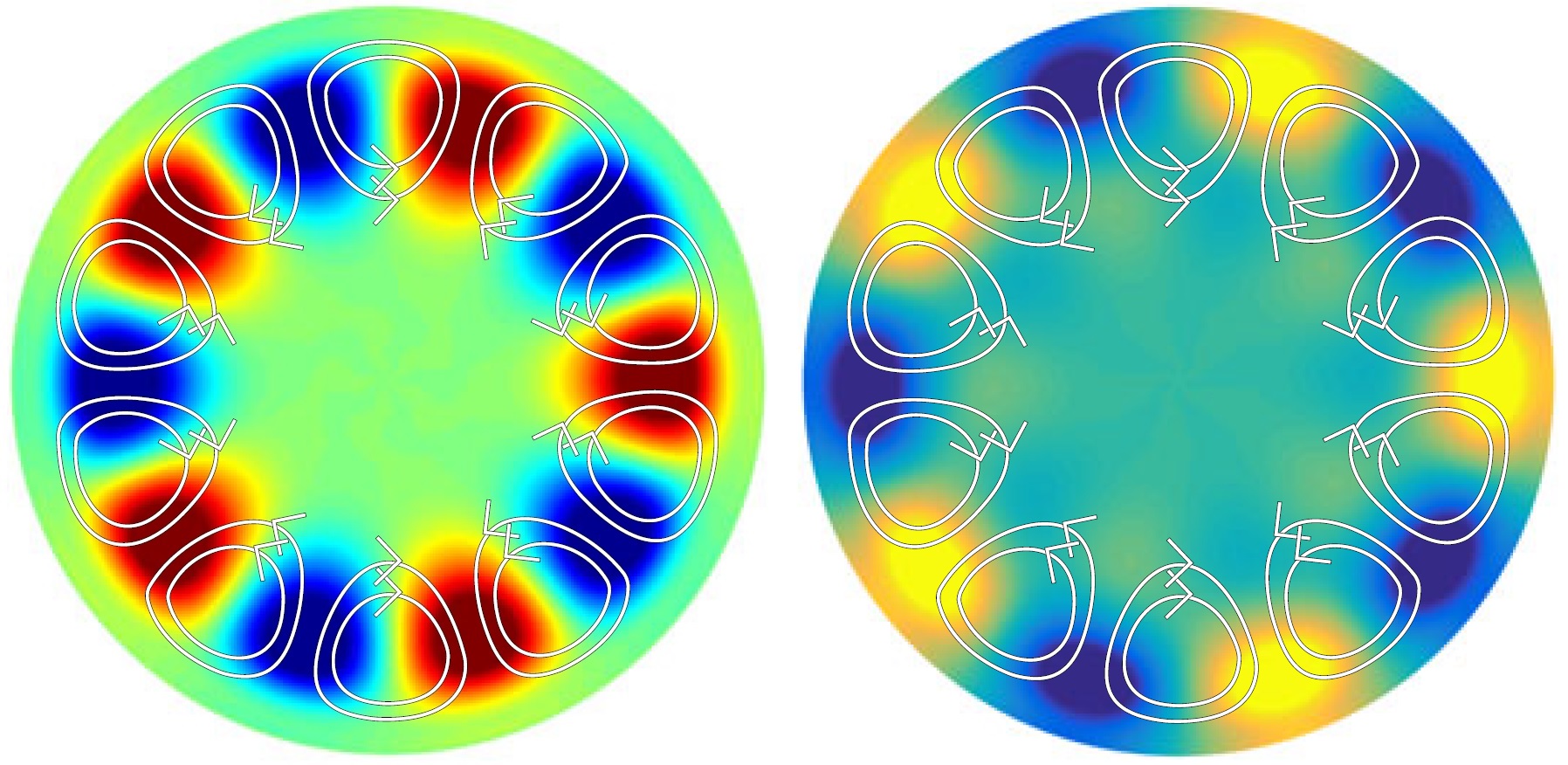}
        \caption{Reconstructed turbulente pipe flow fields from the POD method with azimuthal mode $m=5$ and $Re = 24580$. Left: fluctuation of the streamwise velocity field, right: pressure field [15].}
    \label{m5velocitypressure}
\end{figure}
Using the previously mentioned decomposition, one chooses to observe specific modes, which was done in direct numerical simulations (DNS) --simulations of the Navier-Stokes equations with the spatial part resolved up to the viscous lenght scale-- with $Re = 24580$ \cite{Hellstrom2017}, Fig. \ref{m5velocitypressure} shows the identified CS with azimuthal mode $m=5$ for the streamwise velocity (left) and pressure (right) fields. One can see the illustrated in-plane dynamics between the spots of negative and positive velocity fluctuations, similar to what is observed between the legs of a hairpin vortex. The relative and cumulative energy of those different azimuthal modes are shown in Fig. \ref{modesenergy} for different radial modes $\lambda^{(n)}$. For all radial modes, the energy seems to be concentrated approximately around the same azimuthal mode $m = 3$.

\begin{figure}[h!]
    \centering
        \includegraphics[width=0.8\textwidth]{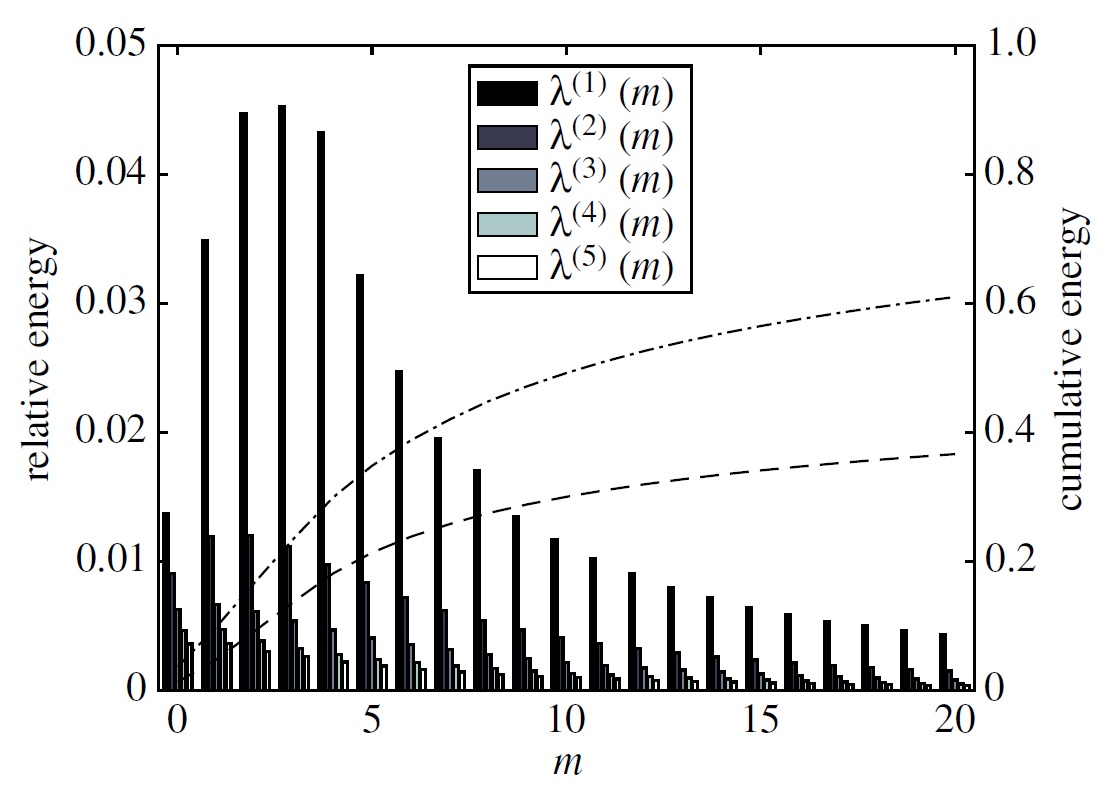}
        \caption{Relative and cumulative energy in turbulente pipe flow ($Re = 24580$) for the first 5 and 20 radial and azimuthal modes respectively [15].}
    \label{modesenergy}
\end{figure}

\subsubsection{Dimensionality Reduction}

\hspace{0.5 cm}The detection of CS by dimensionality reduction lies in the characterization by its symmetry patterns. In pipe flows, this can be achieved by taking advantage of the rotational symmetry along the pipe axis, which implies that those structures should have the same pattern when they are related by a global rotation \cite{Schneider2007}. Using the aforementioned idea, one can define the azimuthal correlation function $C$ of the streamwise velocity field $u_z$, depending on angular increments $\phi$ by
\be 
    C(\phi) = \langle u_z(r_0,\theta_0,z_0) u_z(r_0,\theta_0 +  \phi, z_0) \rangle_{\theta_0} \ ,
    \label{oldcorrelation}
\ee 

\noindent where the zero subscripts are related to the reference point for the correlation function. In order to have an automated procedure to identify the instantaneous structures with $m$-fold symmetry, one could define the instantaneous quantity $Z_m(t)$ as 
\be 
    Z_m(t) = \int_{-\pi}^{\pi} \partial_{\phi}C(\phi) sin(m \phi)d \phi \ ,
\ee 

\noindent which will be peaked when the corresponding symmetry mode of the streamwise velocity field coincides with the chosen $m$. The corresponding snapshot is ``assigned" to a specific state with symmetry $m$ if $Z_m$ is above a threshold which should account for the residual background turbulence. This was done numerically for transitional turbulence ($Re = \{2200, 2350, 2500\}$) \cite{Schneider2007} and the observed CS are shown in Fig. \ref{correlationsurface}, where one sees a $4$-fold structure evolve to a $6$-fold one.

\begin{figure}[h!]
    \centering
        \includegraphics[width=0.8\textwidth]{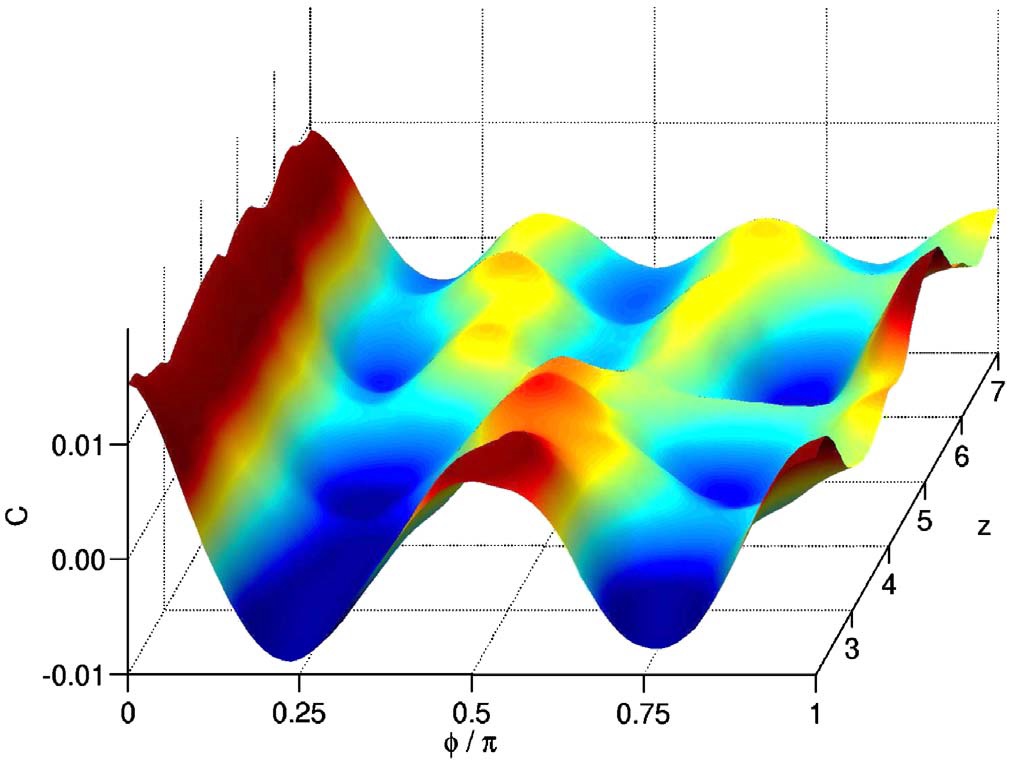}
        \caption{Azimuthal correlation surface [16].}
    \label{correlationsurface}
\end{figure}

The in-plane motions of a snapshot are shown in Fig. \ref{vecandprob} (left), the color map reprersents the streamwise velocity fluctuations, and the vector field represents the radial and azimuthal components of the velocity field. This snapshot exhibits a $4$-fold symmetry, which can be identified both from the pairs of counter-rotating vortices and low-speed streaks from the background. These type of CS are in consonance with those obtained from POD \cite{Hellstrom2017} and with the hairpin structures pattern observed experimentally \cite{Adrian2000}. 

\begin{figure}[h!]
    \centering
        \includegraphics[width=1.0\textwidth]{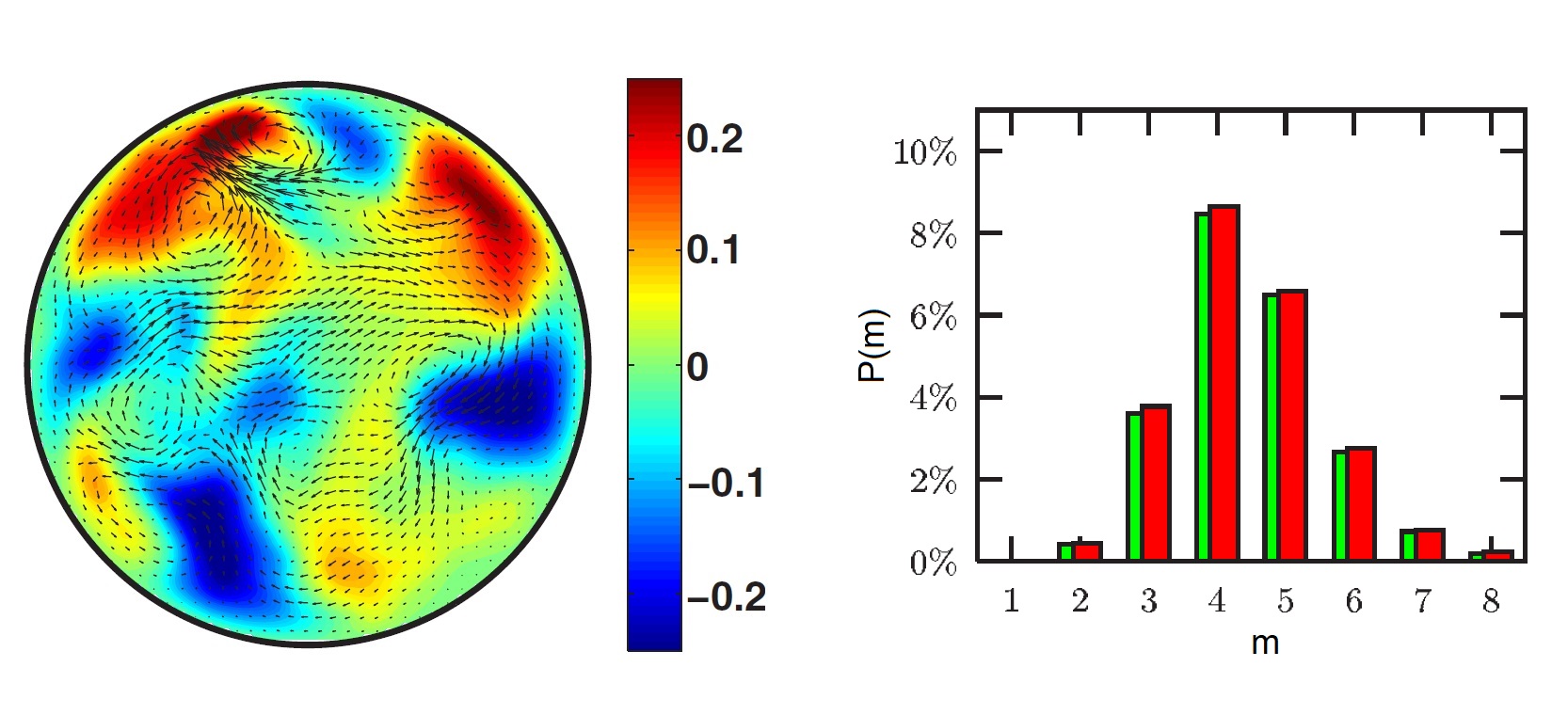}
        \caption{(Left) Instantaneous streamwise velocity fluctuation of a $4$-fold symmetry state with $Re = 2200$ [16]. (Right) Probability distribution of the observed symmetric CS. Red bars are calculed obtained from the simulations and green bars come from an asymptotic prediction from a Markov model among the transition between modes, adapted from [16]. }
    \label{vecandprob}
\end{figure}

One could also study the transition between the labeled structures in order to get more information about its dynamics. This was done in a Markovian fashion in \cite{Schneider2007}, assuming that the transition among the identified modes only depends on the previous mode. The empirical probabilities from the simulation and the modelled ones are shown in Fig. \ref{vecandprob} (right). A non-trivial remark is that the observed probabilities of the labeled symmetric states are peaked more or less in the same region as the energy plot as a function of the azimuthal mode obtained from POD in Fig. \ref{modesenergy}. This could be due to the fact that streamwise velocity fluctuations carry more turbulent kinetic energy than in-plane velocity fluctuations. However, this comparison directly from POD and dimensionality reduction is not so clear.

Experimentally, the identification of CS for highly turbulent flows ($Re = 35000$) \cite{Dennis2014} was done using a slightly different formalism than in \cite{Schneider2007}. First, the correlation function now uses the streamwise velocity fluctuation $\delta u_z$, which will be represented as $u_z$ for simplicity. In this case, one have the velocity-velocity correlation function as
\be 
    R_{uu}(r_0 + \Delta r,\Delta \theta) = \frac{\langle u_z(r_0,\theta_0) u_z(r_0 + \Delta r, \theta_0 + \Delta \theta) \rangle_{\theta_0}}{u^2_{z,rms}} \ ,
    \label{newcorrelation}
\ee 

\noindent where the zero subscripts also are related to the reference correlation point. The denominator is simply $u^2_{z,rms} = \langle u^2_{z,0} \rangle_{\theta_0}$ ($u_{z,0} = u_z(r_0,\theta_0)$), which normalizes the correlation function to unity in the reference point. Additionally, it also depends on the radial distance and increments, as opposed to Eq.~(\ref{oldcorrelation}).
\begin{figure}[h!]
    \centering
        \includegraphics[width=0.9\textwidth]{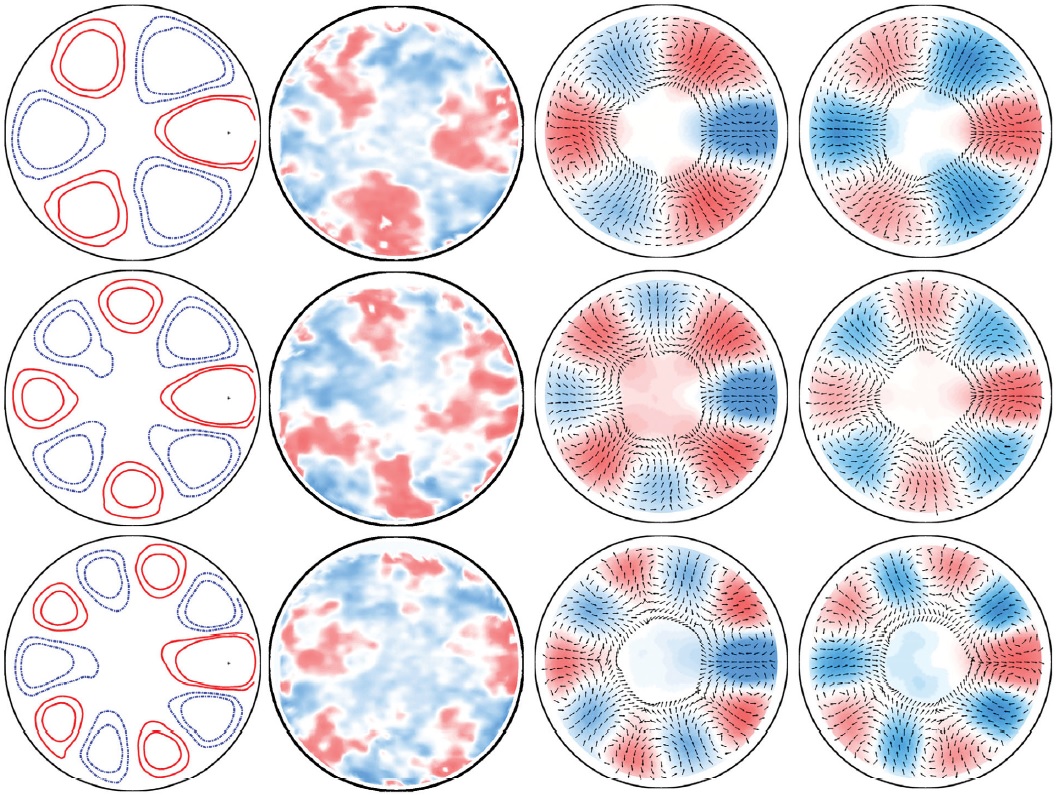}
        \caption{Modes $k_{\theta} = \{3,4,5\}$. From left to right, columns show: $R_{uu}$ correlation function, instantaneous streamwise velocity fluctuation, conditionally average velocity fields with $u_{z,0}<0$, and conditionally average velocity fields with $u_{z,0}>0$ [17]. }
    \label{csdennis}
\end{figure}
The final labeling of the symmetry states is achieved by selecting the modes of highest power in the azimuthal Fourier transform of Eq.~(\ref{newcorrelation}) for a fixed radius.

This approach culminates in a formalism of identification which is free of an arbitrary threshold and labels all the snapshots to a given dominant symmetric mode $k_{\theta}$. Some of the identified CSs are shown in Fig. \ref{csdennis}. It reveals, in principle, how the completely chaotic turbulent system actually exhibits organized and symmetric structures which compose its dynamics. Seven different dominant modes were observed ($k_{\theta} = \{1,...,7\}$) with the most observed modes being $k_{\theta} = \{3,4\}$, whose observed probabilities were respectively $P(k_{\theta} = 3) = 29\%$ and $P(k_{\theta} = 4) = 25\%$. The streamwise extent, in contrast, is significantly different, with the longest occurences observed with $k_{\theta} = \{3,4\}$ being $z_{max}(k_{\theta} = 3) = 4.6R$ and $z_{max}(k_{\theta} = 4) = 2.0R$. 

\begin{figure}[h!]
    \centering
        \includegraphics[width=0.9\textwidth]{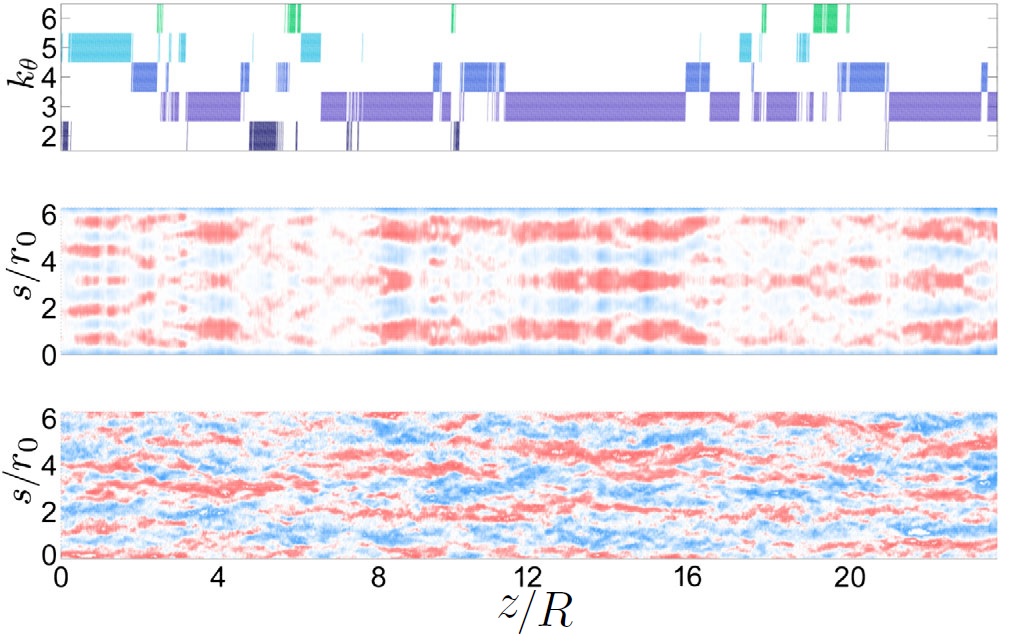}
        \caption{Streamwise extent, Top: dominant azimuthal wave number, middle: arc length of circunferent at the reference radius of the velocity-velocity correlation function, and bottom: instantaneous streamwise velocity fluctuations. Red/blue colors means positive/negative both for correlation and velocity fluctuations. Adapted from [17]. }
    \label{extentdennis}
\end{figure}

The streamwise extent of the CSs for a range of $20$ pipe radius can be seen in Fig. \ref{extentdennis}. From bottom to top, one can see how the dimensionality reduction method encapsulates the symmetry pattern observed in a turbulent pipe flow.

\subsection{Dynamical Systems Point of View}

\hspace{0.5 cm}Laminar pipe flows are known to be linearly stable to perturbations up to a high $Re$ regime \cite{Loulou1997,Avila2023} --experimental evidences could held laminar up to $Re = 10^5$ \cite{Pfenniger1961}, however its susceptibility to change its state from laminar to turbulent is $Re$ dependent.

In this way, Fig. \ref{DSPS} illustrates how the dynamical systems approach aims to understand whether those turbulent states can appear from the former laminar state and what would be the laminar-turbulent boundary in the phase space of solutions.

\begin{figure}[h!]
    \centering
        \includegraphics[width=0.7\textwidth]{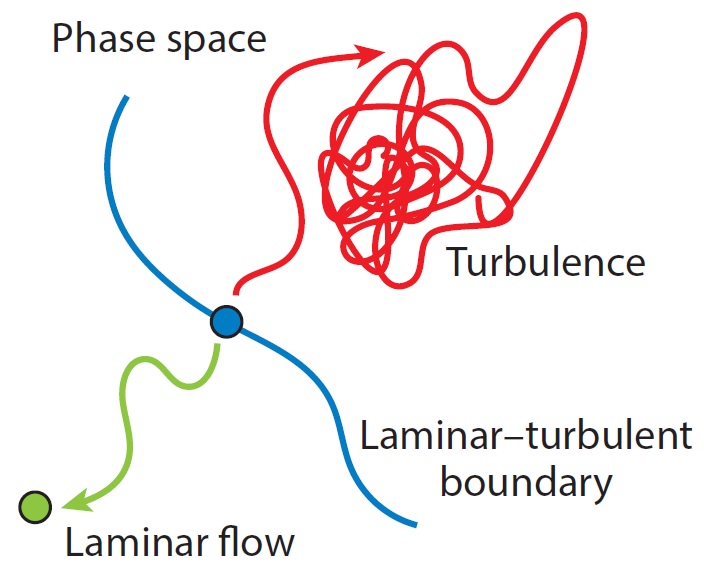}
        \caption{Perspective of NS solutions as a dynamical system. Adapted from [18]. }
    \label{DSPS}
\end{figure}

To make use of the dynamical system formalism, one can rewrite the NS equations as a dynamical system of infinite dimension, 
\be 
    \frac{d \mbf{u'}}{d t} = \mbf{f}_{Re}(\mbf{u'}) \ ,
    \label{DSeq}
\ee 

\noindent where $\mbf{u'} = \mbf{u} - \mbf{u}_{lam}$ is the velocity field perturbation with respect to the laminar solution and the right-hand side (RHS) represents the RHS of the NS equations. For pipe flow, the equilibrium of the DS is a laminar solution $\mbf{u'} = 0$, which satisfies $\mbf{f}_{Re}(0) = 0$. Turbulent fields also satisfy Eq.~(\ref{DSeq}), led by the chaotic attractor --an attracting point in the phase space of solutions which exhibit unpredictable but deterministic dynamics. This formalism also allows one to find travelling wave (TW) `relative' equilibrium solutions, which satisfy $\mbf{u'}(\mbf{x},t) = g(ct)\mbf{u'}(\mbf{x},0)$. The function $g(ct)$ does a spatial streamwise shift by $ct$, where $c$ is the streamwise velocity of the TW. Many of those TW solutions were found --numerically and in some cases, analytically-- in the transition to turbulence in pipe flow, and were used to understand the suppression of turbulence in heated pipes \cite{Marensi2021}.

To illustrate the structure of a TW solution, one can decompose the perturbed velocity field in the reference frame of the wave in three parts \cite{Wedin2004}
\begin{equation}
    \mbf{u'} = \begin{bmatrix} u_r(r,\theta) \\ u_{\theta}(r,\theta) \\ 0 \end{bmatrix}_{rolls}
                + \begin{bmatrix} 0 \\ 0 \\ u_z(r,\theta) \end{bmatrix}_{streaks}
                + \begin{bmatrix} \tilde{u}_r(r,\theta,z-ct) \\ \tilde{u}_{\theta}(r,\theta,z-ct) \\ \tilde{u}_z(r,\theta,z-ct) \end{bmatrix}_{waves} \ ,
\end{equation}

\noindent where the in-plane streamwise independent part of the velocity field is labeled as `roll' and the streamwise independent component of the velocity field in the flow direction is called `streak' --in the same sense as it was presented before.  The streamwise dependent parts of the velocity field are called `waves'.

Inserting the above decomposition into the NS equations and performing the scalar product of $\{ \hat{r}, \hat{\theta} \}$ with its streamwise average, one can neglect the nonlinear terms to obtain the analytical solutions for the streamwise rolls
\bea 
    && u_r(r,\theta) = [J_{m_0+1}(\lambda r) + J_{m_0-1}(\lambda r) - J_{m_0-1}(\lambda)r^{m_0-1}] cos(m_0 \theta ) \ , \label{ur}\\
    && u_{\theta}(r,\theta) = [J_{m_0+1}(\lambda r) - J_{m_0-1}(\lambda r) + J_{m_0-1}(\lambda)r^{m_0-1}] sin(m_0 \theta ) \ , \label{utheta}
\eea

\noindent where $m_0$ is a chosen azimuthal mode, $\lambda$ is associated with a decay rate for the rolls, $J$ is the Bessel function of the first kind and one must have the eigenvalue condition $J_{m_0+1}(\lambda) = 0$. The approximated eigenvalues for the azimuthal modes $m_0 = \{1,2,3,4,5,6\}$ -- assuming only one zero value for the radial functions -- are respectively given by $\lambda_{m_0} = \{5.14,6.38,7.59,8.77,9.94,11.09\}$.

To obtain the equation for the streaks, one must project the streamwise average of the NS equations onto the $\hat{z}$ direction, which -- assuming no wave contribution $\mbf{\tilde{u}}$ -- will give 
\be
    u_r\dfrac{\partial u_z}{\partial r} + \frac{u_{\theta}}{r}\frac{\partial u_z}{\partial \theta} - \frac{1}{Re^2}\nabla^2 u_z - 2ru_r = 0 \ .
    \label{streakeq}
\ee

The numerical solution of Eq.~(\ref{streakeq}) is shown as the contours in Fig. \ref{analinplane}, while the vector fields come from Eqs.~(\ref{ur}, \ref{utheta}). The two figures in the right column of Fig. \ref{analinplane} are really similar to those obtained in \cite{Wedin2004} --Figure 1 (a) and (b), although for $Re = \{1700, 1800\}$, in their case it seems that the full solution was displayed, including the wave part of the vector field.\footnote{This is from the author's perspective.} 

\begin{figure}[h!]
    \centering
        \includegraphics[width=1.0\textwidth]{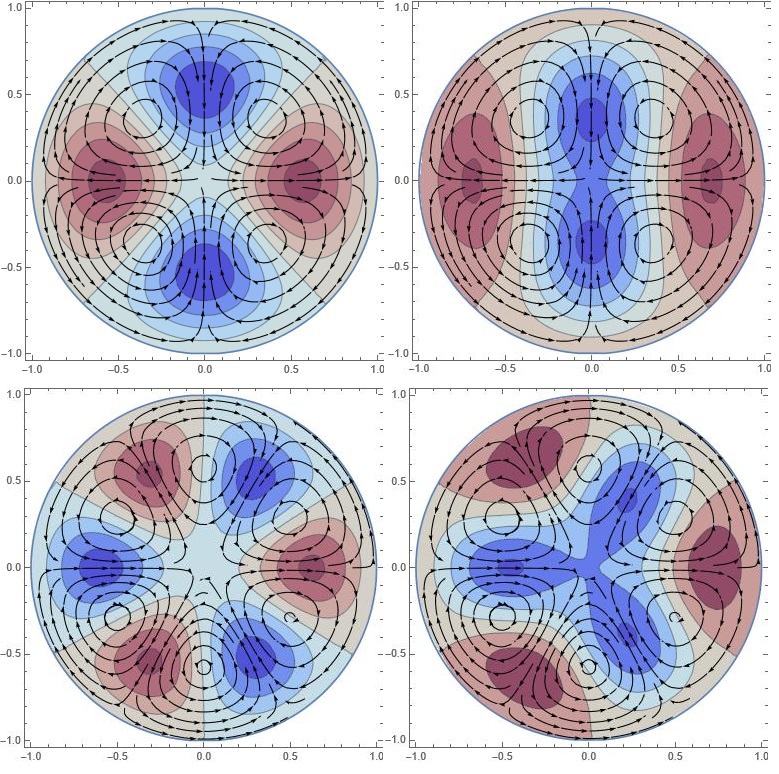}
        \caption{Analytical solution for in-plane vector fields together with numerical solution for streaks. Top left: $m_0 = 2$, $Re = 1$; Top right: $m_0 = 2$, $Re = 10$; Bottom left: $m_0 = 3$, $Re = 1$; And bottom right: $m_0 = 3$, $Re = 10$.}
    \label{analinplane}
\end{figure}

It is worth mentioning how these solutions agree with all the different previously mentioned CS, such as with the near-wall dynamics by a hairpin vortex in Fig. \ref{hairpin}, the structures observed by POD in Fig. \ref{m5velocitypressure}, and the ones observed by the dimensionality reduction mechanism in Fig. \ref{csdennis}.

\section{Magnetohydrodynamic Turbulence}

\hspace{0.5 cm}Magnetohydrodynamic (MHD) flows -- the flow of electrically conducting fluids in the presence of an applied magnetic field -- have many applications of both theoretical and practical interest. These applications are found in, but are not limited to, astrophysics, planetary magnetism, metallurgic casting, and nuclear fusion, among many others \cite{Davidson2001}.

\subsection{MHD Equations}

\hspace{0.5 cm }The MHD equations are a combination of the NS Eqs.~(\ref{NS} and \ref{incompressibility}) in the presence of the Lorentz body force $\mbf{F}_m$ with the induction equation. The latter can be obtained through simple manipulations of Maxwell's equations and Ohm's law, which may be written as \cite{knaepen2004}
\be 
    d_t \mbf{B} = (\mbf{B} \cdot \nabla) \mbf{u} + \eta \nabla^2 \mbf{B} \ , 
    \label{inductioneq}
\ee

\noindent where $\eta$ is the magnetic diffusivity/resistivity. Also, the magnetic field $\mbf{B}$ has to satisfy $\nabla \cdot \mbf{B} = 0$, which expresses the non existence of magnetic monopoles. The Lorentz force --making use of Ampère's law-- can be expressed as
\be
    \mbf{F}_m = \frac{1}{\mu_m} (\nabla \times \mbf{B}) \times \mbf{B} \ , 
    \label{LorentzForce}
\ee 

\noindent where $\mu_m$ is the magnetic permeability and is related to the magnetic resistivity and electric conductivity $\sigma$ by $\eta = 1/(\mu_m\sigma)$.

In this way, the full MHD system of equations can be summarized as 
\bea 
    && \partial_t \mbf{u} = -\nabla p -(\mbf{u} \cdot \nabla) \mbf{u} + \nu \nabla^2 \mbf{u} + \frac{1}{\mu_m\rho}(\nabla \times \mbf{B}) \times \mbf{B} \ , \label{NSE} \\
    && \partial_t \mbf{B} = -(\mbf{u} \cdot \nabla) \mbf{B} + (\mbf{B} \cdot \nabla) \mbf{u} + \eta \nabla^2 \mbf{B} \ , \label{IE} \\
    && \nabla \cdot \mbf{u} = \nabla \cdot \mbf{B} = 0 \ . \label{nablaUB}
\eea

\noindent One can note that the system of MHD equations is coupled, as Eq.~(\ref{NSE}) depends on $\mbf{B}$ and Eq.~(\ref{IE}) depends on $\mbf{u}$.

Analogously to the nondimensionalization of the NS, one could nondimensionalize Eq.~(\ref{IE}) and obtain the so-called magnetic Reynolds number 
\be
    Re_m = \frac{UL}{\eta} \ , 
\ee 

\noindent which measures the ration between convection and diffusion of $\mbf{B}$. Another relevant dimensionaless number is the Hartmann number, which measures the ratio between electromagnetic and viscous forces and can be written as
\be
    Ha = BL\sqrt{\frac{\sigma}{\mu}} \ ,
\ee 

\noindent where $B$ is a characteristic magnetic field intensity. A main feature of MHD flows is associated to the anisotropy imposed by the external magnetic field, which tends to dissipate vortices that are not aligned to it and can lead to a two-dimensional state of the flow, as illustrated in Fig. \ref{3dto2dMHD}.

\begin{figure}[h!]
    \centering
        \includegraphics[width=0.7\textwidth]{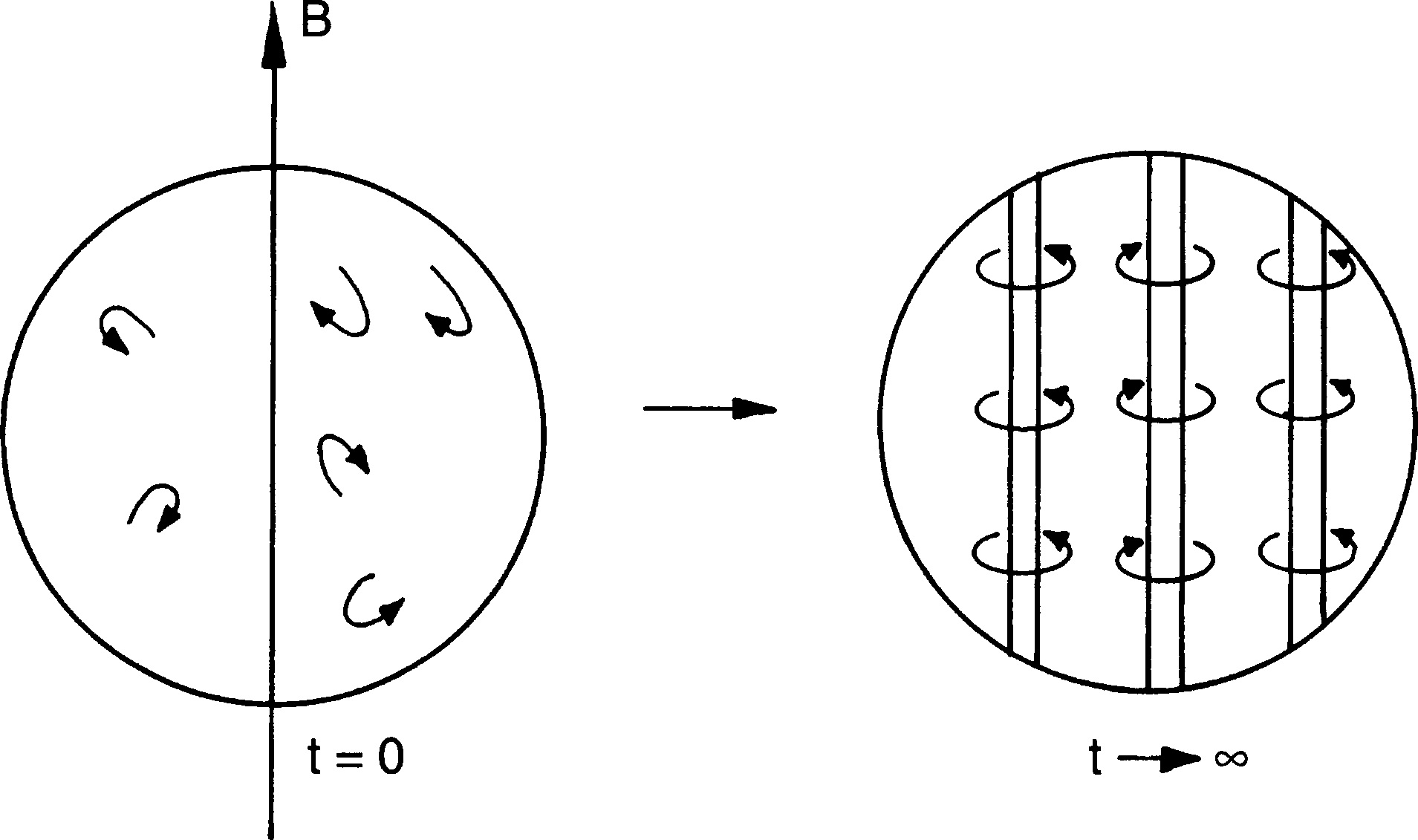}
        \caption{Illustration of isotropic turbulent structures evolving to a two-dimensional state under the influence of an imposed magnetic field [21].}
    \label{3dto2dMHD}
\end{figure}

\subsection{The Quasi-Static Approximation}

\hspace{0.5 cm}Many practical applications involving MHD flows are usually related to the quasi-static (QS) approximation, which means that the induced magnetic field is much smaller than the external one. This approximation is usually valid for low magnetic Reynolds number ($Re_m \ll 1$), and in this scenario one can split the total magnetic field as a superposition of its external applied part $\mbf{B}_{ext}$ and a small fluctuating part $\mbf{b}$ -- where small means that $\vert \mbf{b} \vert \ll \vert \mbf{B}_{ext} \vert$, so
\be 
    \mbf{B} = \mbf{B}_{ext} + \mbf{b}.
\ee 

Replacing the above mentioned splitting of $\mbf{B}$ in Eq.~(\ref{IE}) and assuming that velocity fluctuations are also small -- comparable to magnetic field fluctuations, it is not difficult to show that the QS approximation can be summarized as 
\be 
    \mathcal{O}\left [ (\mbf{u} \cdot \nabla) \mbf{b}  \right ] \approx \mathcal{O}\left [ (\mbf{b} \cdot \nabla) \mbf{u}  \right ] \ll \mathcal{O}\left [ \eta \nabla^2 \mbf{b}  \right ] \ , \mbox{and} \ \ \partial_t \mbf{b} \approx 0 \ .
\ee 

In this case, the induction equation for the induced magnetic field reduces to 
\be 
    \eta \nabla^2 \mbf{b} = (\mbf{u} \cdot \nabla) \mbf{B}_{ext} - (\mbf{B}_{ext} \cdot \nabla) \mbf{u} - \eta \nabla^2 \mbf{B}_{ext} \ ,
    \label{QSeq}
\ee 

\begin{figure}[h!]
    \centering
        \includegraphics[width=0.8\textwidth]{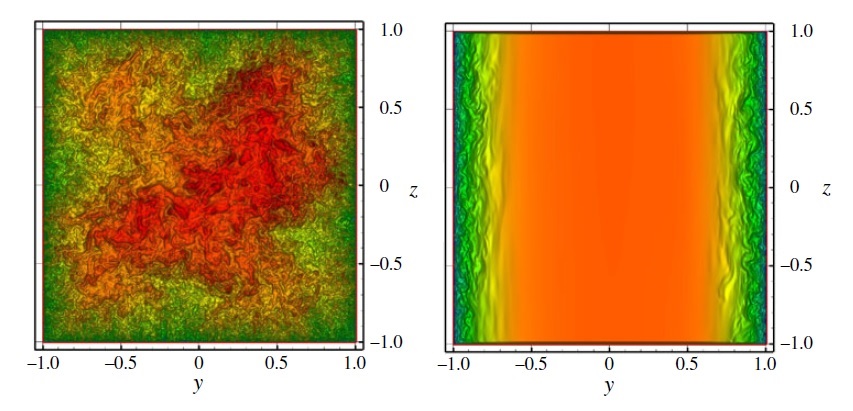}
        \caption{Instantaneous streamwise velocity of a turbulent duct flow at $Re = 10^5$. Colors are associated to velocity intensities that range from 0 (blue) to 1.25 (red) in arbitrary units. Left: $Ha = 0$. Right: $Ha = 300$ and magnetic field applied in $z$-direction . Adapted from [22].}
    \label{turbsupr}
\end{figure}

\noindent which is a Poisson equation for the fluctuation part of the induced magnetic field and, of course, $\nabla \cdot \mbf{b} = 0$.

DNS has demonstrated the dissipation of vortices at wall surfaces perpendicular to the magnetic field in duct flow \cite{Krasnov2012}, which is illustrated in Fig. \ref{turbsupr}. Furthermore, the alignment of the structures with the direction of the magnetic field is also displayed in Fig. \ref{vortexmhd} (Left). The same behavior was observed in turbulent pipe flows \cite{Krasnov2013} and is shown in Fig. \ref{vortexmhd} (Right). 

\begin{figure}[h!]
    \centering
        \includegraphics[width=1.0\textwidth]{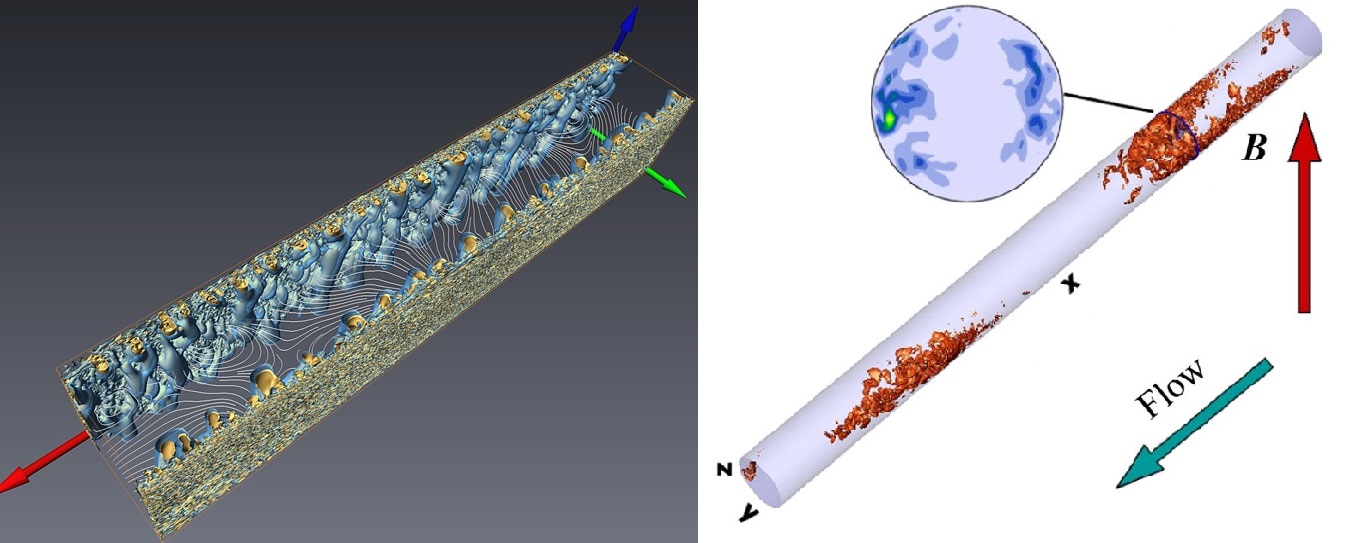}
        \caption{Turbulent structures in MHD flows. Left: Instantaneous isosurfaces of CS captured by the $\lambda_2$-criterion. The magnetic field direction is represented by the blue arrow, $Re = 10^5$, and $Ha = 400$ [22]. Right: Instantenous isosurfaces of the turbulent kinetic energy at $Re = 5000$ and $Ha = 22$ . Adapted from [23].}
    \label{vortexmhd}
\end{figure}

A non-trivial remark is related to the fact that the vortex dissipation due to the magnetic field application can also lead to complete relaminarization of the flow -- that is, a transition from chaotic fully developed turbulence to smooth and stable laminar flow \cite{Krasnov2013} -- depending on the magnetic field intensity. Currently, there is no specific criterion to describe in each regions of the $Re$-$Ha$ parameter space this relaminarization occurs, although there is some evidence that $R = (Ha/Re_{\tau})/\langle Ha/Re_{\tau} \rangle$ might be the controlling parameter for the laminar-turbulence transition \cite{Moriconi2020}.

\end{chapter}

\begin{chapter}{The Lattice Boltzmann Method}
\label{cap4}

\hspace{0.5 cm}A short description of the Lattice Boltzmann Method (LBM) is shown in this chapter, mentioning its basic concepts \cite{Kruger2017} up to a state-of-the-art branch of what has been made in the literature. Many benefits arise from the LBM, which will be made explicit in this chapter. To mention a few, the LBM is easy to parallelize, since the majority of its calculations are done locally, the application of non-trivial boundary conditions is relatively simple, which is one of the main problems related to usual computational fluid dynamics solvers, and the application of general body forces is done with simple modifications to the main algorithm. 

\section{The Boltzmann Equation}
\subsection{Background}

\hspace{0.5 cm}The dynamics of incompressible fluids, as already mentioned in Chap. \ref{cap2}, is described by the NS Eqs.~(\ref{NS} and \ref{incompressibility}). Due to its nonlinearity and nonlocality, the dynamics of this set of equations poses a challenging task, even for numerical simulations. Regarding this and many other difficulties, it seems important to have a good description of the relevant scales of the problem.

The NS equations, although derived from Newton's second law, have the outputs of macroscopic quantities -- which are important in a macroscopic scale $\ell$ and are bounded by the system size $\ell_S$ -- such as the fluid density $\rho(\mbf{x},t)$, the flow velocities $\mbf{u}(\mbf{x},t)$, and the pressure $p(\mbf{x},t)$ of a given fluid element. An important remark is that Newton's second law, in principle, is suitable for predicting the dynamics of individual point particles, which means that it is more appropriate to analyze what happens at a microscopic scale, which has the characteristic length of the atom $\ell_a$.

Looking at the system of multiple-point particles, another scale comes into play, which is the mean free path $\ell_{mfp}$, describing how long, on average, a particle can move until it collides with another particle. These different scales along with their corresponding timescales are shown in Fig. \ref{mesoscalepos}. The scale of the dynamics which is in between the microscopic and macroscopic scale is the so-called mesoscopic scale, which will be relevant to the description that will follow in this chapter.

\begin{figure}[h!]
    \centering
        \includegraphics[width=0.8\textwidth]{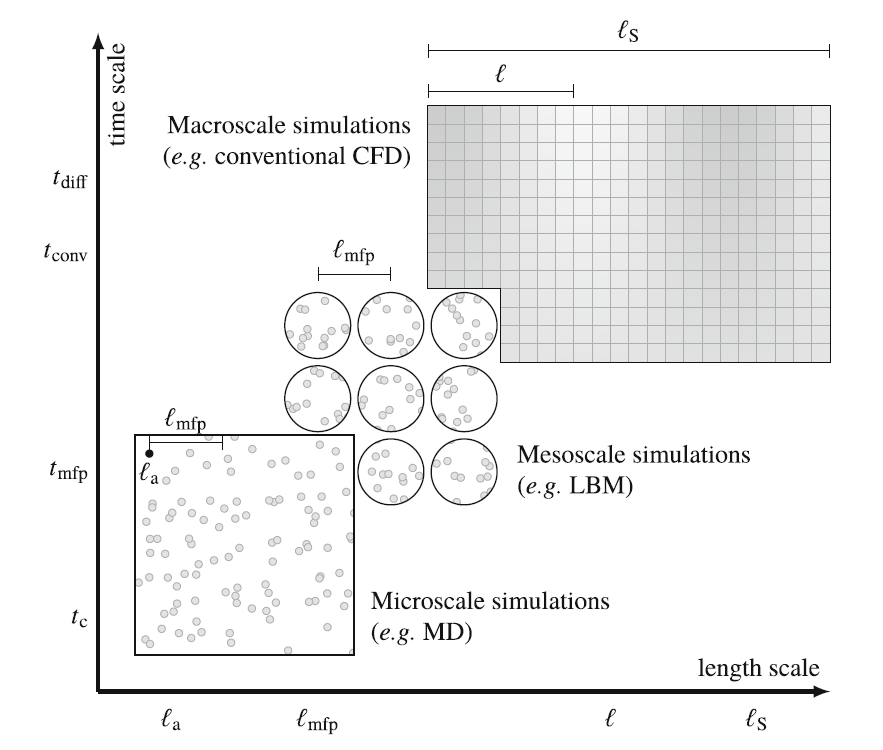}
        \caption{Time and length scales hierarchy for different fluid dynamics problems [25].}
    \label{mesoscalepos}
\end{figure}

In order to have a description of the macroscopic scale, one can define the distribution function $f(\mbf{x},\mbf{v},t)$, which gives, in an Eulerian fashion, the density of particles with velocity $\mbf{v}$ at position $\mbf{x}$ and time $t$. The time evolution of this distribution function, defined in the six dimensional $\mu$-space -- an averaged version of the usual 6N dimensional phase space in statistical mechanics -- can describe the dynamics in a macroscopic scale, and can be achieved by a simple application of the chain-rule, 
\be 
    \frac{df}{dt} = \frac{\partial f}{\partial t} + (\mbf{v}\cdot \nabla_{\mbf{x}}) f + \frac{1}{\rho}(\mbf{F} \cdot \nabla_{\mbf{v}}) f \ .
    \label{preboltz}
\ee 

In the absence of external forces, the distribution's zero and first order moments are respectively the mass density and fluid momentum
\bea 
    && \rho(\mbf{x},t) = \int f(\mbf{x},\mbf{v},t)d^3v \ , \\
    && \rho(\mbf{x},t)\mbf{u}(\mbf{x},t) = \int \mbf{v} f(\mbf{x},\mbf{v},t)d^3v \ ,
\eea  

\noindent which are some of the macroscopic quantities of interest. One of the main advantages of using this kind of macroscopic formalism to study fluid dynamic problems is that pressure comes from the ideal equation of state assuming that the fluid is isothermal, so $p(\mbf{x},t) = \rho(\mbf{x},t) R T$. The latter is a simple and local expression, contrary to its counterpart Eq.~(\ref{poissonP}), which is a non-local Poisson equation which has to be solved for each time $t$.

Asymptotically, this distribution function reaches an equilibrium distribution given by the Maxwell-Boltzmann distribution
\be 
    \lim_{t\to \infty} f(\mbf{x},\mbf{v},t) = f^{eq}(\mbf{v}) = \rho \left (\frac{1}{2\pi R T} \right )^{3/2} e^{-\mbf{v}^2/{2RT}} \ .
\ee 

To properly describe the physics in $\mu$-space, the left-hand side (LHS) of Eq.~(\ref{preboltz}) requires modeling, depending on the properties of the target system. This is usually related physically with collisions in this space, so that $df/dt = (\partial f/ \partial t)_{col} = \boldsymbol{L}(f)$. In this way, one can write
\be 
    \frac{\partial f}{\partial t} + (\mbf{v}\cdot \nabla_{\mbf{x}}) f + \frac{1}{\rho}(\mbf{F} \cdot \nabla_{\mbf{v}}) f = \boldsymbol{L}(f) \ ,
    \label{boltzmanneq}
\ee 

\noindent which is known as the Boltzmann equation. One can easily check that Eq.~(\ref{boltzmanneq}) conserves mass, momentum, and energy.

\subsection{The Elementary Collision Operator}

\hspace{0.5 cm}The most phenomenologically simple and widely used collision operator is the so-called \tbf{BGK}, which was developed by Bhatnagar, Gross, and Krook \cite{Bhatnagar1954},  and is given by 
\be 
    \boldsymbol{L}(f) = -\frac{1}{\tau}(f - f^{eq}) \ ,
    \label{BGK}
\ee 

\noindent where $\tau$ is called the relaxation time. This collision operator, which has the simple form of Hooke's law, tends to ``relax" the distribution $f$ towards the equilibrium distribution $f^{eq}$ according to a relaxation frequency $\omega = 1/\tau$.

\section{The Lattice Boltzmann Equation}

\subsection{From Boltzmann's Equation to the LBM}

\hspace{0.5 cm }The LBM can be obtained by a discretization of the velocities, space, and time on the Boltzmann equation. In this case the distributions can be written as $f(\mbf{x},\mbf{v},t) \to f(\mbf{x},\mbf{c}_i,t) = f_i(\mbf{x},t)$, where $\mbf{c}_i$ is the discretized velocity with index $i$, which will depend on the model and dimension.

In lattice units \footnote{From now one we are going to use lattice units in the rest of this thesis when it comes to the LBM part.} -- i.e. assuming that the spatial and time increments are unity ($\Delta x = \Delta t = 1$) -- is easy to show that the zero and first order moments can be easily calculated, respectively, as
\bea
    && \rho = \sum_i f_i \ , \\
    \label{rholbm}
    && \mbf{u} = \frac{1}{\rho} \sum_i \mbf{c}_i f_i + \frac{\mbf{F}}{2\rho} \ ,
    \label{vellbm}
\eea 

\noindent where the arguments were suppressed for simplicity.

The LBM then, can be summarized in the Lattice Boltzmann equation as 
\be 
    f_i(\mbf{x} + \mbf{c}_i,t+1) - f_i(\mbf{x},t) = \boldsymbol{L}[f_i(\mbf{x},t)] \ ,
    \label{LBE}
\ee 

\noindent where the LHS/RHS is called the streaming/collision step. The streaming step is simply propagating the distributions $f_i$ in the forward time $t+1$ to its neighbours at $\mbf{x}+\mbf{c}_i$.

The discretized Boltzmann equilibrium distribution on a reference frame with velocity $\mbf{u}$ is given by
\be 
    f_i^{eq}(\mbf{x},t) = w_i \rho \left( 1 + \frac{\mbf{u}\cdot\mbf{c}_i}{c_s^2} + \frac{(\mbf{u}\cdot\mbf{c}_i)^2}{2c_s^4} - \frac{\mbf{u}\cdot\mbf{u}}{2c_s^2}  \right ) \ ,
    \label{boltzeqsecondorder}
\ee 

\noindent where it was expanded in a Hermite polynomial base up to second order. The pre-factors $w_i$ are the so-called lattice weights, which depend on the dimension and model chosen, and $c_s$ is the speed of sound on the lattice. For most of the lattices and the rest of this dissertation we have $c_s = 1/\sqrt{3}$. Through the Chapman-Enskog expansions \cite{Chapman1952}, one can show that the recovered macroscopic behaviour from the LBM results is a solution of the NS equations with kinematic viscosity given by 
\be 
    \nu = c_s^2 \left ( \tau - \frac{1}{2} \right ) \ .
\ee 

\begin{figure}[h!]
    \centering
        \includegraphics[width=0.8\textwidth]{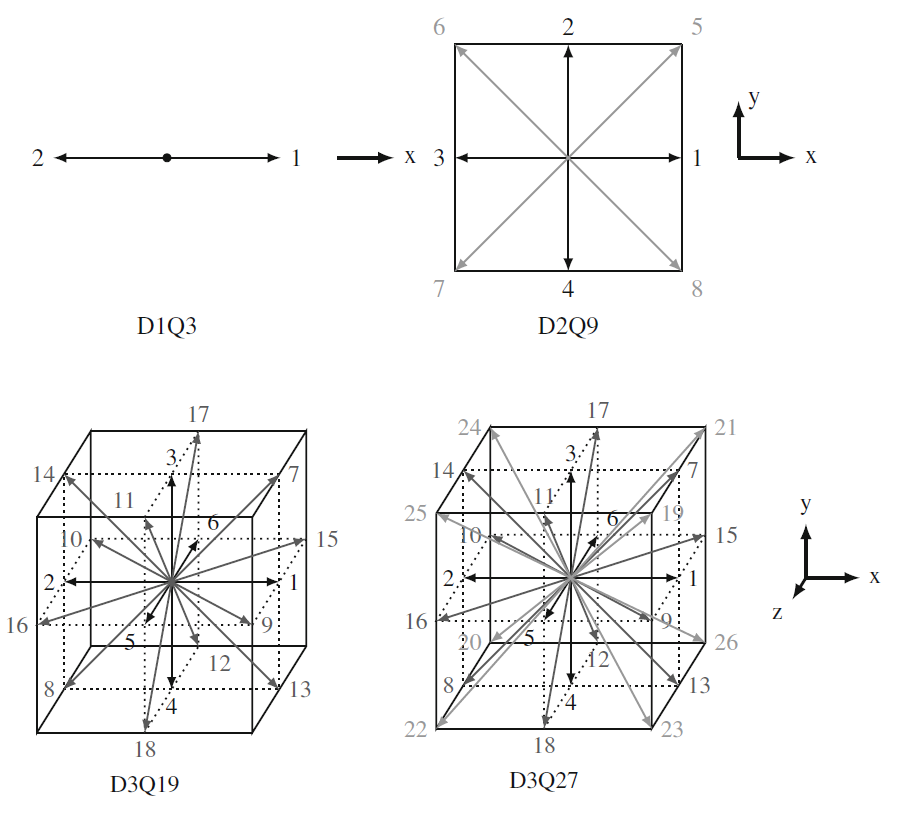}
        \caption{Different velocity sets for one, two, and three dimensional grids. The center point represent the zero-th index. Adapted from [25].}
    \label{velsets}
\end{figure}

The conditions for choosing the specific weights, $w_i$, are such that moments remain isotropic up to the fifth order, which can be summarized as 
\bea
    &&\sum_{i}w_i\left [1,c_{i\alpha}, c_{i\alpha}c_{i\beta}, c_{i\alpha}c_{i\beta}c_{i\gamma}, c_{i\alpha}c_{i\beta}c_{i\gamma}c_{i\mu}, c_{i\alpha}c_{i\beta}c_{i\gamma}c_{i\mu}c_{i\nu} \right ] = \\ 
    &&\hspace{0.6 cm} \left [1, 0, c_s^2\delta_{\alpha\beta}, 0, c_s^4(\delta_{\alpha\beta}\delta_{\gamma\mu} + \delta_{\alpha\gamma}\delta_{\beta\mu} + \delta_{\alpha\mu}\delta_{\beta\gamma}), 0\right ] \ . \nonumber
\eea

Some of the most common velocity sets are illustrated in Fig. \ref{velsets}, where d/q in the acronym DdQq represents the dimension/number of velocity vectors of the model.

\subsection{Boundary Conditions}

\hspace{0.5 cm}As mentioned in Chap. \ref{cap2}, viscous flows in the presence of physical boundaries are described by the NS equations together with the no-slip condition, which impose zero velocity at the walls. One of the main advantages from the LBM is related to its facility on the application of such boundary conditions (BC). Those can be straightforwardly applied through the so-called bounce back approach, illustrated in Fig. \ref{bounceback1}.

\begin{figure}[h!]
    \centering
        \includegraphics[width=0.9\textwidth]{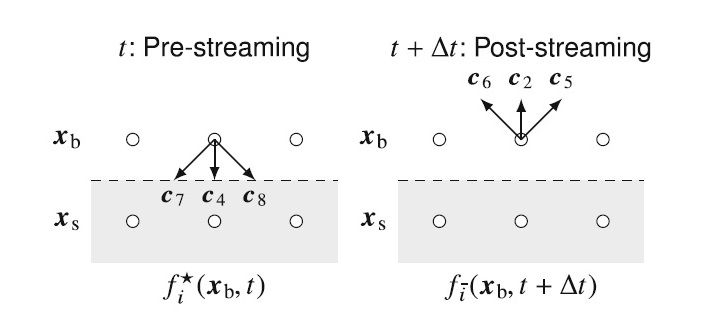}
        \caption{Visualization of the inversion of incoming distributions towards the wall in the D2Q9 scheme [25].}
    \label{bounceback1}
\end{figure}

The main idea behind the bounce back approach is to invert the incoming distributions towards the wall, in a way that the sum of all distributions on the boundary -- which will be used to calculate the velocity fields as the first order moment of $f_i$ -- will be zero, imposing the no-slip BC. An illustration of the macroscopic behaviour of the bounce back scheme is shown in Fig. \ref{bounceback2}.

\begin{figure}[h!]
    \centering
        \includegraphics[width=0.9\textwidth]{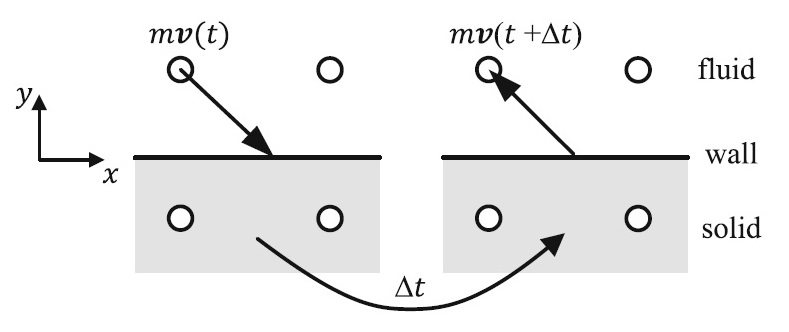}
        \caption{Macroscopic illustration of the bounce back scheme [25].}
    \label{bounceback2}
\end{figure}

Regarding non-cartesian boundaries, as shown in Fig. \ref{BCYu}, one can use several schemes represented by different interpolations of the bounce back scheme \cite{Yu2003}.

\begin{figure}[h!]
    \centering
        \includegraphics[width=0.65\textwidth]{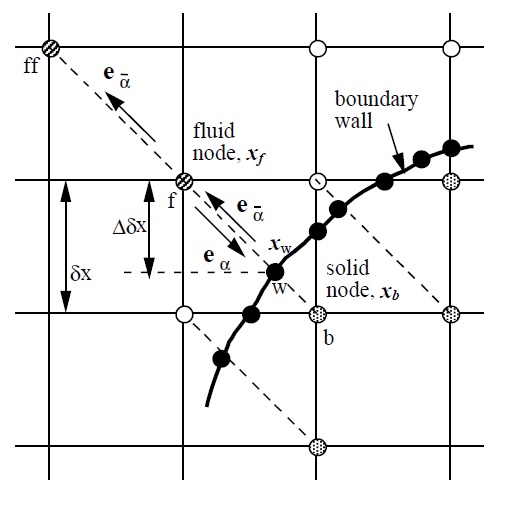}
        \caption{Illustration of the important points for a non-cartesian boundary [27].}
    \label{BCYu}
\end{figure}

The BC from Bouzidi et al. \cite{Bouzidi2001}, extensively used in the literature, was developed from the aforementioned principle and resulted in a second order scheme, given by
\bea 
&&f_{\overline{\alpha}}(\mbf{x}_f,t+1) = \frac{1}{2\Delta}\tilde{f}_{\alpha}(\mbf{x}_f,t) + \frac{2\Delta - 1}{2\Delta}\tilde{f}_{\overline{\alpha}}(\mbf{x}_f,t)  \mbox{   for  } \Delta \geq \frac{1}{2} \ , \  \label{fbc1}\\
&&f_{\overline{\alpha}}(\mbf{x}_f,t+1) = 2\Delta \tilde{f}_{\alpha}(\mbf{x}_f,t) + (1 - 2\Delta) \tilde{f}_{\alpha}(\mbf{x}_{f} - \mbf{c}_{\alpha},t)  \mbox{   for  } \Delta < \frac{1}{2} \ , \ \label{fbc2}
\eea 

\noindent where $\Delta$ is a parameter which takes into account the distance from the boundary to the lattice point within the fluid domain, and is given by
\be
    \Delta \equiv \frac{\vert \boldsymbol{x}_f - \boldsymbol{x}_w \vert}{\vert \boldsymbol{x}_f - \boldsymbol{x}_b \vert} \ . \ \label{Delta}
\ee

\noindent In Eqs.~(\ref{fbc1} -- \ref{Delta}) $\mbf{x}_f$, $\mbf{x}_w$, and $\mbf{x}_b$ represent the fluid, wall, and boundary lattice points. Also, $\overline{\alpha}$ denotes the opposite direction to $\mbf{c}_{\alpha}$ and $\tilde{f}$ represent the post-collision distribution.

\subsection{Advanced Collision Operators}

\hspace{0.5 cm} The RHS of Eq. \ref{boltzmanneq} is related to the collisions in the space of distributions $f_i$, as the previously mentioned. The BGK collision operator, also known as single relaxation time (SRT), has advantage of being local and easily computed due to its simplicity, but lacks stability when subjected to small values of the relaxation time, which means small viscosity/high Reynolds numbers. This issue can be minimized to a great extent using advanced collision operators, such as the multiple relaxation time (MRT) model \cite{dhumieres1992}, in which the collisions happen in the space of hydrodynamic moments and have different values for each moment that can be tuned individually.

In this case, Eq.~(\ref{LBE}) is replaced by
\be 
    \vert f (\mbf{x} + \mbf{c}_i,t+1)\rangle - \vert f (\mbf{x},t) \rangle = -\mbf{M}^{-1}\mbf{\Lambda}\left [\vert m (\mbf{x},t) \rangle -\vert m^{eq}(\mbf{x},t) \rangle \right ] + \left(\mbf{I} - \frac{\Lambda}{2}\right)\mbf{M}^{-1}\vert\mathcal{F} \rangle \ ,
    \label{MRT}
\ee

\noindent where $\vert f \rangle = (f_0,f_1,...,f_i)^T$, $\mbf{M}$ is the transformation matrix from the space of distributions to the space of moments $\vert m \rangle = \mbf{M}\vert f \rangle = (m_0,m_1,...,m_i)^T$, $\mbf{\Lambda}$ is the relaxation matrix for the different moments, and $\vert\mathcal{F} \rangle$ is the mesoscopic external force vector, which is related to the force term in the Boltzmann equation.

The main advantage behind the MRT model is that only the moments associated to the stress tensor are set to relax with relaxation time $\omega$, while the others, are usually set to unity, in a way that they all relax to the equilibrium value at each time step, resulting in an increase of stability.

A related issue regarding stability for the three dimensional LBM, the main focus of this dissertation, is the expansion order of the Boltzmann equilibrium distribution. It was already shown that the expansion up to the second order presented in Eq.~(\ref{boltzeqsecondorder}) does not lead to Galilean invariant post-collision moments and forcing terms in the D3Q27 model, impacting on the stability of the resulting model \cite{coreixas_chopard_latt2019,malaspinas2015,Coreixas_wissocq_puigt_boussuge_sagaut2017,coreixas_phd_thesis,derosis2017,derosis_luo2019}. This can be tackled by a sixth-order expansion, given by
\begin{eqnarray}
            &&f_i^{eq} =  \omega_i \rho \Bigg  \{ 1 + \frac{\mathbf{c}_i \cdot \mathbf{u}}{c_s^2} + \frac{1}{2c_s^4} \Bigg [ \mathcal{H}_{ixx}^{(2)}u_x^2 + \mathcal{H}_{iyy}^{(2)}u_y^2 + \mathcal{H}_{izz}^{(2)}u_z^2 + 2 \Bigg ( \mathcal{H}_{ixy}^{(2)}u_x u_y + \mathcal{H}_{ixz}^{(2)}u_x u_z + \nonumber \\
            &+&\mathcal{H}_{iyz}^{(2)}u_y u_z \Bigg ) \Bigg ] + \frac{1}{2c_s^6} \Bigg [ \mathcal{H}_{ixxy}^{(3)}u_x^2 u_y + \mathcal{H}_{ixxz}^{(3)}u_x^2 u_z + \mathcal{H}_{ixyy}^{(3)}u_x u_y^2 + \mathcal{H}_{ixzz}^{(3)}u_x u_z^2 + \mathcal{H}_{iyzz}^{(3)}u_y u_z^2 +  \nonumber \\
            &+&\mathcal{H}_{iyyz}^{(3)}u_y^2 u_z +2 \mathcal{H}_{ixyz}^{(3)}u_x u_y u_z \Bigg ] + \frac{1}{4c_s^8} \Bigg [ \mathcal{H}_{ixxyy}^{(4)}u_x^2 u_y^2 + \mathcal{H}_{ixxzz}^{(4)}u_x^2 u_z^2 + \mathcal{H}_{iyyzz}^{(4)}u_y^2 u_z^2 + \nonumber \\
            &+&2 \Bigg( \mathcal{H}_{ixyzz}^{(4)}u_x u_y u_z^2 + \mathcal{H}_{ixyyz}^{(4)}u_x u_y^2 u_z + \mathcal{H}_{ixxyz}^{(4)}u_x^2 u_y u_z \Bigg ) \Bigg ] + \frac{1}{4c_s^{10}} \Bigg [ \mathcal{H}_{ixxyzz}^{(5)}u_x^2 u_y u_z^2 + \nonumber \\ 
            &+& \mathcal{H}_{ixxyyz}^{(5)}u_x^2 u_y^2 u_z + \mathcal{H}_{ixyyzz}^{(5)}u_x u_y^2 u_z^2 \Bigg ] + \frac{1}{8c_s^{12}}\mathcal{H}_{ixxyyzz}^{(6)}u_x^2 u_y^2 u_z^2  \Bigg \} \ , \ 
            \label{BoltzmannEquilibriumDistribution}
\end{eqnarray}

\noindent where $\mathcal{H}_{i}^{(n)}$ is the Hermite polynomial of order $n$. 

The lattice velocities for the D3Q27 model are given by
\bea
    && \vert c_{ix} \rangle =  (0, 1,-1, 0, 0, 0, 0, 1,-1, 1,-1, 1,-1, 1,-1, 0, 0,0, 0, 1,-1, 1,-1, 1,-1, 1,-1)^T  \ , \  \nonumber \\
    && \vert c_{iy} \rangle =  (0, 0, 0, 1,-1, 0, 0, 1, 1,-1,-1, 0, 0, 0, 0, 1,-1, 1,-1, 1, 1,-1,-1, 1, 1,-1,-1)^T \ , \ \nonumber \\
    && \vert c_{iz} \rangle = (0, 0, 0, 0, 0, 1,-1, 0, 0, 0, 0, 1, 1,-1,-1, 1, 1,-1,-1, 1, 1, 1, 1,-1,-1,-1,-1)^T \ , \  \nonumber \\
\eea

\noindent and the lattice weights by $\omega_1 =...= \omega_6 = \omega_0/4 \ , \ \omega_7 =...= \omega_{18} = \omega_0/16 \ , \ \omega_{19} =...= \omega_{26} = \omega_0 / 64$
with $\omega_0 = 8/27$.

An improved version of the MRT collision operator is the so-called central moments (CM) \cite{Rosis2019}, where all lattice velocities are shifted locally to the comoving reference frame of the fluid velocity \cite{Geier2006}, and are given by
\be
\overline{c}_{ix} =  c_{ix} - u_x  \ , \  
\overline{c}_{iy}  =  c_{iy} - u_y \ , \  
\overline{c}_{iz} = c_{iz} - u_z  \ . \ \label{cm}
\ee

Similarly to the MRT, the CM can be obtained by
\be 
    k_i \equiv \langle T_i \vert f \rangle \ ,
\ee 

\noindent where the transformation matrix \textit{T} is defined by \cite{derosis2017}
\begin{eqnarray}
&& \vert T_{0} \rangle = \vert 1,...,1 \rangle \ , \ 
 \vert T_{1} \rangle = \vert \overline{c}_{ix} \rangle \ , \  
 \vert T_{2} \rangle = \vert \overline{c}_{iy} \rangle \ , \ 
 \vert T_{3} \rangle = \vert \overline{c}_{iz} \rangle \ , \ 
\vert T_{4} \rangle = \vert \overline{c}_{ix}\overline{c}_{iy} \rangle \ , \  \nonumber \\
&& \vert T_{5} \rangle = \vert \overline{c}_{ix}\overline{c}_{iz} \rangle \ , \ 
\vert T_{6} \rangle = \vert \overline{c}_{iy}\overline{c}_{iz} \rangle \ , \ 
\vert T_{7} \rangle = \vert \overline{c}_{ix}^2 - \overline{c}_{iy}^2 \rangle \ , \ 
\vert T_{8} \rangle = \vert \overline{c}_{ix}^2 - \overline{c}_{iz}^2 \rangle \ , \ \nonumber \\
&& \vert T_{9} \rangle = \vert \overline{c}_{ix}^2 + \overline{c}_{iy}^2 + \overline{c}_{iz}^2 \rangle \ , \  \vert T_{10} \rangle = \vert \overline{c}_{ix} \overline{c}_{iy}^2 + \overline{c}_{ix} \overline{c}_{iz}^2 \rangle \ , \ \vert T_{11} \rangle = \vert \overline{c}_{ix}^2 \overline{c}_{iy} + \overline{c}_{iy} \overline{c}_{iz}^2 \rangle \ , \  \nonumber \\
&& \vert T_{12} \rangle = \vert \overline{c}_{ix}^2\overline{c}_{iz} + \overline{c}_{iy}^2\overline{c}_{iz} \rangle \ , \   \vert T_{13} \rangle = \vert \overline{c}_{ix} \overline{c}_{iy}^2 - \overline{c}_{ix} \overline{c}_{iz}^2 \rangle \ , \  \vert T_{14} \rangle = \vert \overline{c}_{ix}^2 \overline{c}_{iy} - \overline{c}_{iy} \overline{c}_{iz}^2 \rangle \ , \  \nonumber \\
&& \vert T_{15} \rangle = \vert \overline{c}_{ix}^2\overline{c}_{iz} - \overline{c}_{iy}^2\overline{c}_{iz} \rangle \ , \ \vert T_{16} \rangle = \vert \overline{c}_{ix} \overline{c}_{iy} \overline{c}_{iz} \rangle \ , \  \vert T_{17} \rangle = \vert \overline{c}_{ix}^2 \overline{c}_{iy}^2 + \overline{c}_{ix}^2 \overline{c}_{iz}^2 + \overline{c}_{iy}^2 \overline{c}_{iz}^2 \rangle \ , \  \nonumber \\
&& \vert T_{18} \rangle = \vert \overline{c}_{ix}^2 \overline{c}_{iy}^2 + \overline{c}_{ix}^2 \overline{c}_{iz}^2 - \overline{c}_{iy}^2 \overline{c}_{iz}^2 \rangle \ , \   \vert T_{19} \rangle = \vert \overline{c}_{ix}^2 \overline{c}_{iy}^2 - \overline{c}_{ix}^2\overline{c}_{iz}^2 \rangle \ , \  \vert T_{20} \rangle = \vert \overline{c}_{ix}^2 \overline{c}_{iy} \overline{c}_{iz}  \rangle \ , \  \nonumber \\
&& \vert T_{21} \rangle = \vert \overline{c}_{ix} \overline{c}_{iy}^2 \overline{c}_{iz} \rangle \ , \   \vert T_{22} \rangle = \vert \overline{c}_{ix} \overline{c}_{iy} \overline{c}_{iz}^2 \rangle \ , \   \vert T_{23} \rangle = \vert \overline{c}_{ix} \overline{c}_{iy}^2 \overline{c}_{iz}^2  \rangle \ , \  \vert T_{24} \rangle = \vert \overline{c}_{ix}^2 \overline{c}_{iy} \overline{c}_{iz}^2 \rangle \ , \  \nonumber \\
&& \vert T_{25} \rangle = \vert \overline{c}_{ix}^2 \overline{c}_{iy}^2 \overline{c}_{iz} \rangle \ , \  \vert T_{26} \rangle = \vert \overline{c}_{ix}^2 \overline{c}_{iy}^2 \overline{c}_{iz}^2  \rangle \ . \
\label{Ts}
\end{eqnarray}

In this formalism, the post-collision CMs are given by
\be
    \vert k^* \rangle = (\mbf{I} - \mbf{\Lambda})\vert k \rangle + \mbf{\Lambda} \vert k^{eq} \rangle + \Bigg ( \mbf{I} - \frac{\mbf{\Lambda}}{2} \Bigg )\vert R \rangle \ , \ \label{k*}
\ee

\noindent where $\vert R \rangle =  T \vert \mathcal{F}\rangle$. A commonly used good approximation for the mesoscopic forcing term \cite{guo2002} is given by
\be
   \vert \mathcal{F} \rangle = -\mbf{F} \cdot \nabla_{\bf{c}} | f^{eq} \rangle \ . \ 
    \label{Forcing_term}
\ee

The relaxation matrix, as it was mentioned previously, is constructed to allow moments associated to the stress tensor approach local equilibrium with relaxation frequency $\omega$ and is given by
\be
    \Lambda = diag\left [ 1,1,1,1,\omega,\omega,\omega,\omega,\omega,1,...,1 \right ] \ .
\ee 

After some algebra, it is possible to show that the post-collision moments are given by
\begin{eqnarray}
&& k_0^* = \rho \ , \  k_1^* = F_x/2 \ , \  k_2^* = F_y/2 \ , \  k_3^* = F_z/2 \ , \ k_4^* = (1-\omega)k_4 \ , \  k_5^* = (1-\omega)k_5 \ , \ \nonumber \\
&& k_6^* = (1-\omega)k_6 \ , \  k_7^* = (1-\omega)k_7 \ , \  k_8^* = (1-\omega)k_8 \ , \  k_9^* = 3\rho c_s^2 \ , \ k_{10}^* = F_x c_s^2 \ , \  \nonumber \\
&& k_{11}^* = F_y c_s^2 \ , \  k_{12}^* = F_z c_s^2 \ , \ k_{13}^* =  k_{14}^* = k_{15}^* =  k_{16}^* = 0 \ , \  k_{17}^* = \rho c_s^2 \ , \ k_{18}^* = \rho c_s^4 \ , \ 
\nonumber  \\
&& k_{19}^* = k_{20}^* =  k_{21}^* = k_{22}^* = 0 \ , \ k_{23}^* = F_x c_s^4/2 \ , \ 
k_{24}^* = F_y c_s^4/2 \ , \  
k_{25}^* = F_z c_s^4/2 \ , \ \nonumber \\
&& k_{26}^* = \rho c_s^6 \ , \
\end{eqnarray}
where
\bea
&& k_4 = \sum_i f_i \overline{c}_{ix}\overline{c}_{iy} \ ,  \  k_5 = \sum_i f_i \overline{c}_{ix}\overline{c}_{iz} \ , \  k_6 = \sum_i f_i \overline{c}_{iy}\overline{c}_{iz} \ , \ \nonumber \\
&&k_7 = \sum_i f_i (\overline{c}_{ix}^2 - \overline{c}_{iy}^2) \ , \  k_8 = \sum_i f_i (\overline{c}_{ix}^2 - \overline{c}_{iz}^2) \ . \
\eea

By construction, one can see that all post-collision moments are Galilean invariant quantities. The post-collision populations are easily obtained through,
\be
\vert f^* \rangle = T^{-1}\vert k^* \rangle \ . \ \label{k*f*}
\ee

The last remaining step is to perform the streaming, given by
\be
    f_i(\mbf{x}+ \mbf{c}_i,t+1) = f_i^* (\mbf{x},t).
    \label{stream}
\ee 

This exposition resumes one of the most advanced collision operators in the LBM, the CM approach.

\section{LBM for MHD}

\hspace{0.5 cm} As previously brought up, MHD systems are treated by Eqs.~(\ref{NSE}, \ref{IE}, and \ref{nablaUB}). It was also mentioned that the LBM can recover macroscopically the NS equations. To simulate the complete MHD system, we are only left with the induction equation, which, as one can see, has some similarities with the NS equations. 

In theory, this equation can be simulated in several ways, such as finite differences, finite volume, and finite elements, among others. The main issue is related to the many computational problems faced with the NS equations, which are going to be present for the induction equation as well.

Having these issues in mind, Dellar \cite{DELLAR2002} developed a LBM method with a BGK collision operator for the induction equation. This was done by unifying a MHD kinetic approach with a general construction of BGK collision models for many different systems that obey some conservation laws \cite{Croisille1995 , Bouchut1999}. 

\begin{figure}[h!]
    \centering
        \includegraphics[width=0.60\textwidth]{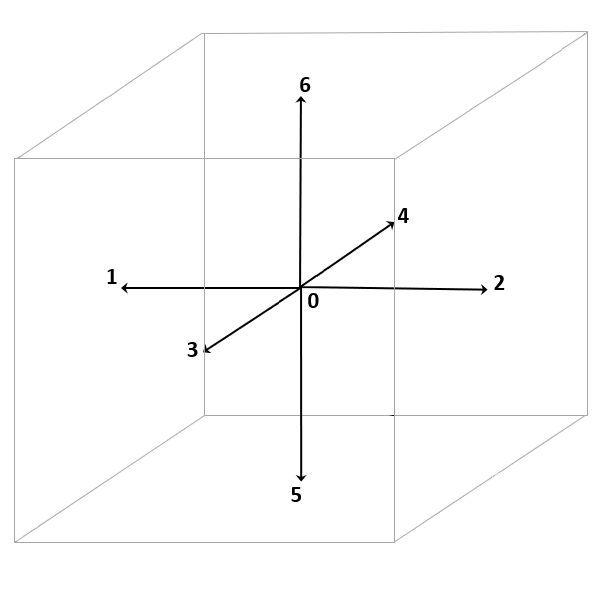}
        \caption{D3Q7 lattice cell.}
    \label{D3Q7lattice}
\end{figure}

Alternatively to its hydrodynamic counterpart, the magnetic LBM is based on vector valued distributions $\mbf{g}_i$, instead of scalar ones. As we are considering a detailed velocity set for the hydrodynamic part, a D3Q7 model is usually enough for the magnetic part, illustrated in Fig. \ref{D3Q7lattice}, with the magnetic field being the zeroth order moment of the aforementioned distributions, such as

\be 
    \mbf{B}(\mbf{x},t) = \sum_{i = 0}^6 \mbf{g}_i(\mbf{x},t) \ , 
    \label{macroscopicB}
\ee 

\noindent where $\mbf{g}_i(\mbf{x},t) \equiv \mbf{g}(\boldsymbol{\xi}_i,\mbf{x},t)$, with $\boldsymbol{\xi}$ being the magnetic lattice velocities, given by
\bea
    && \vert \xi_{ix} \rangle =  (0, 1, -1, 0, 0, 0, 0)^T  \ , \  \nonumber \\
    && \vert \xi_{iy} \rangle =  (0, 0, 0, 1,-1, 0, 0)^T \ , \ \nonumber \\
    && \vert \xi_{iz} \rangle = (0, 0, 0, 0, 0, 1,-1)^T \ . \  \nonumber \\
\eea

Then, the LBM for the induction equation with a BGK collision model can be summarized as
\be
    \mbf{g}_i(\mbf{x} + \boldsymbol{\xi}_i, t + 1) - \mbf{g}_i(\mbf{x}, t) = - \frac{1}{\tau_m}[\mbf{g}_i(\mbf{x}, t) - \mbf{g}^{eq}_i(\mbf{x}, t)] \ , \ 
\label{bgkmagnetic}
\ee

\noindent where $\tau_m$ is the magnetic relaxation time, which is related with the magnetic diffusivity $\eta$ as
\be 
    \tau_m = \frac{\eta}{c_m^2} + \frac{1}{2} \ . 
    \label{magrelaxtime}
\ee 

The parameter $c_m$ is the sound speed of the magnetic grid. The equilibrium distributions are related to the correct change in the momentum equations, taking into account the Maxwell stress tensor, and is given by
\be
    \mbf{g}^{eq}_i = w^m_i \left \{ \mbf{B} + c_m^{-2} [ ( \boldsymbol{\xi}_i \cdot \mbf{u}) \mbf{B}-   ( \boldsymbol{\xi}_i \cdot \mbf{B}) \mbf{u} ] \right \} \ , \
    \label{gEq}
\ee

where $w^m_i$ are the magnetic lattice weights, and are given by
\be
    w^m_0 = 1/4 \ , \  w^m_1 = ... = w^m_6 = 1/8 \ .
\ee

The coupling of the magnetic and velocity field effects can be done through the application of the Lorentz force (\ref{LorentzForce}) directly through Eqs.~(\ref{vellbm}), (\ref{k*}), and (\ref{Forcing_term}). However, a local approach can be achieved by correcting the stress tensor. This can be done if the equilibrium distributions are corrected by the term
\be
    f_{i,mag}^{eq} = \frac{\omega_i}{2 c_s^4} ( \mbf{c}_i \otimes \mbf{c}_i - c_s^2 \mbf{I} ) : \left [ M - c_s^2 \mbf{I}(Tr M)\right ] \ , 
\ee 

\noindent where $\otimes$ denote the tensorial product, $:$ is the Frobenius inner product, $\mbf{I}$ is the identity matrix, and $M$ is the Maxwell stress tensor, given by
\be 
    M_{\alpha \beta} = \frac{1}{2} \delta_{\alpha \beta} \vert \mbf{B} \vert^2 - B_{\alpha} B_{\beta}.
\ee 

In $D$ dimensions, $f_{i,mag}^{eq}$ may be written as
\be
    f_{i,mag}^{eq} = \frac{\omega_i}{2 c_s^4} \left [ \frac{1}{2} \vert\mbf{c}_i\vert^2 \vert \mbf{B} \vert^2  - (\mbf{c}_i \cdot \mbf{B})^2 + c_s^2(Tr M)(D c_s^2 -  \vert \mbf{c}_i \vert^2 - 1) \right ] \ , 
\ee 

\noindent where $D = \{ 2,3 \}$. In $D$ dimensions, however, the trace of Maxwell' stress tensor is given by
\be
    Tr M = \vert \mbf{B}\vert^2 \left ( \frac{D}{2} - 1 \right ) \ .
\ee 

So, after a straightforward algebra, $f_{i,mag}^{eq}$ can be written as 
\be
    f_{i,mag}^{eq} = \frac{\omega_i}{2 c_s^4} \Bigg [ \frac{\vert \mathbf{c}_i \vert^2 \vert \mathbf{B} \vert^2}{D} - (\mathbf{c}_i \cdot \mathbf{B})^2  \Bigg ] \ . \ 
\ee
 
\end{chapter}

\begin{chapter}{Statistical Analysis of Experimental Data}
\label{cap5}

\hspace{5 mm} The statistical analysis of experimental data on turbulent pipe flows performed in the \emph{Núcleo Interdisciplinas de Dinâmica de Fluidos} (NIDF) - UFRJ is presented in this chapter. Several datasets --of turbulent flows with water at $Re = \{ 5300,12000,17800,24400 \ $
$,29000 \}$ and MHD turbulent flows with salt water at $Re = \{ 5300,7100 \}$ and $Ha = \{ 0, 8.1\}$-- were analyzed and the CS were labeled by their azimuthal dominant mode \cite{JackelPOF2023,JackelICHMT2023}. Also, the transition between the identified modes were analyzed as an stochastic process at $Re = 24400$ \cite{JackelPRF2023}.

\section{Experimental Setup and Validations}

\hspace{0.5 cm}To investigate the dynamics of the CS in turbulent pipe flow, an experiment was performed at NIDF - UFRJ, with dimensions comparable to many industrial applications, such as in the oil industry. The experiment was performed on a pipe with a diameter of approximately 15,2 cm, a total length of 10 meters and with a single-phase water flow. The entrance length to observe fully turbulent flows was estimated as \cite{Bhatti1987}
\be 
    L = 1.359D(Re)^{1/4} \ , 
\ee 

\noindent which is satisfied for all Reynolds numbers investigated in this dissertation, since the larger entrance length is $L \approx 2.70$ m ($Re = 29000$) and the observation section is located at $L = 5.8$ m.

Turbulent data was collected by means of stereoscopic PIV (SPIV), which is able to reconstruct two-dimensional three-component velocity vector fields. To achieve this, the flow was filled in with silver-coated hollow glass spheres with a mean size of 17 microns and Stokes number --which measures how inertial the particle might be-- of $10^{-4}$, which is small enough to consider that the particles can be treated as tracers, following the flow with great accuracy \cite{Tropea2007}.

The snapshot measurements through the SPIV system were carried out at a frequency of 15 Hz, and approximately 20000 vector fields were analyzed for each Reynolds number. The total number of snapshots is a sum of independent runs with 2000 snapshots each --limited by the equipment. The experimental setup is illustrated in Fig. \ref{expsetup}.

\begin{figure}[h!]
    \centering
        \includegraphics[width=1.0\textwidth]{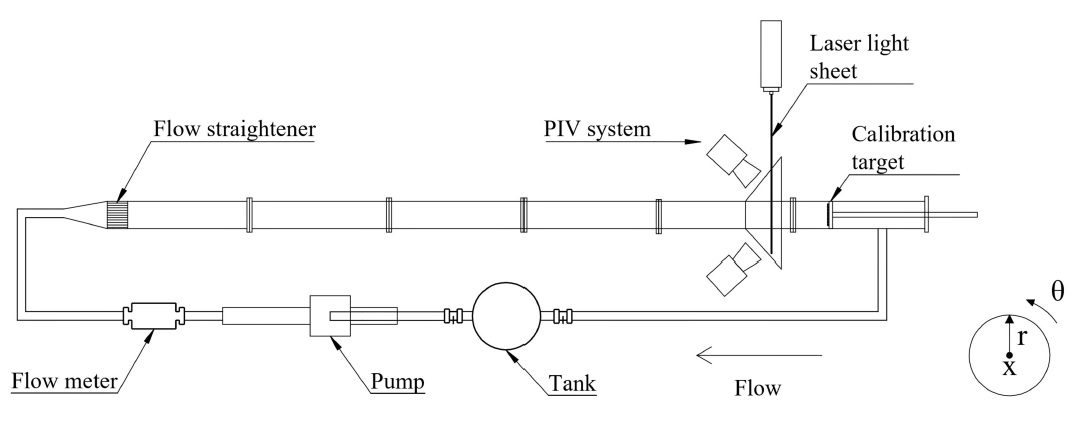}
        \caption{Experimental setup (not to scale) and coordinate system of the pipe rig with SPIV system. The flow direction is clockwise [45].}
    \label{expsetup}
\end{figure}

For the turbulent MHD flow, a modular magnetic frame was employed, composed of 3D-printed rings of paramagnetic resin. Each section in the magnetic frame is made of a hexagonal arrangement of neodymium permanent magnets ( 5.1 x 5.1 x 2.5 cm ) with an intensity of approximately 0.5 Tesla on its center. The chosen fluid --in order to maximize conductivity while still transparent to be observed by the SPIV system-- was NaCl, with a conductivity of approximately 24.5 $Sm^{-1}$. The magnetic frame was mounted in front of the SPIV system, as shown in Fig. \ref{magneticframeexp}. A detailed description of the experimental setups can be found in Refs. \cite{JackelPOF2023, JackelICHMT2023,JackelUFRJ2023}.

\begin{figure}[h!]
    \centering
        \includegraphics[width=0.75\textwidth]{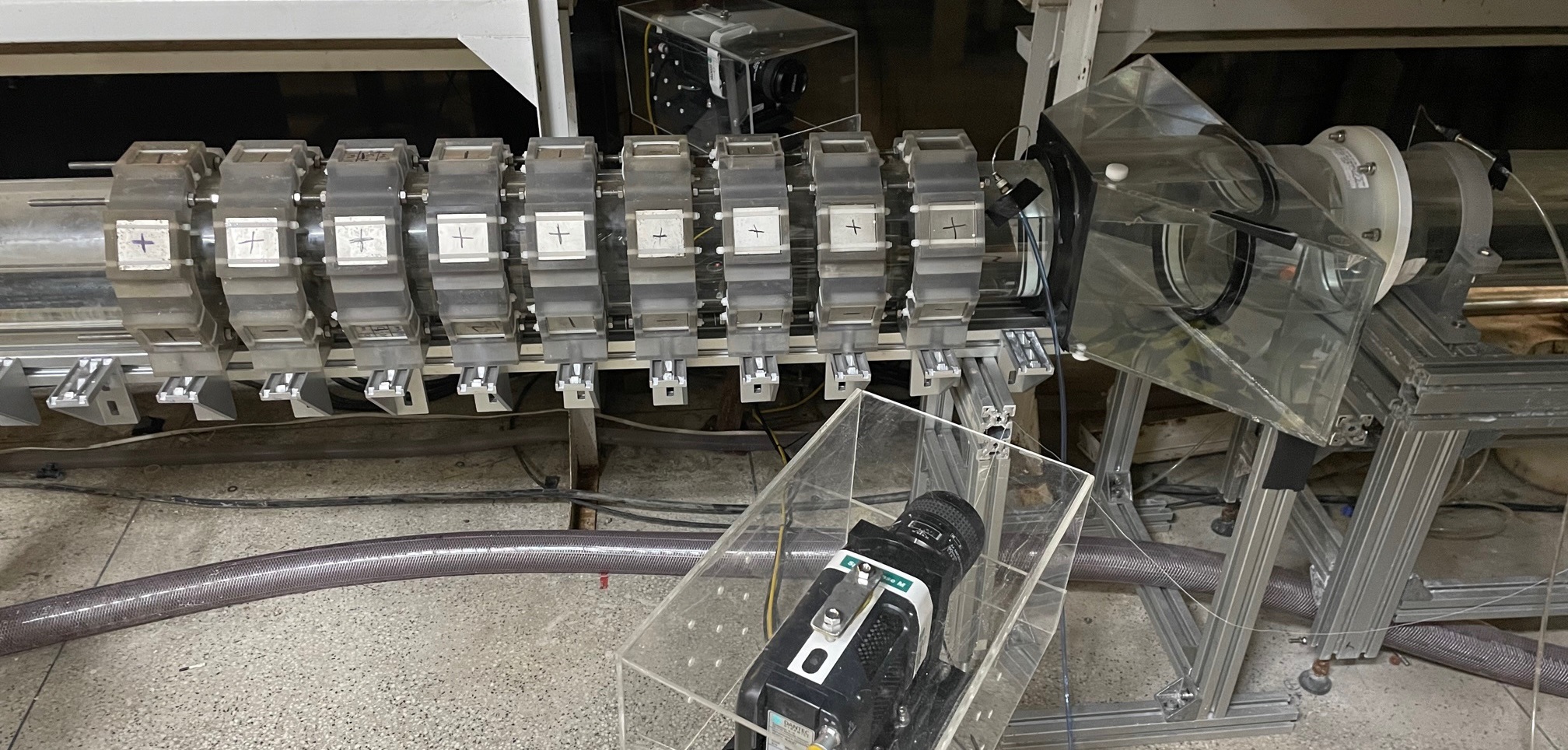}
        \caption{Modular magnetic frame mounted in front of the SPIV visualization section [46].}
    \label{magneticframeexp}
\end{figure}

\vspace{0.7 cm}

The target used to capture the flow has a uniform grid of 78 x 78 points. A snapshot containing all three components of the turbulent velocity field is depicted in Fig. \ref{vectorfieldexp} for $Re = 24414$. At least four CS can be noticed by eye, which might represent hairpin or quasi-streamwise vortices.

\begin{figure}[h!]
    \centering
        \includegraphics[width=0.75\textwidth]{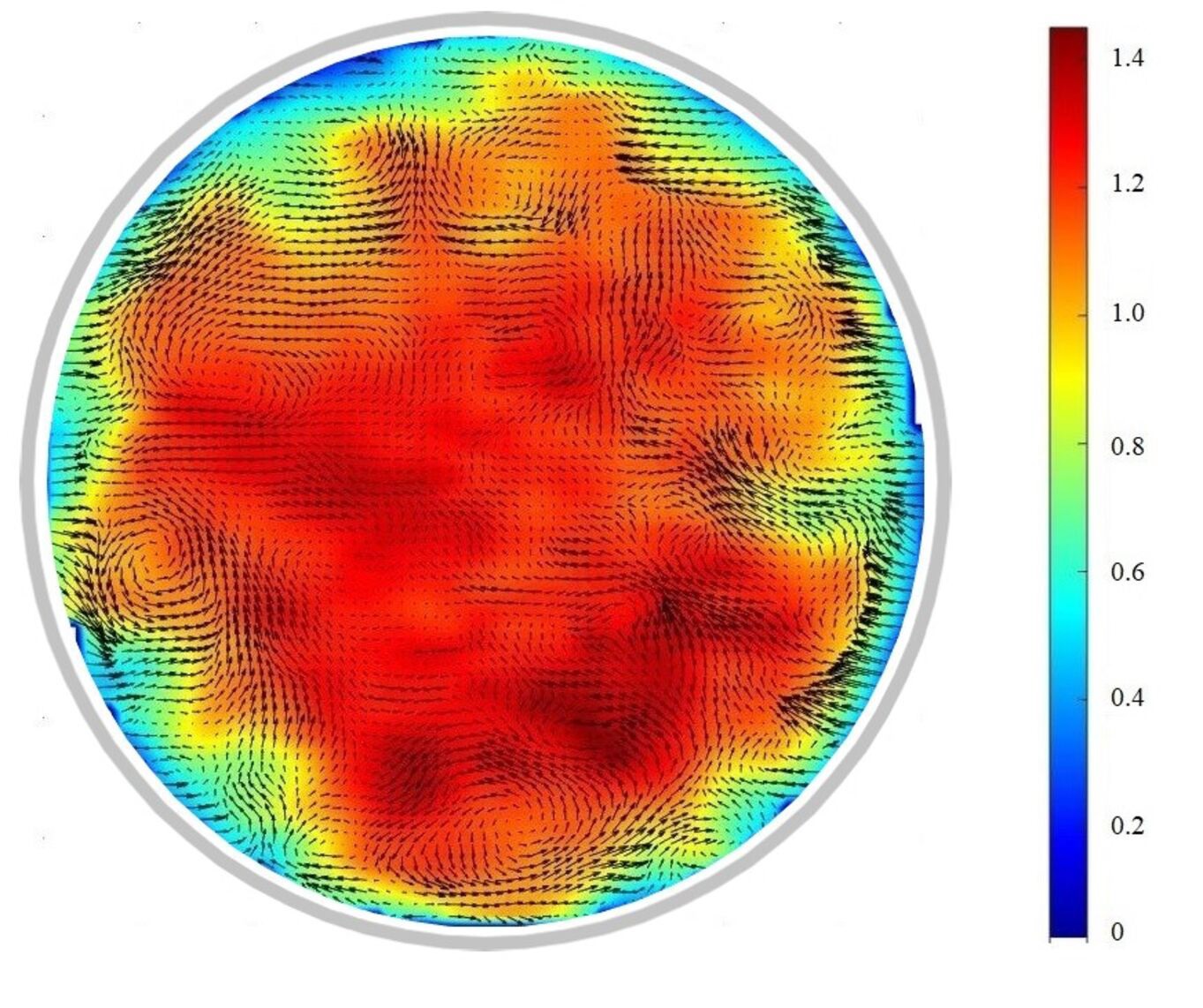}
        \caption{Instantaneous vector field for the flow at Re = 24414. The color bar indicates the magnitude of the streamwise velocity component normalized by the bulk velocity. [45]}
    \label{vectorfieldexp}
\end{figure}

The mean velocity profile was computed through a time average as in Eq.~(\ref{timeaverage}) plus an azimuthal average, as
\be 
    \langle \mathcal{F}(r,\theta) \rangle_{\theta} = \mathcal{G}(r) = \frac{1}{2\pi}\sum^{2\pi}_{\theta=0}\mathcal{F}(r,\theta) \ ,
    \label{azimuthalaverage}
\ee 

\noindent where a bilinear interpolation was used to obtain the velocities in points that don't coincide with the rectangular grid. The mean streamwise velocity normalized by the rms velocity is shown in Fig \ref{velocityprofileexp} --in inner units, which will be defined later-- for the Reynolds numbers of 4928 and 29089. Furthermore, PIV measurements from Eggels et al. \cite{Eggels1994} are also displayed for comparison.

\begin{figure}[h!]
    \centering
        \includegraphics[width=0.9\textwidth]{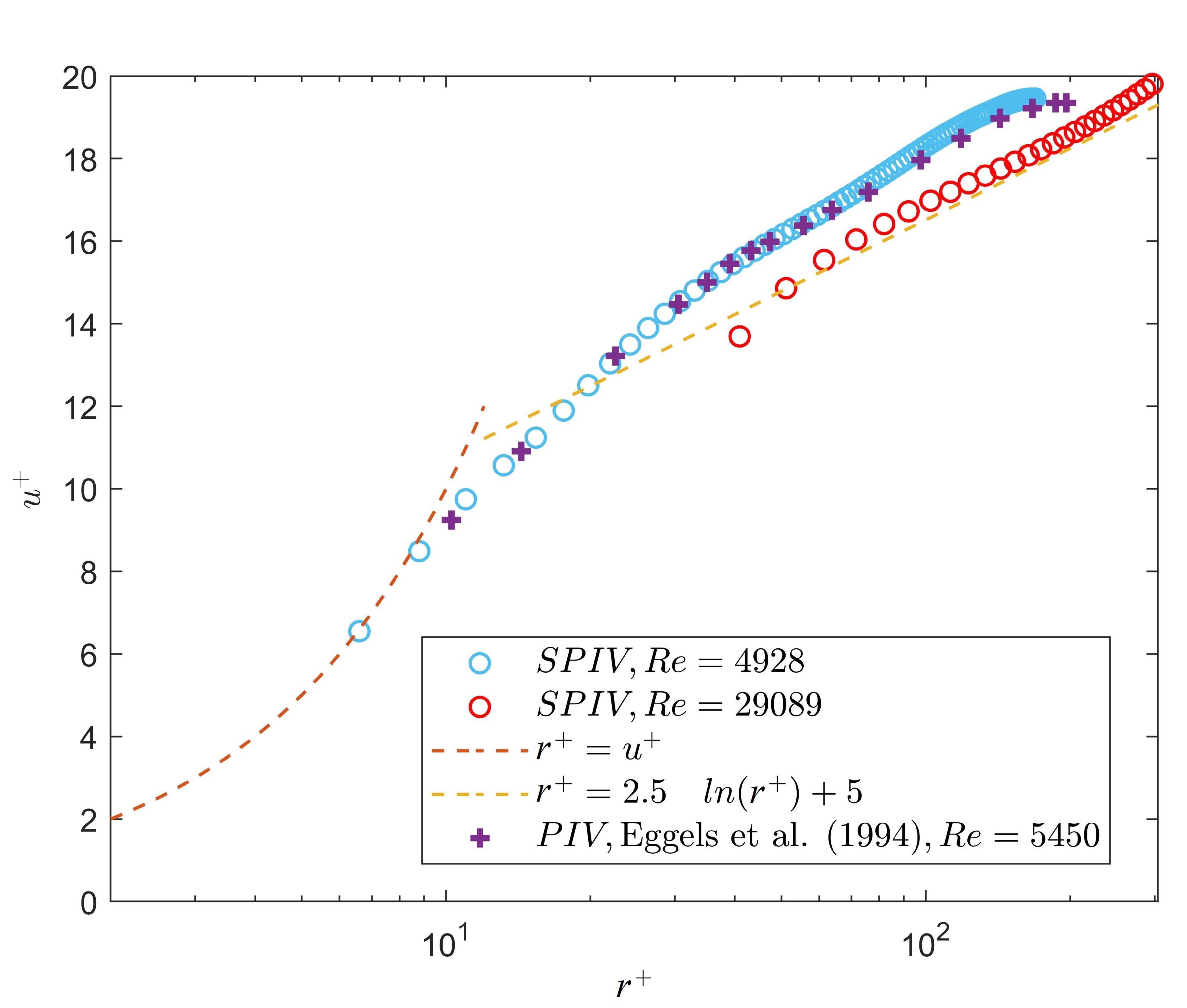}
        \caption{Streamwise velocity profile in inner units at a Reynolds number of 4928 and 29089 obtained with SPIV, compared with the results of Eggels et al [45,51].}
    \label{velocityprofileexp}
\end{figure}

\vspace{0.7 cm}

\section{Identification of CS}

\hspace{0.5 cm}The snapshot labeling was performed by taking the highest peak in the spectral power density
\be 
    I(k_n) = \int_0^{2\pi} d \theta e^{i k_n \theta}f_{uu}(r_0,\theta) \ ,
    \label{spectrum2}
\ee 

\noindent where $f_{uu}(r_0,\theta) = u^2_{z,rms}R_{uu}(r_0, \theta)$, $k_n = n \in \mathbb{Z}^{+}$ is the azimuthal wave number, and $r_0 = 0.78R$ is the reference radius --regions which are closer to the wall, above 0.8R, usually are not well resolved by SPIV, probably due to the strong gradients. 

Using the above procedure, for each snapshot there is always a discrete dominant mode (highest peak), although, for the complete time series, how the first peak is compared to the second highest strongly oscillates. To illustrate a clear example with a prominent peak, Fig. \ref{spectrum} shows the spectral density for a particular snapshot with $Re = 24414$ and dominant mode $k_n = 4$. Also, this mode carries 32\% of the turbulent streamwise kinetic energy in the shell $0.6 \leq r/R \leq 0.8$.

\begin{figure}[h!]
    \centering
        \includegraphics[width=0.75\textwidth]{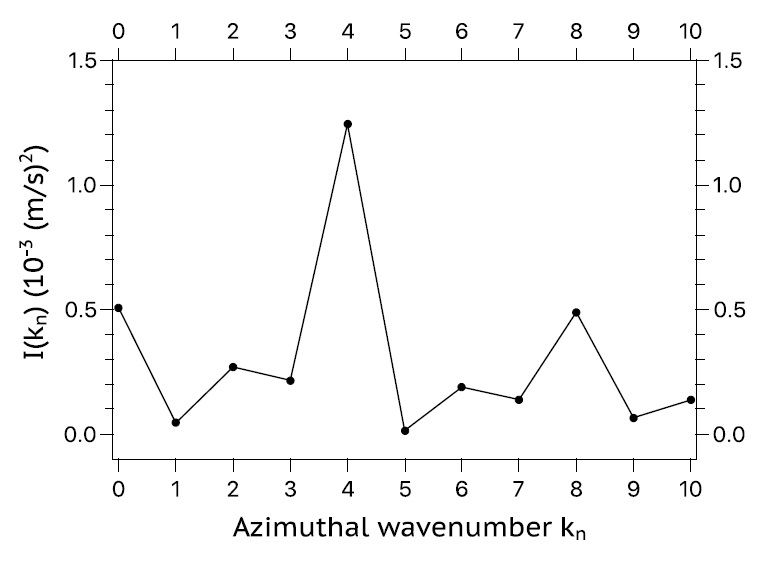}
        \caption{Power spectrum for a given sPIV snapshot of the flow at $Re = 24414$, as evaluated from Eq.~(\ref{spectrum2}). The streamwise velocity field is in this instance dominated by the wave number $k^* = 4$ [47].}
    \label{spectrum}
\end{figure}

Using this procedure it was possible to observe the remarkable number of 10 different azimuthal wave numbers for all Reynolds numbers analyzed, with and without magnetic field. Their distribution alongside the mean for all Reynolds numbers is shown in Fig. \ref{distributionofmodes} in the absence of magnetic fields. Another remarkable fact is that all Reynolds numbers have very similar distributions whose shapes somehow agrees qualitatively well with other results, such as POD shown in Fig. \ref{modesenergy} and with previous studies using slightly different dimensionality reduction approaches, as shown in Fig. \ref{vecandprob}.

\begin{figure}[h!]
    \centering
        \includegraphics[width=0.8\textwidth]{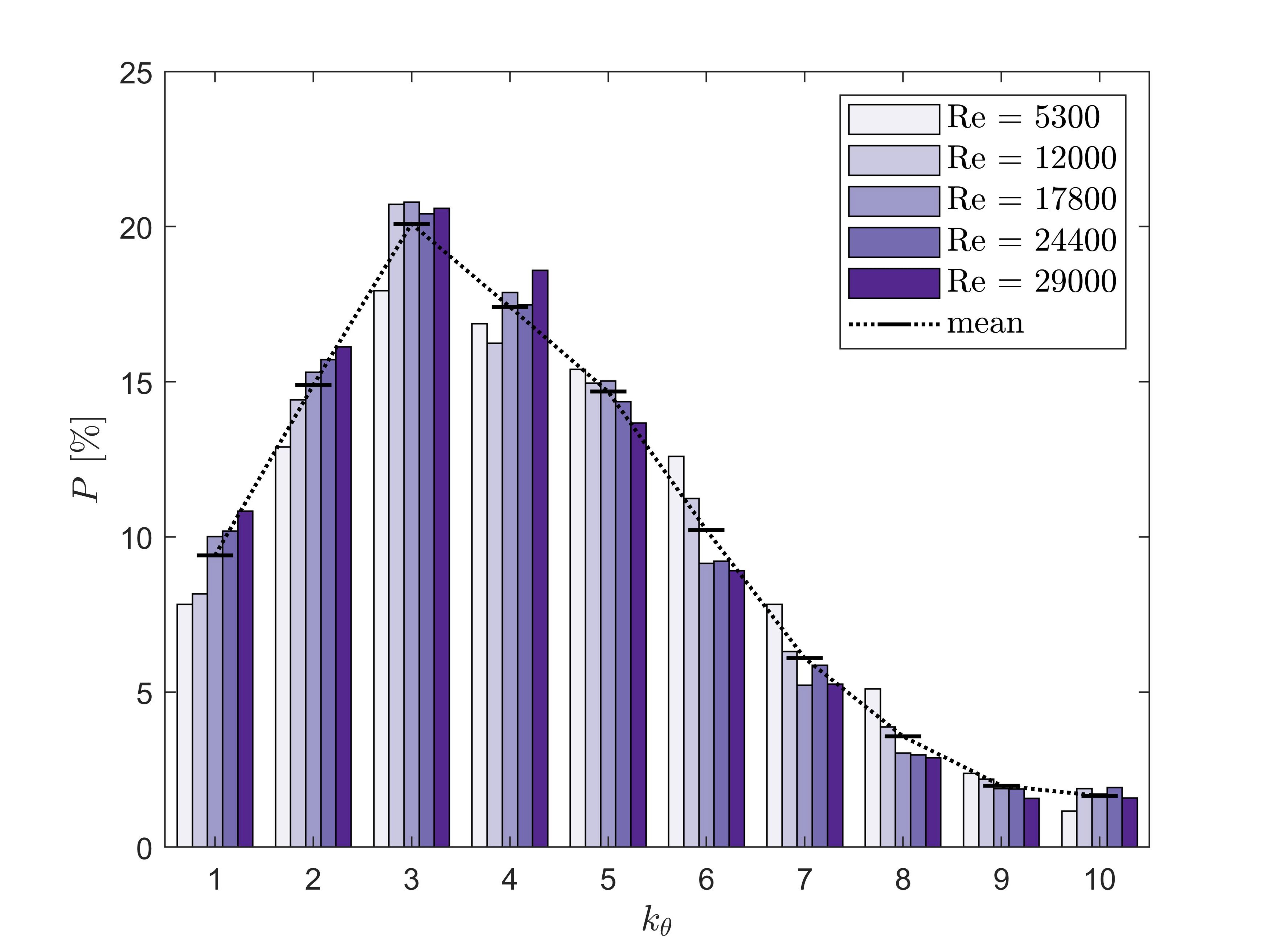}
        \caption{Normalized distribution of dominant wave number states based on snapshot observations [45].}
    \label{distributionofmodes}
\end{figure}

The ensemble average of the streamwise velocity-velocity correlation function defined in Eq.~(\ref{newcorrelation}) conditioned to the same azimuthal wave numbers are shown in Fig. \ref{coherent_states} by means of a countor plot for $Re = 17800$. Similarly to the CS observed in Chap. \ref{cap3}, these regions are associated with pairs of counter-rotating vortices, where the negative/positive regions of the correlation function (blue/red) are related to low/high speed streaks, with the fluctuating part of the streamwise velocity field being 

\begin{figure}[h!]
    \centering
        \includegraphics[width=1.0\textwidth]{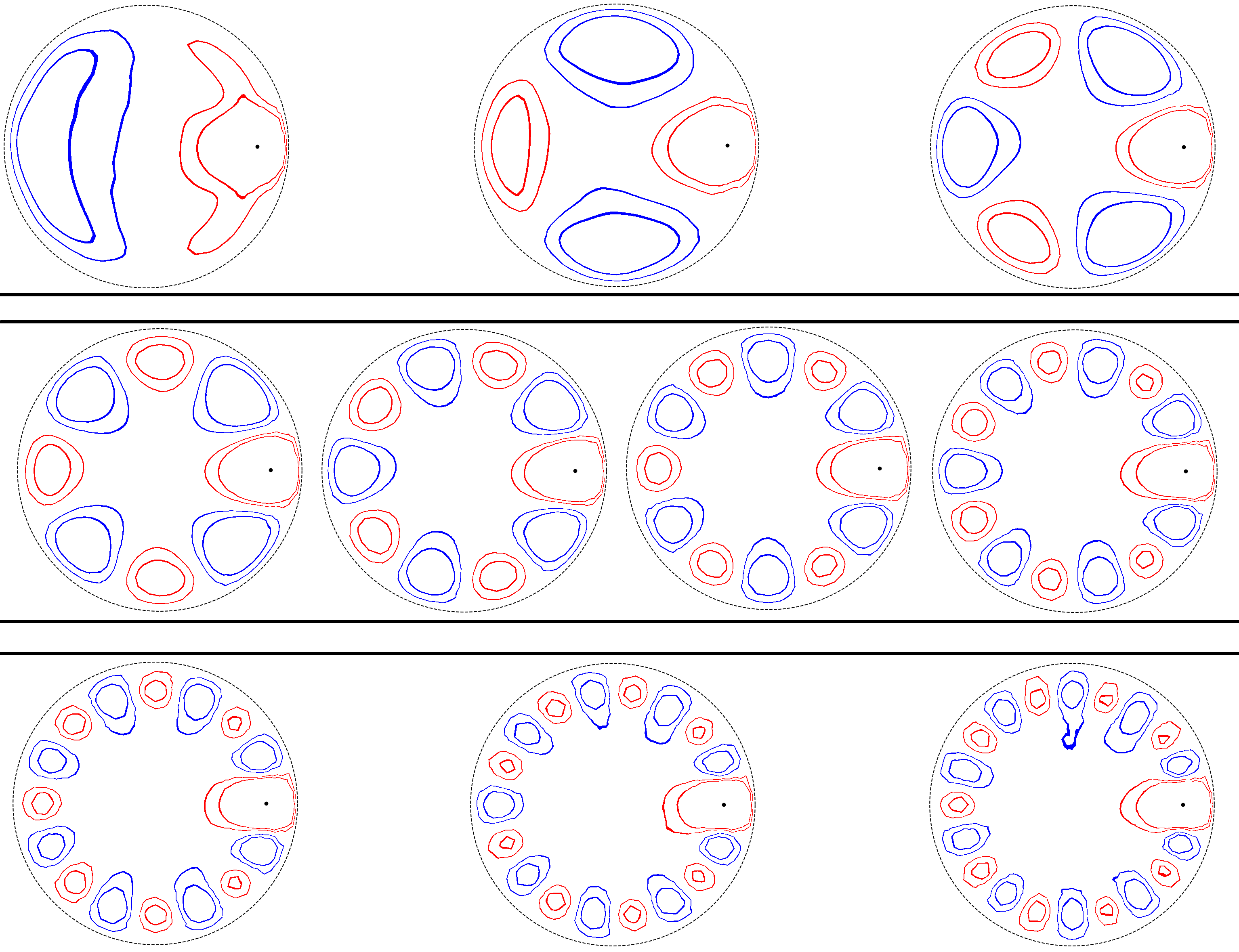}
        \caption{Organizational turbulent states with azimuthal wave numbers of 1 to 10 of $Re=17800$. The images show the spatial correlation function $R_{uu}$, the red level curves correspond to $R_{uu} = 0.05$ and 0.1, and the blue ones to the opposite sign. The reference point is illustrated as a black dot.}
    \label{coherent_states}
\end{figure}

\begin{figure}[h!]
    \centering
        \includegraphics[width=1.0\textwidth]{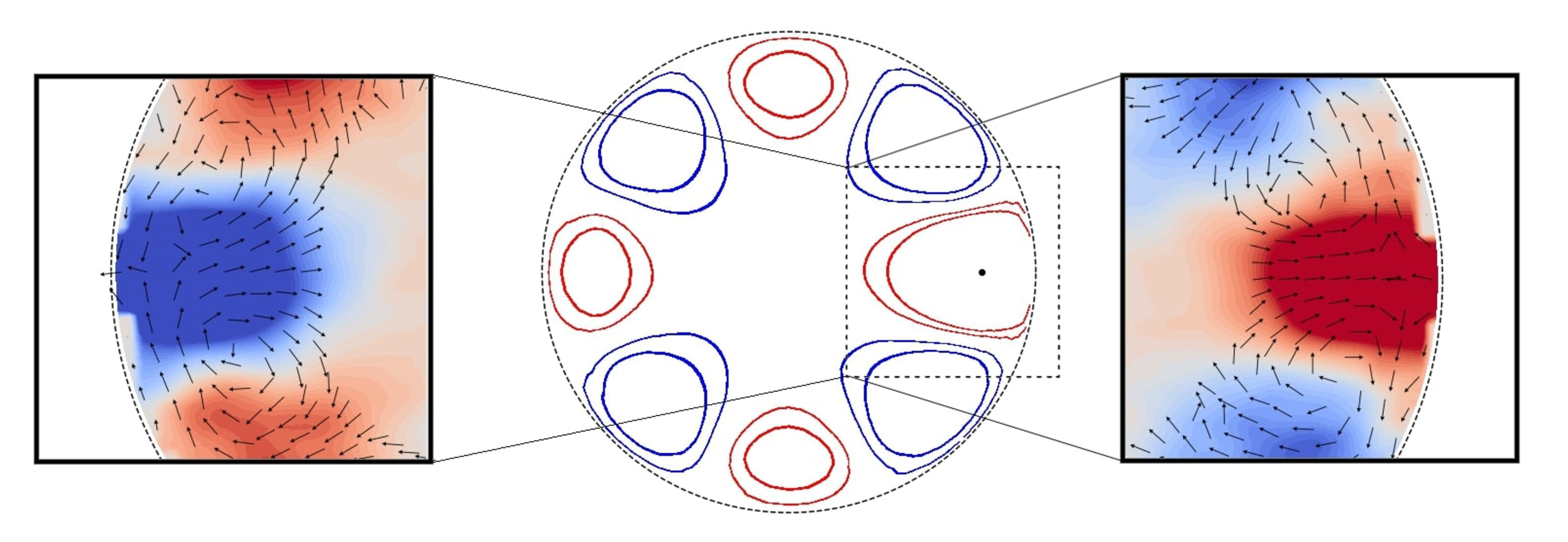}
        \caption{Conditionally averaged vector field (left) and corresponding principal coherent patterns (right) related to regions of positive (red) and negative velocity fluctuations (blue) of wave number state 4 of Re = 17800 [45].}
    \label{correlationandvectors}
\end{figure}

\noindent below/above the mean. The ensemble average vector fields (in-plane + streamwise velocity fluctuation) conditioned to the azimuthal wave number $k_{\theta} = 4$ are shown in Fig. \ref{correlationandvectors}. The streamwise extent of the azimuthal correlation is shown in Fig. \ref{correlationsurface2} for a length of 5 pipe radii, where one clearly sees the transition between the dominant modes. For symmetry reasons, the plot is shown for angular increments from $0$ to $\pi$.

\begin{figure}[h!]
    \centering
        \includegraphics[width=0.8\textwidth]{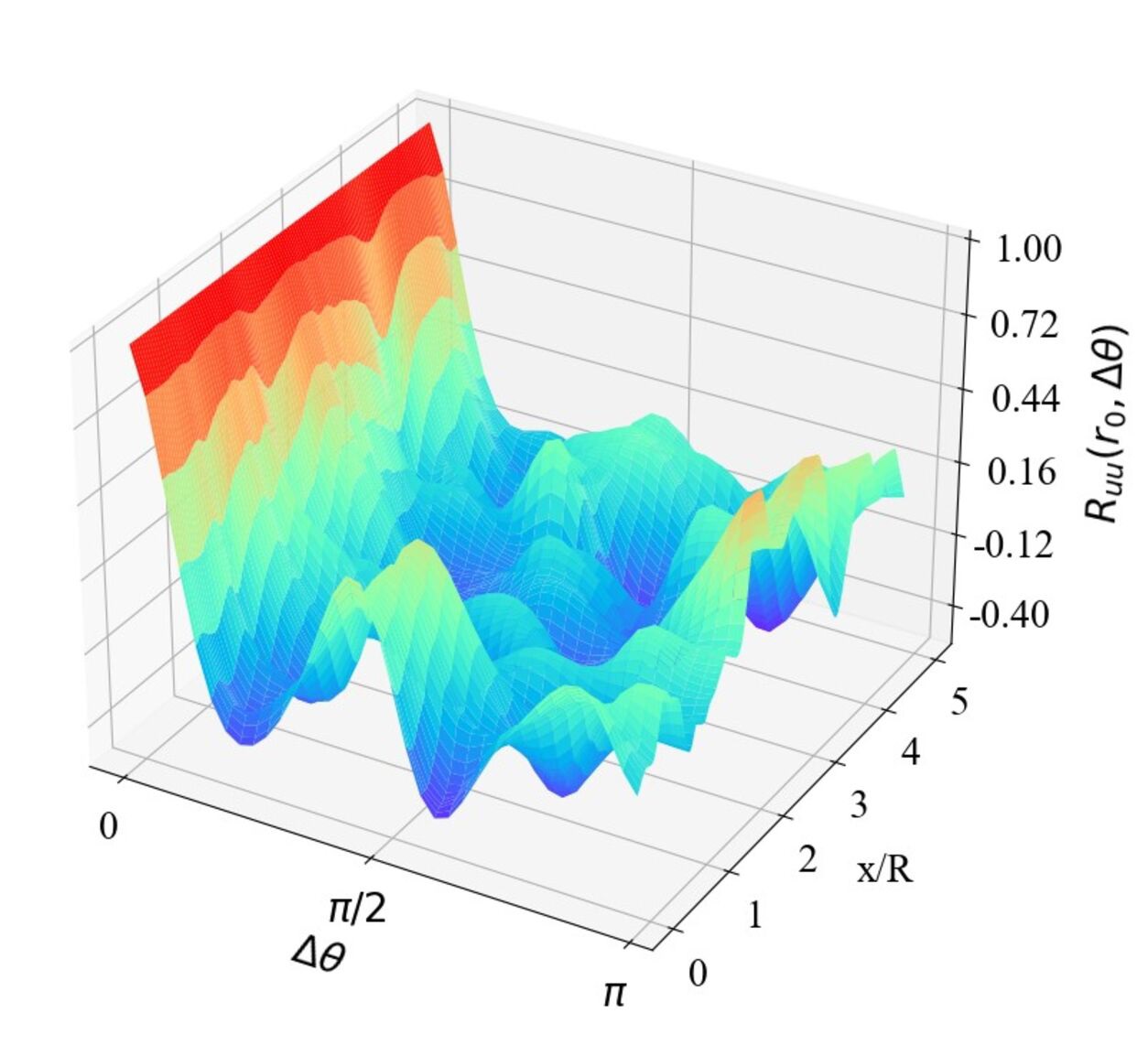}
        \caption{Streamwise extent of the azimuthal correlation over a length of 5 radii, here for the Reynolds number of 5300. Positive peaks indicate a correlation, negative peaks anti-correlation. Note that for reasons of symmetry we only plot along an azimuth from zero to $\pi$ [45].}
    \label{correlationsurface2}
\end{figure}

In order to have a more concise picture of the dynamics of the identified modes, their evolution as a function of the distance is shown in Fig. \ref{snapshottransitions} using the Taylor frozen hypothesis \cite{Taylor1938,Dennis2008}. The transition between the discrete modes is clear and displays some degree of coherency, as expected, once they are related to CS within the flow.

\vspace{0.5 cm}

\begin{figure}[h!]
    \centering
        \includegraphics[width=1.0\textwidth]{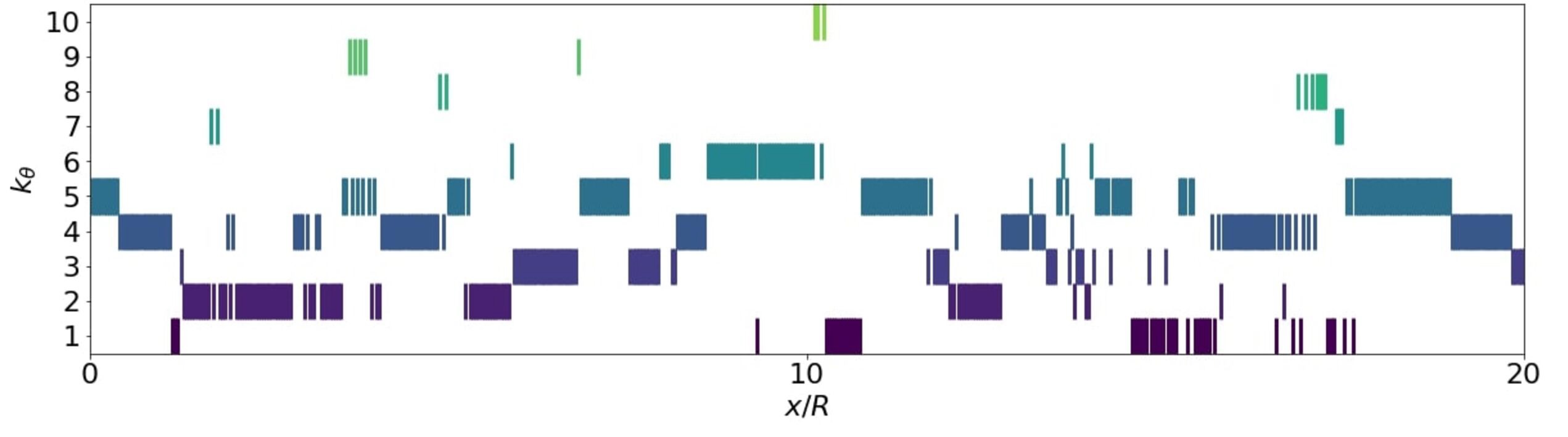}
        \caption{Streamwise extent of coherent states along the main flow direction in an arbitrary section of 20 pipe radii at $Re = 5300$ [45].}
    \label{snapshottransitions}
\end{figure}

The MHD turbulent CS observed were also identified through this formalism. A non-trivial remark is that although the magnetic frame was designed to maximize the magnetic field intensity near the walls and the number of regions where the magnetic field is perpendicular to the pipe surface, its intensity decreases considerably fast, and hence they could not affect the structures substantially. Many of the previous results were similar, however, the observed streamwise length of the structures increased by a significant amount for both Reynolds numbers, a result depicted in Fig. \ref{modelengthmag}. This could be related to a decrease in the number of transitions between modes, with the magnetic field forcing the CS to stay in the current mode for longer periods.

\begin{figure}[h!]
    \centering
        \includegraphics[width=1.0\textwidth]{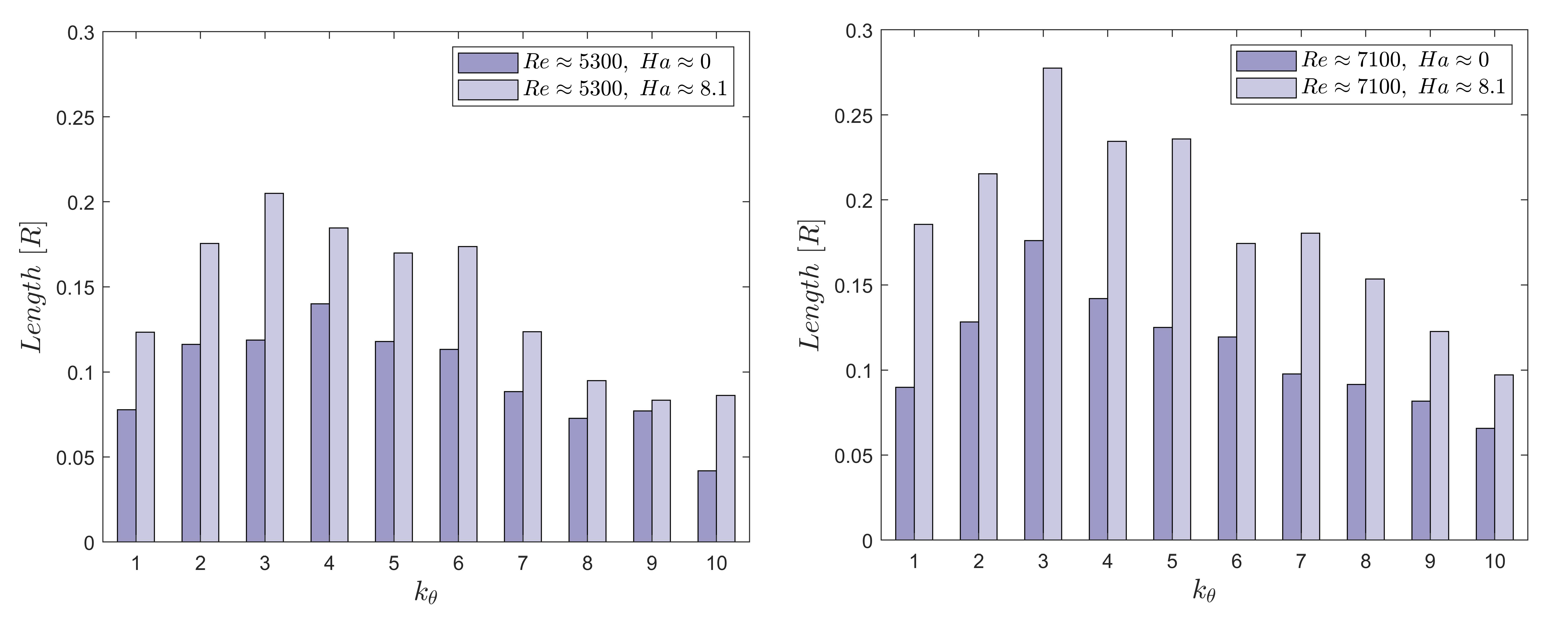}
        \caption{Comparison of average lengths of dominant wave number structures in radii for Re $\approx$ 5300 (left) and 7100 (right) with a Hartmann number of zero (plain turbulence) and $\sim 8.1$ (magnetically affected turbulence) [46].}
    \label{modelengthmag}
\end{figure}

\section{Stochastic Mode Transitions}

\hspace{0.5 cm}In order to investigate the transition dynamics between the modes, a data set of 20000 snapshots at $Re = 24415$, captured with a frequency of 10 Hz, was interpreted as a stochastic process $\mathcal{S}$ at equally spaced time intervals $\Delta = 0.1$ s, defined by
\be
    \mathcal{S} \equiv  \{ k^*(t), k^*(t + \Delta), k^*(t+2\Delta), \ ... \  \} \ ,
    \label{stochasticseries}
\ee

\noindent where $k^*(t)$ is the dominant azimuthal wave number at instant $t$. An example of a given transition between two labelled CS observed in our dataset is shown in Fig. \ref{transition}.

\begin{figure}[h!]
    \centering
        \includegraphics[width=0.8\textwidth]{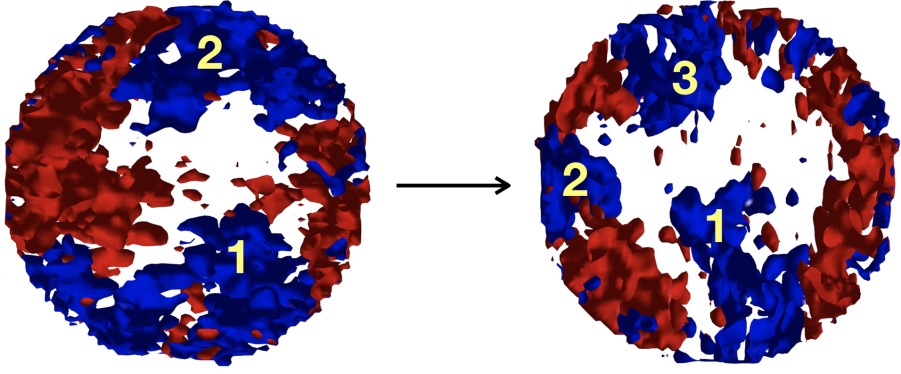}
        \caption{Example of a transition between organizational states, as sampled out from our measurements, which are associated with two and three low-speed streaks. Blue and red refer, respectively, to negative and positive streamwise velocity fluctuations around the mean [47].}
    \label{transition}
\end{figure}

The first feature investigated was whether this process is Markovian. This was achieved by checking if the Chapman-Kolmogorov (CK) equation holds, a necessary, but not sufficient condition for a stochastic process to be Markovian \cite{cirlan1975}. The CK equation was checked by constructing a 10 x 10 transition probability matrix among the states and looking at its eigensystem. This transition matrix is constructed based on the transition from its immediate state at time $t + \Delta$, as a result, its eigensystem will always have an eigenvalue of unity, where the respective eigenvector is associated with the asymptotic probability distribution --which is an application of the Perron-Frobenius theorem \cite{Mishra2020}.

Afterward, a decimated stochastic process separated by the time interval $h\Delta$ was created, and its eigensystem was also evaluated. The CK equation would impose that the absolute value of the eigenvalues from the decimated series would be equal to the original ones but raised to the power of $h$. This was investigated and is shown in Fig. \ref{CKexperiment}. As it can be seen, the stochastic process doesn't satisfy the CK equation, indicating that $\mathcal{S}$ is non Markovian, in contrast to previous literature accounts \cite{Schneider2007}, where Markovianity was assumed from the agreement between the asymptotic eigenstate and the empirical distribution of modes. For large enough intervals, the process is expected to be Markovian, with its memory effects weakly correlated. 

\begin{figure}[h!]
    \centering
        \includegraphics[width=0.8\textwidth]{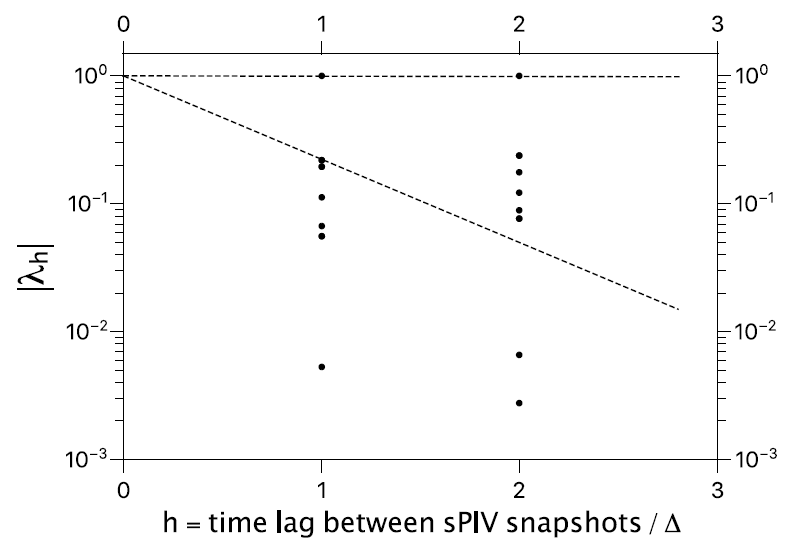}
        \caption{Eigenvalues of the probability transition matrices for the original process ($h = 1$) and a decimated one ($h = 2$). The dashed lines should intercept eigenvalue pairs if $\mathcal{S}$ were a Markovian process. In other words, we find that the transition probability matrix of the decimated process is not given as the square of the transition probability matrix of the original process, as would be expected for a Markovian process [47].}
    \label{CKexperiment}
\end{figure}

To investigate more quantitatively this non-Markovian behavior of the transition dynamics, two temporal correlation functions are defined, as follows. To start, we define the state vector $\mbf{V}(t)$ and the state transition matrix $\mbf{M}(t)$, which can be obtained from $\mathcal{S}$ as
\be
    V_m(t) = \begin{cases} 1, & \mbox{if } k^*(t) = m \\ 0, & \mbox{otherwise  }\end{cases} \ ,
\ee 

\noindent and
\be
    M_{m',m}(t) = \begin{cases} 1, & \mbox{if } k^*(t) = m \ \mbox{  and  } \  k^*(t + \Delta) = m' \\ 0, & \mbox{otherwise}\end{cases} ,
\ee

\noindent where $0 \leq m,m' \leq k^*_{max}$ are associated with the observed azimuthal wave number at a given instant of time. The desired time correlation functions can be defined as
\bea 
\Tilde{F}(t-t') &\equiv& \langle \mathbf{V}(t) \cdot \mathbf{V}(t') \rangle -  \langle \mathbf{V} \rangle ^2 \ , \ \label{Fcorrelationexp} \\
\Tilde{G}(t-t') &\equiv& Tr [ \langle \mathbf{M}^T(t) \mathbf{M}(t') \rangle -  \langle \mathbf{M} \rangle^T \langle \mathbf{M} \rangle ] \ , \ 
\eea 

\noindent with their respective normalized versions being
\be 
    F(t-t') \equiv \frac{\Tilde{F}(t-t')}{\Tilde{F}(0)} \ , \  G(t-t') \equiv \frac{\Tilde{G}(t-t')}{\Tilde{G}(0)} \ . \
    \label{normalizedFG}
\ee

These correlation functions describe the tendency of a given mode to return to its current state and the correlation of transitions which are separated by a time interval of $\vert t - t' \vert$. Their behavior are plotted in Fig. \ref{FandGExp}, which exhibits a power-law scaling with exponent $-1$ for both functions up to $\delta t \approx 20 \Delta = 2$ s $\approx 2D/U$ --i.e., scaling with the flow outer units, suggesting that memory effects might be present near subsequent CS, since their mean lifetime is approximately $\delta t/10$ \cite{JackelPRF2023}. For longer time separations, the correlation drops abruptly due to undersampling, a fact that was verified by considering a 10\% larger data set, which resulted on this drop being pushed to larger time lags. However, to considerably increase this time window, the data sets should be orders of magnitude larger, thus a compromise must be sought.

Although this behavior seems surprising, it agrees with the non-Markovianity of the stochastic series, once that if the process were Markovian, this types of correlation functions should decay exponentially --much faster than any algebric decaying.

\begin{figure}[h!]
    \centering
        \includegraphics[width=0.8\textwidth]{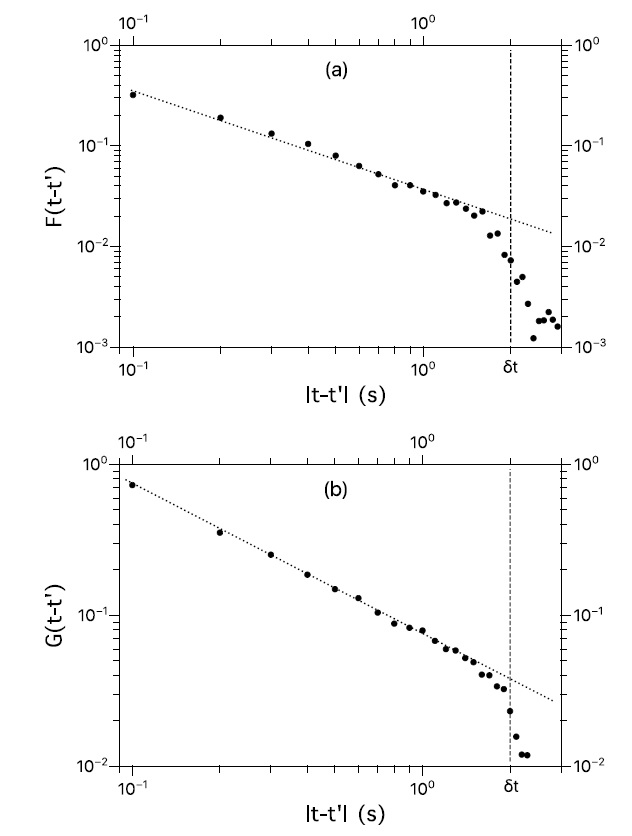}
        \caption{The time-dependent correlation functions defined in Eq.~(\ref{normalizedFG}) are noticed to decay as power laws for $\vert t - t' \vert \leq  \delta t \approx 2$ s. The dotted lines in (a) and (b) have scaling exponent $-1$ for both $F(t - t')$ and $G(t - t' )$ [47].}
    \label{FandGExp}
\end{figure}

\vspace{1.0 cm}

\subsection{Markovian Model for Microstates}

\hspace{0.5 cm}Even though the fact that the stochastic process $\mathcal{S}$ has a non-Markovian behavior, we can still create a Markovian process for its ``microstates" which can generate the same macroscopic state associated with a given azimuthal wave number.

In order to describe the above statement more concretely, we assume that the creation of a given macroscopic state $k^* = m$ can be made from the combination of
\be 
    \Omega(k^*_{max},m) = \binom{k^*_{max}}{m}
\ee 

\noindent different ways, which represent the microstate degeneracy to build a given macrostate. The main physical idea behind this separation between micro and macro states is to take into account the different ways in which a given macrostate observed could be formed. The positions of the $m$ filled low speed streaks can be any channel of the equally spaced in $k^*_{max}$ regions in the pipe's azimuthal direction. Figure \ref{modelillustration} shows an observed snapshot of streamwise velocity fluctuation with identified dominant mode $m = 4$ on the left --same snapshot used for the spectrum shown in Fig. \ref{spectrum}-- and on the right how the pipe is partitioned with the correspondent streak channels active. 

\begin{figure}[h!]
    \centering
        \includegraphics[width=0.8\textwidth]{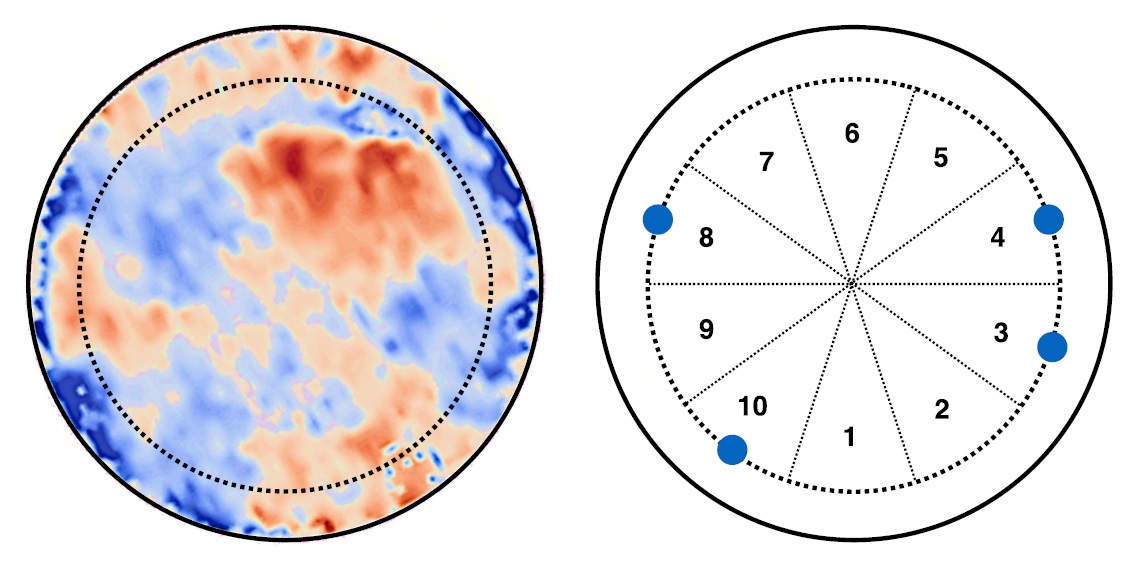}
        \caption{The sPIV snapshot on the left gives the power spectrum of Fig. \ref{spectrum}. The reference radial distance used in Eq.~(\ref{spectrum2}), $r_0$, is the radius of the dotted circle, which essentially crosses the streamwise velocity profile in four disconnected regions that contain low-speed streaks (blue spots). Partitioning the snapshot into $k^*_{max} = 10$ slices, we schematically represent on the right the positions of the low-speed streaks as small blue dots (the active streak channels are hence labeled as 3, 4, 8, and 10) [47].}
    \label{modelillustration}
\end{figure}

The phase space of the microstates for the desired Markovian model are defined by all possible sets of $k^*_{max}$  streak bits, $X \equiv \{ s_1,s_2,...,s_{k^*_{max}} \} $, where 
\be 
    s_i(t) = \begin{cases} 1, & \mbox{if the i-th streak channel is active } \\ 0, & \mbox{otherwise  }\end{cases} \ ,
\ee 

\noindent and the macroscopic state $m$ at time $t$ is simply $m(t) = \sum_{i = 1}^{k^*_{max}}s_i(t)$.

The evolution of the microstates $X$ occurs with independent persistence probabilities $q_m$ and $p_m$ of a given streak bit to remain inactive and active, respectively. Their counterpart are defined by $1-q_m$ and $1-p_m$, which are associated with the probabilities of a given inactive state to become active and an active state to become inactive, respectively.

Table \ref{table:1} summarizes the main parameters of the model when a given mode transition $m \to m'$ occurs, where $m = n_3 + n_4$ is the sum of the active states before the transition and $m' = n_2 + n_4$ is the number of active states after the transition.

\begin{table}[h!]
\centering
\begin{tabular}{|c c c|} 
 \hline
 \hline
 Transition type \hspace{0.5 cm} & \# of streak channels \hspace{0.5 cm} & Transition probability\\ [0.5ex] 
 \hline
 $0 \to 0$ & $n_1$ & $q_m$  \\ 
 \hline
 $0 \to 1$ & $n_2$ & $1-q_m$  \\
 \hline
 $1 \to 0$ & $n_3$ & $1-p_m$ \\
 \hline
 $1 \to 1$ & $n_4$ & $p_m$ \\
 \hline
 \hline
\end{tabular}
\caption{Definition of the four possible transition types for the streak channel states together with the
notations for their occurrence numbers and individual transition probabilities. $m = n_3 + n_4$ labels the azimuthal mode [47].}
\label{table:1}
\end{table}

Using the aforementioned definitions, it is not difficult to show that the transition probability between two macrostates $T_{m'm}$, taking into account the degeneracy factors, will be given by
\bea
    T_{m'm} =&&\binom{k^*_{max}}{m}^{-1} \sum_{n_1 = 0}^{k^*_{max}} \sum_{n_2 = 0}^{k^*_{max}} \sum_{n_3 = 0}^{k^*_{max}} \sum_{n_4 = 0}^{k^*_{max}} \delta\left (\sum_{i=1}^4 n_i, k^*_{max} \right ) \delta(n_3 + n_4, m) \delta(n_2 + n_4, m')  \nonumber  \\
    &\times& \binom{k^*_{max}}{n_1} \binom{k^*_{max} - n_1}{n_2} \binom{k^*_{max} - n_1 - n_2}{n_3}q_m^{n_1}(1-q_m)^{n_2}(1-p_m)^{n_3}p_m^{n_4}  \ , \ 
    \label{Transition_matrix}
\eea 

\noindent where $\delta(a,b) := \delta_{ab}$ are the discrete Kroenecker delta functions, which are represented in this form for simplicity.

Fixing $k^*_{max} = 10$, in agreement with the empirical data, the Markovian model for microstate transitions has 20 undetermined parameters ($q_0,...,q_9$) and ($p_1,...,p_{10}$). These are computed through the minimization of the quadratic error
\be
    d(\{ q_m \}, \{ p_m \}) \equiv \vert\vert \mathbb{P} - \mathbb{P}_{\infty} \vert\vert^2 \ ,
    \label{minimizationofd}
\ee 

\noindent where $\mathbb{P}$ and $\mathbb{P}_{\infty}$ are the asymptotic probability eigenvector from the model and from the experimental observations, respectively. The optimization scheme described in Eq.~(\ref{minimizationofd}) is overdetermined, since there are 20 independent parameters to model the nine independent asymptotic probabilities. To overcome this issue, it was assumed that: (i) the persistence probabilities are all equal, not small, and mode-independent, i.e., $p = p_2 = ... = p_{10}$ and (ii) the transitions from mode $m=1$ to $m=0$ are suppressed by imposing $p_1 = 1$ --transitions to the mode $m=0$ from other modes are possible, but they are not significant since they are of $\mathcal{O}\left [ (1-p)^2 \right ]$.

The resulting set of parameters is left with 11 independent entries $(q_0,...,q_9,p)$ which are determined by a Monte Carlo procedure to obtain the set of $q_m$ for different fixed values of $p$. As it can be seen in Fig. \ref{minimizationofdfig}, the quadratic error decreases considerably fast for $p\geq 0.85$. The list of the obtained probabilities for the fixed values of $p = 0.86$ and $p=0.95$ is shown in Table \ref{table:2}, which were the ones used to compute the comparisons to the empirical observables. 

\begin{figure}[h!]
    \centering
        \includegraphics[width=0.9\textwidth]{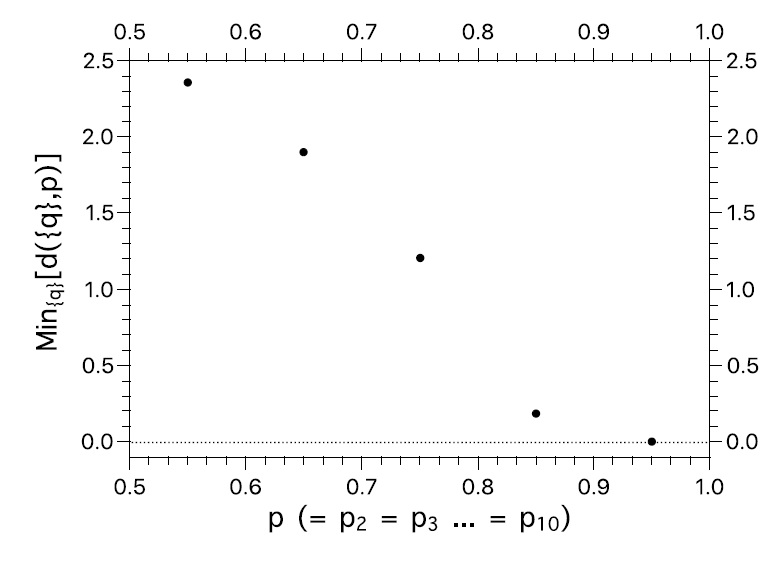}
        \caption{Minimization of the quadratic distance $d(\{ q \},p)$ for various values of $p$ [47].}
    \label{minimizationofdfig}
\end{figure}

\begin{table}[h!]
\centering
\begin{tabular}{|c c c c c c c c c c c|} 
 \hline
 \hline
 $p$ & $q_0$ & $q_1$ & $q_2$ & $q_3$ & $q_4$ & $q_5$ & $q_6$ & $q_7$ & $q_8$ & $q_9$ \\ [0.5ex] 
 \hline
 0.86 & 0.53 & 0.96 & 0.95 & 0.92 & 0.92 & 0.85 & 0.95 & 0.75 & 0.86 & 1.0  \\ 
 \hline
 0.95 & 0.22 & 0.98 & 0.98 & 0.97 & 0.97 & 0.96 & 0.97 & 0.93 & 0.94 & 0.49  \\
 \hline
 \hline
\end{tabular}
\caption{List of probabilities $q_m$ which describe the persistence of inactive streak channels for the cases $p = 0.86$ and $p = 0.95$ [47].}
\label{table:2}
\end{table}

\vspace{0.5 cm}

The comparison between the asymptotic probabilities of the model and the empirical ones are shown in Fig. \ref{modeledprobs} with an excellent agreement. The fixed values of $p$ with its respective sets of $q_m$ were used to compute the time correlation functions $F(t-t')$ and $G(t-t')$ and are presented in Fig. \ref{FandGExpPlusModel}. As mentioned, the Markovian model in the microstates is able to  describe the non-Markovian behaviour reproducing the algebric decaying of the proposed correlation functions.

The model is able to produce a stochastic series, which is large enough to compute the correlation functions beyond the crossover timescale $\delta t$. 

\begin{figure}[h!]
    \centering
        \includegraphics[width=0.8\textwidth]{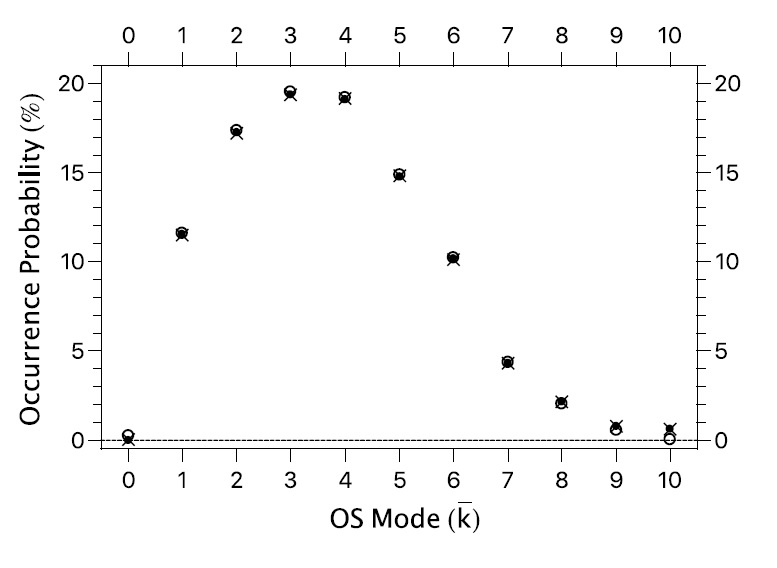}
        \caption{Occurrence probability of coherent structures modes obtained from the experiment (dots) and from the stochastic model (open circles: $p = 0.86$; crosses: $p = 0.95$), defined by the transition matrix elements (\ref{Transition_matrix}) [47].}
    \label{modeledprobs}
\end{figure}

\begin{figure}[h!]
    \centering
        \includegraphics[width=0.8\textwidth]{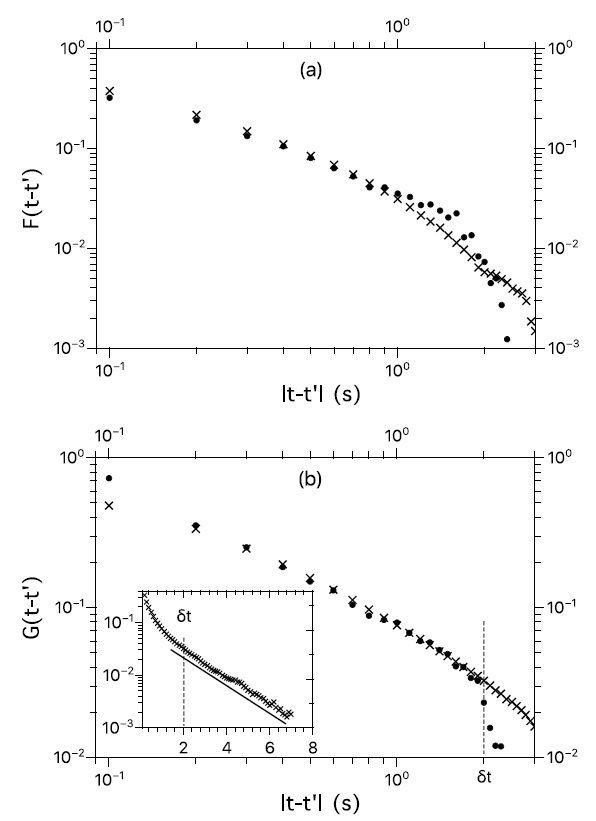}
        \caption{Empirical (dots) and modeled (crosses) correlation functions $F(t - t')$ and $G(t - t')$. Crosses refer in (a) and (b), respectively, to modeling parameters $p = 0.86$ and $p = 0.95$. The semilog plot in the inset of (b) indicates the simple exponential form of $G(t - t')$ at large enough $\vert t - t' \vert$ [47].}
    \label{FandGExpPlusModel}
\end{figure}

As is illustrated in the inset of Fig. \ref{FandGExpPlusModel} the correlation function drops exponentially, as it was expected, since the non-Markovian behavior should transition to the Markovian one for large time separations, as the memory effects become smaller. However, a more extensive study with larger data sets is required to confirm these observations, which by now, are set as a prediction for the long time correlation associated with these mode transitions.

As a result, the physical picture that appears from this work is that these azimuthal modes are packed as chains of low-speed streaks and vortical structures, which are strongly correlated within sizes that scale with the pipe's diameter, although they can be larger and are merged along the entire pipe flow. This could be seen as evidence of the features presented by the hairpin vortex packets \cite{Adrian2000,Khoury2013,Wu2015}, since that the wall's topology would encapsulate their formation and set a complex interaction and organization.

\end{chapter}

\begin{chapter}{Lattice Boltzmann Simulations}
\label{cap6}

\hspace{0.5 cm}We describe in this chapter simulations of MHD flows and pipe flow turbulence which we carried out employing the LBM. A MRT collision operator is set for the magnetic vector-valued distributions together with a distance dependent boundary condition. The QS approximation is simulated and compared with analytical solutions for Laminar MHD pipe flows \cite{Magacho2023,Tavares2023}. Lastly, the simulation and statistical analysis of a turbulent pipe flow with $Re = 5300$ are shown \cite{Magacho2024}.

\section{Magnetic MRT Collision Model}

\hspace{5 mm} The BGK collision model for the magnetic vector valued distributions has essentially the same advantages and disadvantages as its hydrodynamic counterpart. The main difference is related to the fact that in the magnetic case, in order to simulate the QS regime, in general, one wants to increase the magnetic diffusivity and consequently the magnetic relaxation time given by Eq.~(\ref{magrelaxtime}). The observed fact is that this collision model also lacks stability for relaxation times considerably different from unity.

Having these considerations in mind, a MRT model for the magnetic induction LBM is proposed, taking advantage of the similarities with the MRT collision model developed for convection diffusion (CD) systems \cite{Yoshida2010}, since, in fact, Eq.~(\ref{IE}) can be seen as a CD equation in the presence of a source term. To start, define the magnetic vector-valued moment distributions $\mbf{m}_i$, as
\be 
    \boldsymbol{m}_i  = \langle M_i \vert \boldsymbol{g} \rangle \ , \ \label{mi}
\ee 

\noindent where $\vert \mbf{g} \rangle = (\mbf{g}_0,\mbf{g}_1,...,\mbf{g}_6)^T$, and 
\be
    M = 
    \begin{bmatrix}
    \langle 1 \vert \\
    \langle \xi_x \vert \\
    \langle \xi_y \vert \\
    \langle \xi_z \vert \\
    \langle 6 - 7\boldsymbol{\xi}^2 \vert \\
    \langle 3 \xi_x^2 - \boldsymbol{\xi}^2 \vert \\
    \langle \xi_y^2 - \xi_z^2 \vert 
    \end{bmatrix}
    =
    \begin{bmatrix}
    1 & 1 & 1 & 1 & 1 & 1 & 1\\
    0 & 1 & -1 & 0 & 0 & 0 & 0\\
    0 & 0 & 0 & 1 & -1 & 0 & 0\\
    0 & 0 & 0 & 0 & 0 & 1 & -1\\
    6 & -1 & -1 & -1 & -1 & -1 & -1\\
    0 & 2 & 2 & -1 & -1 & -1 & -1\\
    0 & 0 & 0 & 1 & 1 & -1 & -1
    \end{bmatrix} \ . \
\ee 

Now, Eq.~(\ref{bgkmagnetic}) is replaced by
\be
\boldsymbol{g}_i(\mathbf{x} + \boldsymbol{\xi}_i, t + 1) - \boldsymbol{g}_i(\mathbf{x}, t)=- \sum_{j=0}^6 (M^{-1} S M)_{ij} 
[ \boldsymbol{g}_j(\mathbf{x}, t) - \boldsymbol{g}^{eq}_j(\mathbf{x, t}) ] \ , \
\label{premrtmag}
\ee

\noindent where the MRT collision matrix $S$, is defined as
\be
    S^{-1} = 
    \begin{bmatrix}
    \tau_0 & 0 & 0 & 0 & 0 & 0 & 0\\
    0 & \tau_{xx} & \tau_{xy} & \tau_{xz} & 0 & 0 & 0\\
    0 & \tau_{yx} & \tau_{yy} & \tau_{yz} & 0 & 0 & 0\\
    0 & \tau_{zx} & \tau_{zy} & \tau_{zz} & 0 & 0 & 0\\
    0 & 0 & 0 & 0 & \tau_{4} & 0 & 0\\
    0 & 0 & 0 & 0 & 0 & \tau_{5} & 0\\
    0 & 0 & 0 & 0 & 0 & 0 & \tau_{6}
    \end{bmatrix} \label{matrix} \ . \
\ee

Restricting to isotropic magnetic diffusion,
\be
\tau_{xx} =  \tau_{yy} =  \tau_{zz} = \tau_m \ , \ \tau_{xy} = \tau_{yx} =\tau_{xz} = \tau_{zx} = \tau_{yz} =\tau_{zy} = 0 \ , \ \label{taus}
\ee
where $\tau_m$ is the usual collision relaxation time, and, by convention,
\be
\tau_0 = \tau_4 =\tau_5 = \tau_6 = 1 \ . \
\ee

The latter imposition is associated with the suppression of the zero-th and high-order moments by imposing their value to the equilibrium exactly in each time step. The magnetic BGK collision model is recovery for the specific case of $\tau_0 = \tau_4 =\tau_5 = \tau_6 = \tau_m$, of course.

Applying $M$ to both sides of Eq.~(\ref{premrtmag}), the MRT model for the magnetic moments will be given by
\be
\boldsymbol{m}_i (\boldsymbol{x} + \boldsymbol{\xi_i}, t + 1)  
- \boldsymbol{m}_i (\boldsymbol{x}, t)  =  - \sum_{j=0}^6 S_{ij}  [ \boldsymbol{m}_j (\boldsymbol{x}, t) - \boldsymbol{m}^{eq}_j  (\boldsymbol{x}, t) ]
\ . \  \label{moments}
\ee

The multiplet of equilibrium moments, $\boldsymbol{m}_j^{eq}$, straightforwardly computed from Eq.~(\ref{gEq}), is
\be
\vert \boldsymbol{m}^{eq} \rangle = ( \mathbf{B} \ , \ u_x \mathbf{B} - B_x \mathbf{u} \ , \ 
u_y \mathbf{B} - B_y \mathbf{u} \ , \  u_z \mathbf{B} - B_z \mathbf{u} \ , \ 3\mathbf{B}/4 \ , \ 0 \ , \ 0 )^T \ . \ \label{meqB}
\ee

After the collision and streaming steps of Eq.~(\ref{moments}), the magnetic vector-valued distributions can be obtained by the inverse transformation of Eq.~(\ref{mi}) and the magnetic field can be computed through Eq.~(\ref{macroscopicB}).

So far, this MRT model for the magnetic vector valued distribution only has the effect of increasing its stability for higher values of relaxation time. To simulate the desired QS regime described by Eq.~(\ref{QSeq}) within the LBM, the time derivative of the induced magnetic field cannot be neglected. In this case, the desired equations can be written as
\be
    \partial_t \mathbf{b} =  - (\mathbf{u} \cdot \mathbf{\nabla})\mathbf{B}_{ext} + (\mathbf{B}_{ext} \cdot \mathbf{\nabla})\mathbf{u} + \eta \nabla^2 (\mathbf{b} + \mathbf{B}_{ext}) \ . \
    \label{IELBM}
\ee

The effect of suppressing the time derivative should come as a consequence of the higher stability when increasing $\eta$. Mutatis mutandis, the MRT for the induced part of the magnetic field in the QS regime still follows the same structure as in Eq.~(\ref{moments}), but with Eq.~(\ref{meqB}) replaced by
\be
    \vert \boldsymbol{m}^{eq} \rangle = ( \mathbf{b} \ , \ u_x \mathbf{B}_{ext} - B_{x,ext} \mathbf{u} \ , \  u_y \mathbf{B}_{ext} - B_{y,ext} \mathbf{u} \ , \ u_z \mathbf{B}_{ext} - B_{z,ext} \mathbf{u} \ , \ 3\mathbf{b}/4 \ , \ 0 \ , \ 0)^T.
\ee 

Another great advantage of working with the induced magnetic field --instead of the electric potential, which is also used in the literature-- is related to the application of Dirichlet boundary conditions.  For this purpose, Appendix \ref{apendicea} shows the derivation of a distance dependent Dirichlet boundary condition for the vector valued distributions, based on the extension from a convection diffusion one \cite{Li2013}, which is given by
\bea 
\boldsymbol{g}_{\overline{i},\alpha}(\boldsymbol{x}_f, t + 1) &=&  2(\Delta -  1)\tilde{\boldsymbol{g}}_{i,\alpha}(\boldsymbol{x}_f,t) - \bigg ( \frac{(2 \Delta - 1)^2}{2 \Delta + 1} \bigg )\tilde{\boldsymbol{g}}_{i,\alpha}(\boldsymbol{x}_{ff},t) +  
\nonumber \\
&+& 2 \bigg ( \frac{2 \Delta -1}{2 \Delta + 1} \bigg ) \tilde{\boldsymbol{g}}_{\overline{i},\alpha}(\boldsymbol{x}_f,t) + \frac{1}{3} \bigg ( \frac{3 - 2 \Delta}{2 \Delta + 1} \bigg ) b_\alpha |_{wall} \ , \
    \label{boundarycondition_gmag}
\eea

\noindent where $\boldsymbol{x}_f$ and $\boldsymbol{x}_{ff}$ are the flow points which are nearest and next to nearest of the boundary. The case of interest in this dissertation is associated with insulated walls, which can be translated to ($\mbf{b}\vert_{walls} = 0$).

\section{Simulation Results for the MRT Model}

\hspace{0.5 cm}To validate the developed CM-MRT scheme, we analyze its performance in two well-known benchmark problems, such as the Orszag-Tang (OT) vortex model \cite{Orszag1979,Mininni2006,Jadhav2021} and the Gold's problem \cite{gold1962}. The latter is associated with a laminar pipe flow subject to a uniform transverse magnetic field. Furthermore, we explored the numerical performance for the case of a pipe flow under the presence of a non-homogenous magnetic field.

\subsection{3D Orszag-Tang Vortex}

The dynamics of the OT vortex problem is related to all the terms in Eq.~(\ref{IE}). So, to simulate the desired system in a CM-MRT fashion, Eqs.~(\ref{moments}) and (\ref{meqB}) have to be used, without taking into account the QS effects. The initial velocity and magnetic fields in the OT model are given by
\bea 
\mathbf{u}(\boldsymbol{x},t=0) =&& \hspace{-0.5 cm} 2u_0[ sin(Y),sin(X),0] \ , \  \\ 
\mathbf{B}(\boldsymbol{x},t=0) =&& \hspace{-0.5 cm} 0.8B_0[-2sin(2Y) + sin(Z), \nonumber \\ 
&&\hspace{-0.5 cm} 2sin(X) + sin(Z),sin(X) + sin(Y)] \ , \
\eea 

\noindent where $X_i = 2\pi x_i/n_{x_i}$ and $u_0 = B_0 = 0.0203$. Figure \ref{6.1} shows the maximum value of the current density $J = \nabla \times \mathbf{B}$ as a function of time for $R_m=(10,100,200)$ and $Re = (100,400,2000)$. Simulations are performed with a computational grid of size $32^3$.

As depicted in Fig. \ref{6.1}, the current density dynamics is well-behaved for all the nine studied cases. The absolute value of vorticity and current density for the most dynamical case with $R_m=200$ and $Re = 2000$ at $t=1000$ is shown in Fig. \ref{6.2}. The algorithm proved to be able to capture the complex dynamics in the OT vortex problem associated with high velocity gradients and the occurrence of magnetic reconnections.

\begin{figure}[h!]
    \centering
        \includegraphics[width=0.8\textwidth]{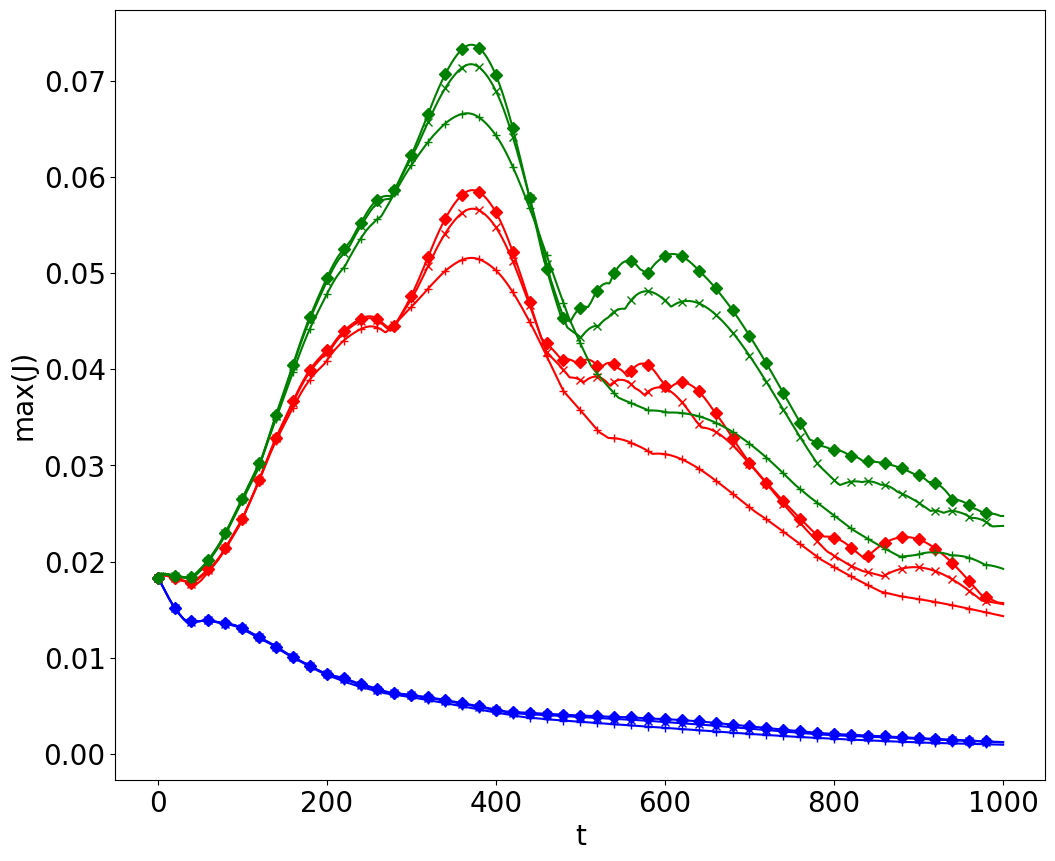}
        \caption{Maximum value of the current density as a function of the time. Blue curves are associated with $R_m=10$, red to $R_m=100$, and green to $R_m=200$. The three studied Reynolds number cases are $Re=100$ (plus), $Re=400$ (cross), and $Re=2000$ (diamond) [58].}
    \label{6.1}
\end{figure}

\begin{figure}[h!]
    \centering
        \includegraphics[width=1.0\textwidth]{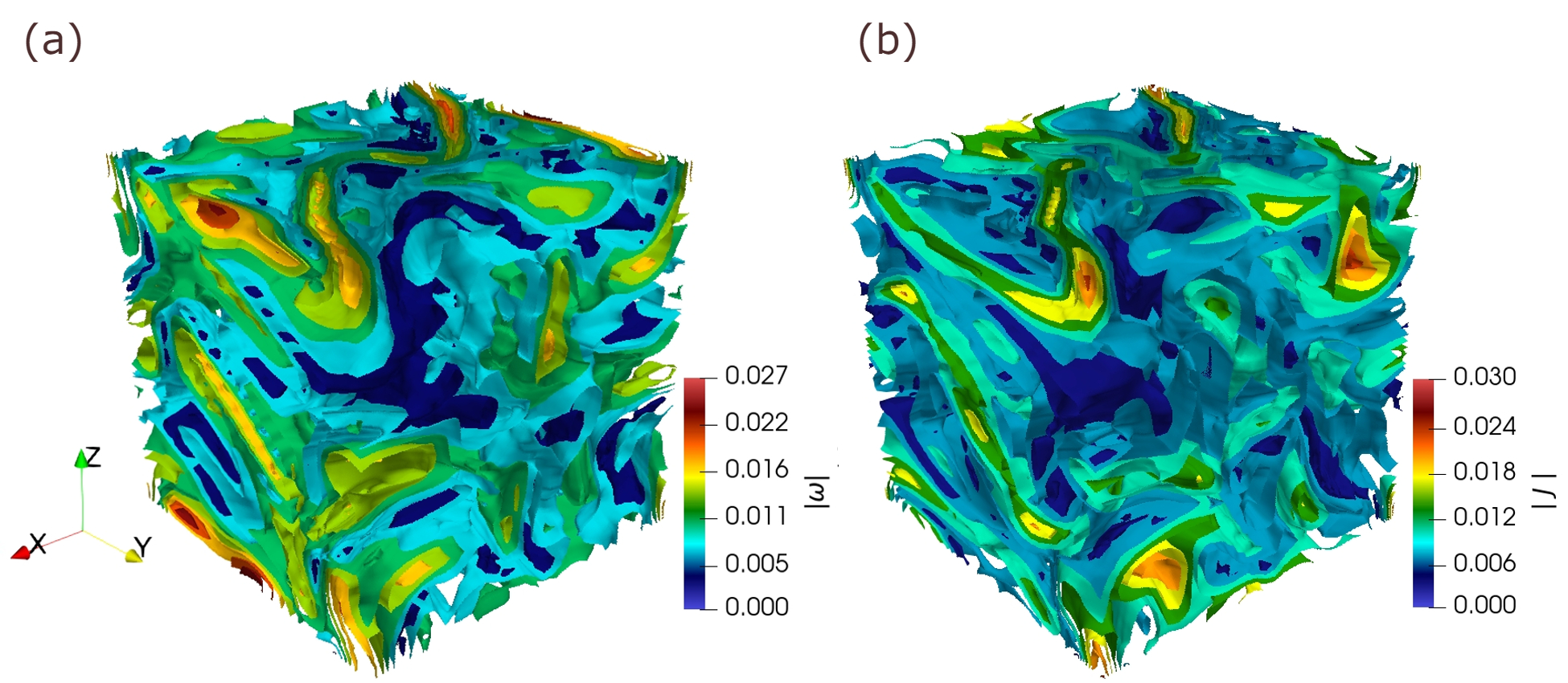}
        \caption{Absolute values of (a) vorticity and (b) current density. Both cases have $Re=2000$ and $R_m=200$ [58].}
    \label{6.2}
\end{figure}

\subsection{Pipe Flow in the Presence of a Transverse Uniform Magnetic Field}

\hspace{0.5 cm}The flow dynamics of an electrically conducting fluid in a pipe with insulating walls and radius $R$ in the $z$ direction --the pipe' symmetry axis-- in the presence of a transverse magnetic field $\mbf{B} = B_0\hat{y}$, and driven by a constant pressure gradient $h = \partial p/ \partial z$ is dictated by Eqs.~(\ref{NSE}), (\ref{IE}), and (\ref{nablaUB}). This setup is known by Gold's analytical solution for laminar regimes and are describe by 
\bea 
&&u_z(r,\theta) = \frac{R^2 h}{2 {\hbox{Ha}} \nu} \left [ \sum_{n = -\infty}^{\infty} \{e^{-\alpha r cos(\theta)} + (-1)^n e^{\alpha r cos (\theta)} \}\frac{I'_{n}(\alpha)}{I_n(\alpha)}I_n(\alpha r) e^{in\theta} \right ] \ , \ \label{goldu} \\
&&b_z(r,\theta) = B_z(r,\theta) = \frac{R^2 h}{2 {\hbox{Ha}} (\eta \nu)^{1/2}} \times \nonumber \\
&& \left [ \sum_{n = -\infty}^{\infty} \{ e^{-\alpha r cos (\theta)} - (-1)^n e^{\alpha r cos (\theta)} \} \frac{I'_{n}(\alpha)}{I_n(\alpha)}I_n(\alpha r) e^{in\theta} - 2r cos (\theta) \right ]
\label{goldb}
\ , \
\eea 

\noindent where $r$ and $\theta$ are the polar coordinates, and $I_n$ is the n-th order modified Bessel function.

To simulate the aforementioned system, a lattice of $n_x \times n_y \times n_z = 40 \times 40 \times 5$ is used with a periodic boundary condition. The viscosity and initial uniform velocity and magnetic fields are $\nu = 0.04$, $\mbf{u}_0$, and $\mbf{b} = 0$ in lattice units. The Reynolds number is $Re \approx 40$, the explored Hartmann numbers, pipe radius, and pressure gradient are, respectively, $0\leq Ha\leq 15$, $R = 19.5$, and $h = u_0 \nu Ha/R^2$. Also, simulations were performed with magnetic Reynolds numbers varying in the range $0.2 \leq R_m \leq 39$.

The magnetic energy balance, obtained with the scalar product of $\mbf{b}$ with Eq.~(\ref{IELBM}), is shown in Fig. \ref{6.3} (a) for different values of magnetic Reynolds numbers at $Ha=5$. Figure \ref{6.3} (b) shows how the suppression of $\partial_t \mbf{b}^2$ occurs as the magnetic Reynolds number is decreased. Below $R_m = 0.2$, instabilities start to occur, related to the high values of magnetic relaxation time $\tau_m \approx 40$. Compared with the BGK collision operator, the MRT model developed here can go at least two orders of magnitude lower for the same grid resolution, as BGK lacks below $R_m \approx 10$. Figure \ref{6.4} shows that the magnetic energy is being conserved during our simulations for the highest magnetic Reynolds number ($R_m = 39$), the case which exhibits the largest time dependent variations.

\begin{figure}[h!]
    \centering
        \includegraphics[width=0.7\textwidth]{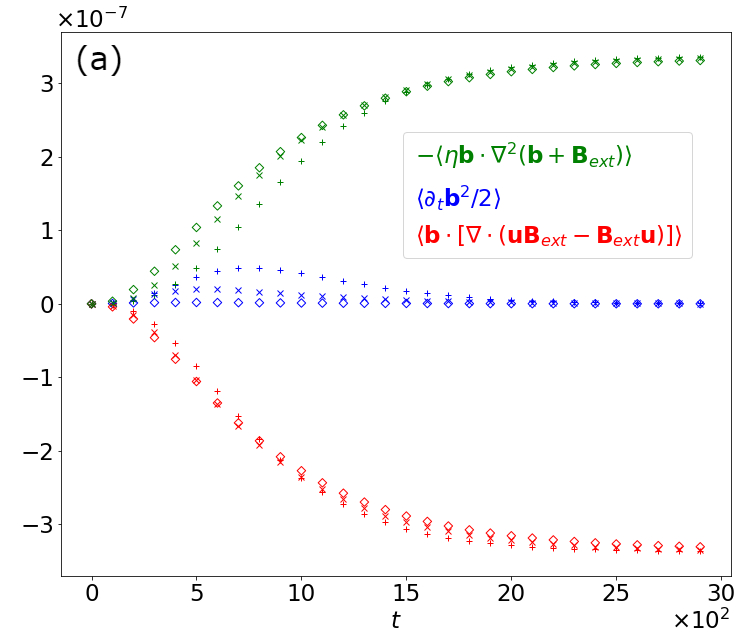}
        \includegraphics[width=0.7\textwidth]{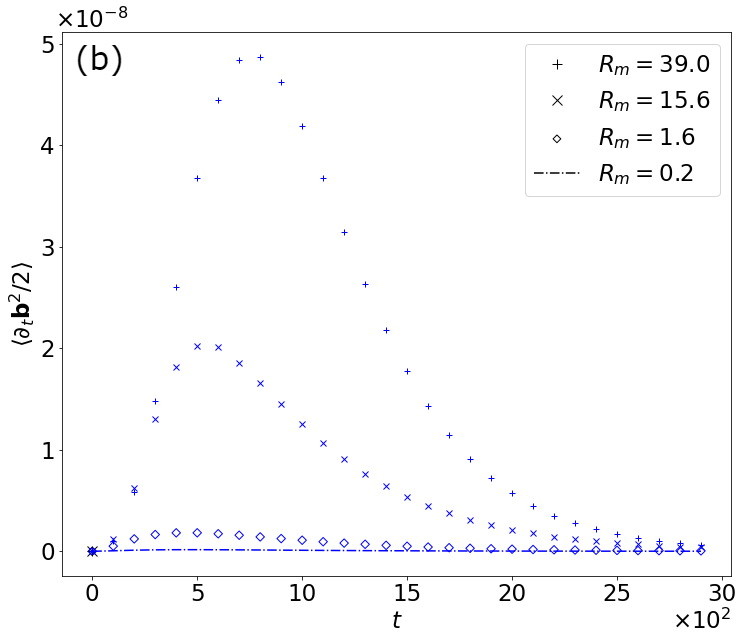}
        \caption{Results of the lattice-Boltzmann simulations at $Ha=5$. (a) Production terms for the magnetic energy balance at $R_m=39$ (blue plus), $R_m = 15.6$ (blue cross), and $R_m = 1.6$ (blue diamond). (b) A closer look at the time evolution of $\partial_t \mbf{b}^2/2$ for various $R_m$. [58].}
    \label{6.3}
\end{figure}

\begin{figure}[h!]
    \centering
        \includegraphics[width=0.9\textwidth]{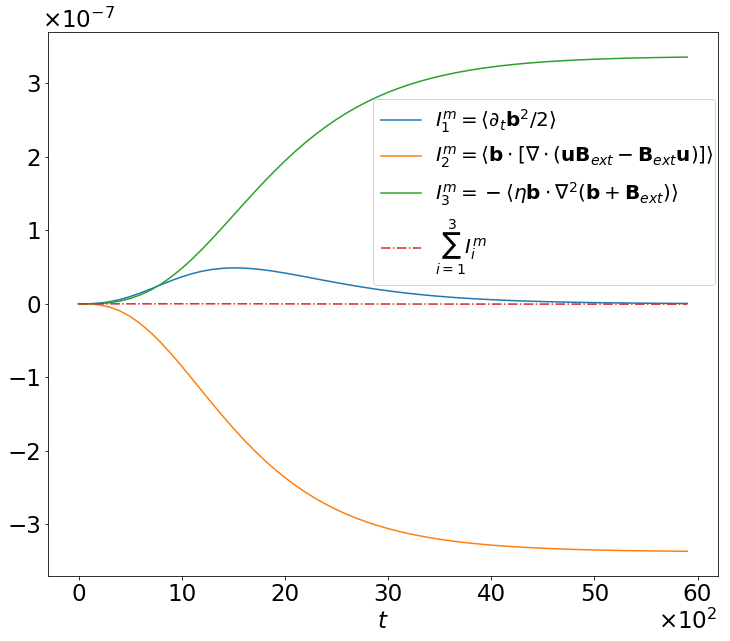}
        \caption{Magnetic energy balance for $Ha=5$ and $R_m=39$ [58].}
    \label{6.4}
\end{figure}

Furthermore, the dependence of the results when varying the Hartmann number was also investigated. It was found that the proposed scheme displays an excellent convergence with a computational grid of $n_x \times n_y \times nz = 80 \times 80 \times 5 $ and a magnetic Reynolds number of $R_m = 15$ ($\tau_m = 4.5$). The streamwise velocity and induced magnetic fields, together with their analytical solutions given by Eqs.~(\ref{goldu}) and (\ref{goldb}), are shown in Figs. \ref{6.5} (a) and (b), respectively. The relative error computed with a norm $L^2$ was approximately $1\%$ for all tested cases.

\begin{figure}[h!]
    \centering
        \includegraphics[width=0.9\textwidth]{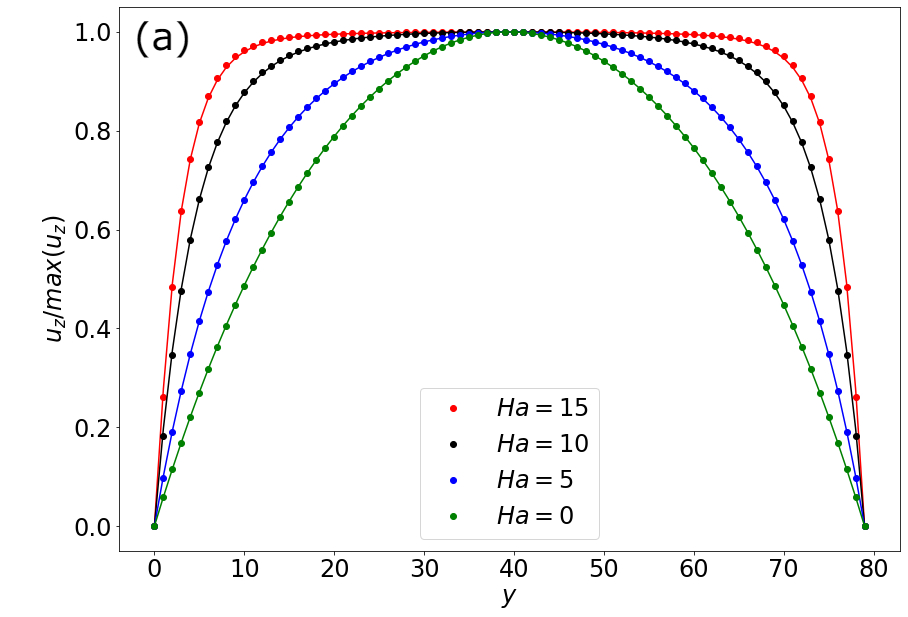}
        \includegraphics[width=0.9\textwidth]{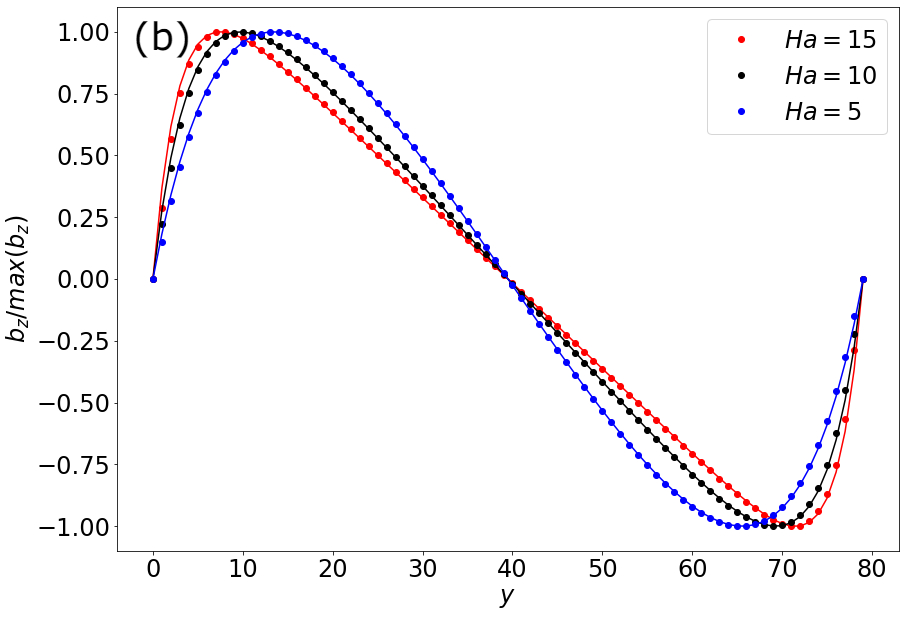}
        \caption{Results for simulations at $R_m=15$. (a) Streamwise velocity field $u_z$, normalized by its centerline value. (b) The induced magnetic field $b_z$. Solid lines represent the Gold solution (\ref{goldu}) and (\ref{goldb}) [58].}
    \label{6.5}
\end{figure}

\subsection{Pipe Flow in the Presence of a Non-Uniform Magnetic Field}

In order to test the proposed scheme in a more complex situation, we carried out a simulation of a pipe flow subject to a non-uniform magnetic field, composed by a hexagonal structure of ideally infinite magnetic slabs with alternating poles, as shown in Fig. \ref{6.6} (a). The magnetic field of a given magnetic slab can be obtained straightforwardly by Biot-Savart's law \cite{Griffithsbook}, and is given by
\bea 
    B_x(x,y) &=& \ln \left [ \frac{(x+L)^2 + y^2}{(x-L)^2 + y^2} \right ] \ , \ \label{Bslab1} \\
    B_y(x,y) &=& 2\left ( \arctan{\frac{x-L}{y}} - \arctan{ \frac{x+L}{y}  } \right ) \ , \ \label{Bslab2} \\
    B_z(x,y) &=& 0 \ , \ \label{Bslab3}
\eea 

\noindent where the $z$ component is due to symmetry, $L$ is the width of the slab's rectangular cross section with an aspect ratio 2. The in-plane magnetic vector field is shown in Fig. \ref{6.6} (b). We chose $L = R/6$ and the total magnetic field is obtained by the superposition of the fields given by rotations of (\ref{Bslab1}) and (\ref{Bslab1}) in the $xy$ plane.

\begin{figure}[h!]
    \centering
        \includegraphics[width=0.9\textwidth]{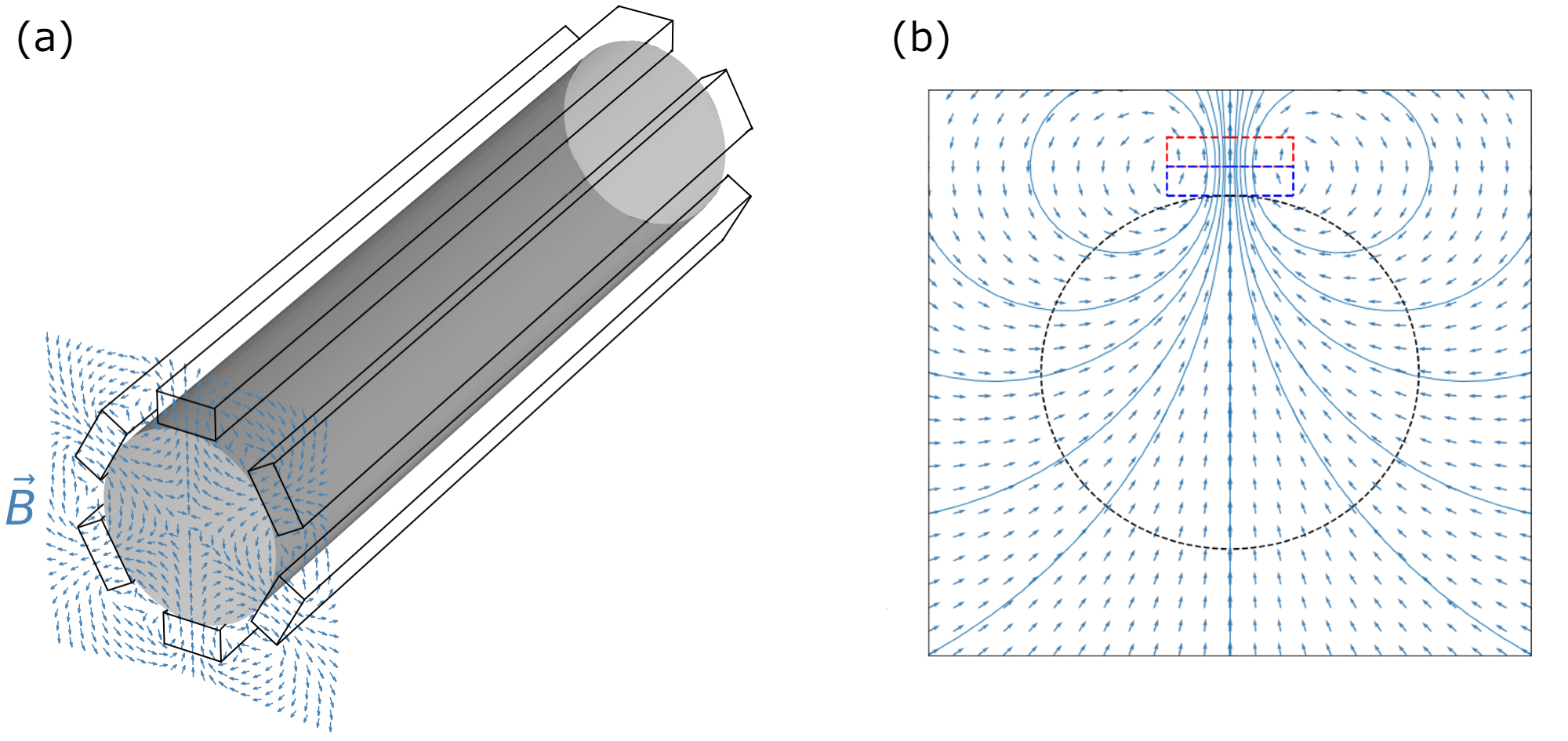}
        \caption{(a) Schematics of the pipe flow setup with a representation of the positions of the six covering magnetic slabs. (b) The non-uniform magnetic field produced by just one of the magnetic slabs is represented in a cross-sectional plane, as modeled in Eqs. (\ref{Bslab1}) and (\ref{Bslab2}) [58].}
    \label{6.6}
\end{figure}

Simulations were performed with the same computational grid as the uniform case at $Re = 80$ and $R_m = 3.16$. The estimated Hartmann number, using the maximum value of the magnetic field on the pipe' surface, was $Ha \approx 10$. As this specific case doesn't have an analytical solution for comparison, we stick to symmetry properties of the fields and the accuracy of magnetic and kinetic energy balances.

One can easily check that the NS equations on the presence of a Lorentz force (\ref{NSE}) is symmetric under the change $\mbf{B} \to -\mbf{B}$, which implies that the velocity field $u_z = u_z(x,y)$ must be symmetric under rotations of $\pi/3$ around the z axis, which could be observed from the contour lines of the streamwise velocity field shown in Fig. \ref{6.7} (a).  Also, one can see from the magnetic induction equation (\ref{IELBM}) that the induced field $\mbf{b}$ is anti-symmetric under the change $\mbf{B}_{ext} \to -\mbf{B}_{ext}$ --associated to discrete rotations of $\pi/3$ as well, which was also confirmed from the LBM simulation shown in Fig. \ref{6.7} (b).

\begin{figure}[h!]
    \centering
        \includegraphics[width=1.0\textwidth]{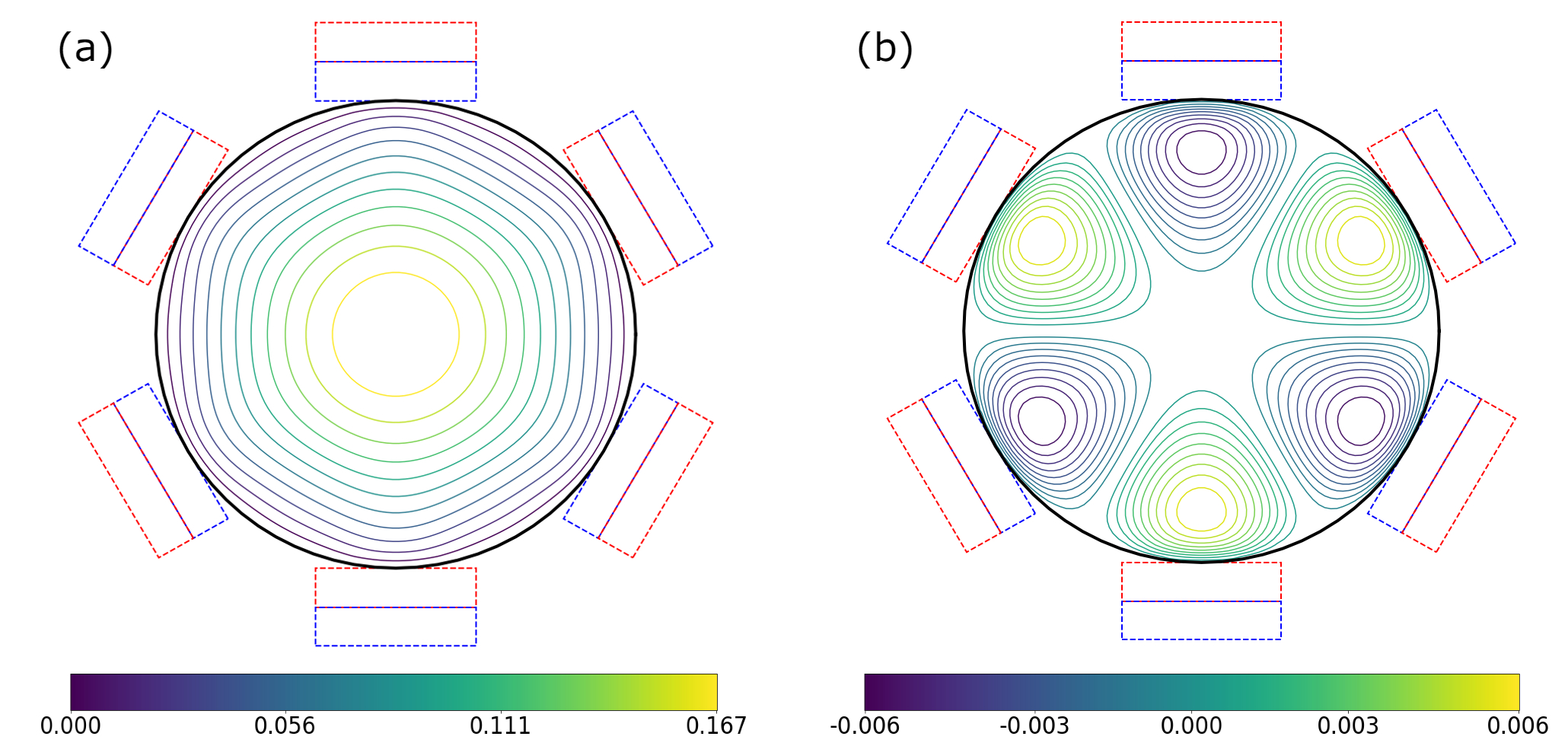}
        \caption{Level curves of (a) the velocity field and (b) the induced magnetic field. Both of them are parallel to the pipe’s symmetry axis. The color bars indicate the values of the velocity and magnetic fields [58].}
    \label{6.7}
\end{figure}

The kinetic and magnetic energy balances are shown in Fig. \ref{6.8} (a) and (b), respectively. As it can be seen, the CM–MRT lattice-Boltzmann simulations respect energy conservation all during the dynamic evolution, up to the asymptotic stationary regime.

\begin{figure}[h!]
    \centering
        \includegraphics[width=0.75\textwidth]{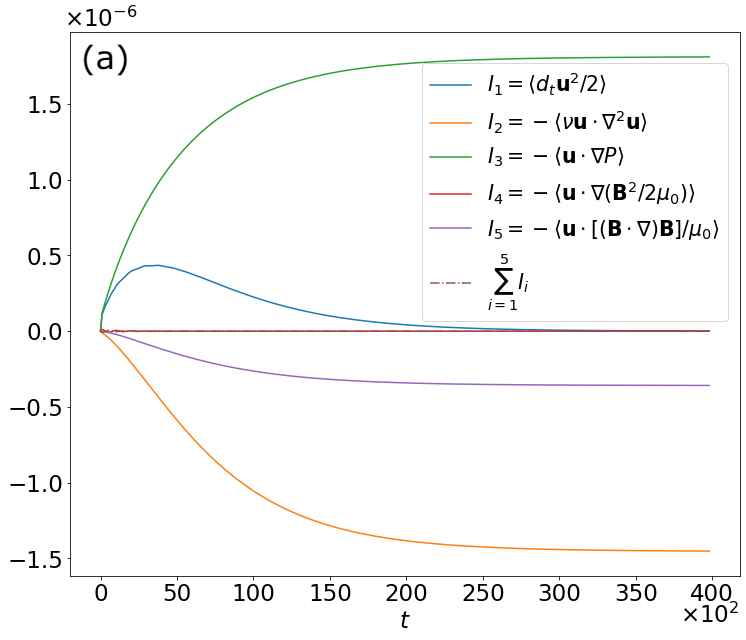}
        \includegraphics[width=0.75\textwidth]{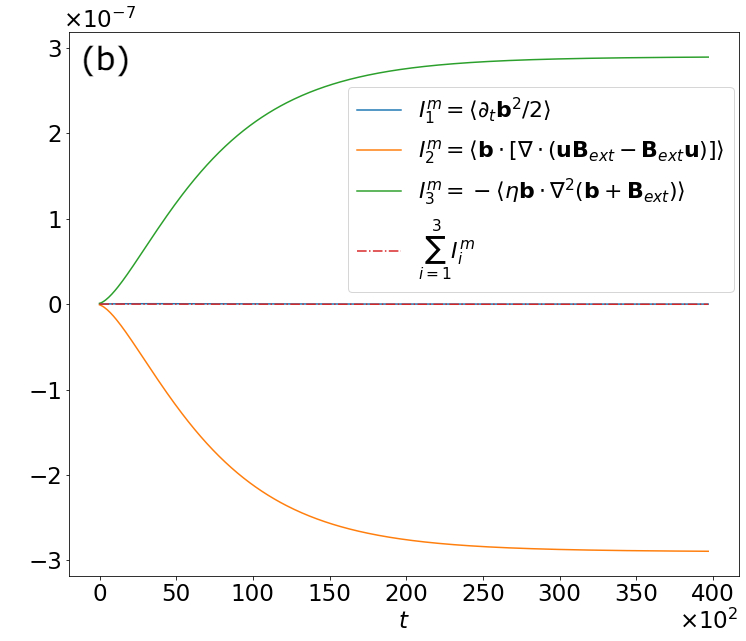}
        \caption{Energy balance analysis for (a) the Navier–Stokes Eq.~(\ref{NSE}) and (b) the magnetic induction Eq.~(\ref{IELBM}) [58].}
    \label{6.8}
\end{figure}

Figure \ref{6.9} shows a comparison between the energy balances provided by the approaches based on the CM–MRT and the usual CM–BGK lattice-Boltzmann strategies, both using the same parameter setup described above ($Ha \approx 10, Re = 80 ,R_m = 3.16 $). It turns out that the simulation carried out within the CM–BGK lines blows up after a few numerical iterations, while the proposed CM–MRT method is actually able to simulate, with excellent accuracy, the MHD evolution up to the stationary regime.

\begin{figure}[h!]
    \centering
        \includegraphics[width=0.8\textwidth]{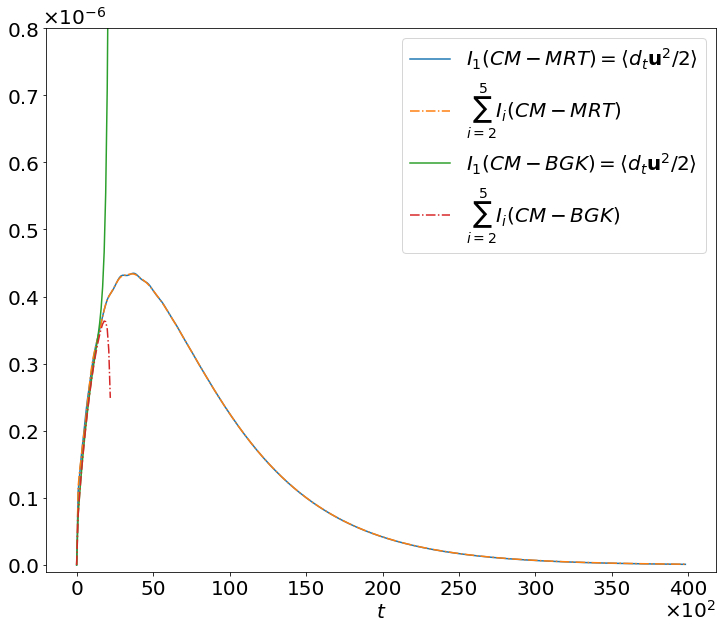}
        \caption{Comparison of the energy balance between the CM--MRT and the CM--BGK performances [58].}
    \label{6.9}
\end{figure}

Lastly, to display the near-wall velocity profiles along lines that are parallel ($\theta = \pi/2$) and perpendicular ($\theta = 0$) to the magnetic field lines, Fig. \ref{6.10} is showing how the streamwise velocity field vary in these directions up to $r'/R \approx 0.2$, where $r'$ is the distance from the pipe wall.

\begin{figure}[h!]
    \centering
        \includegraphics[width=0.8\textwidth]{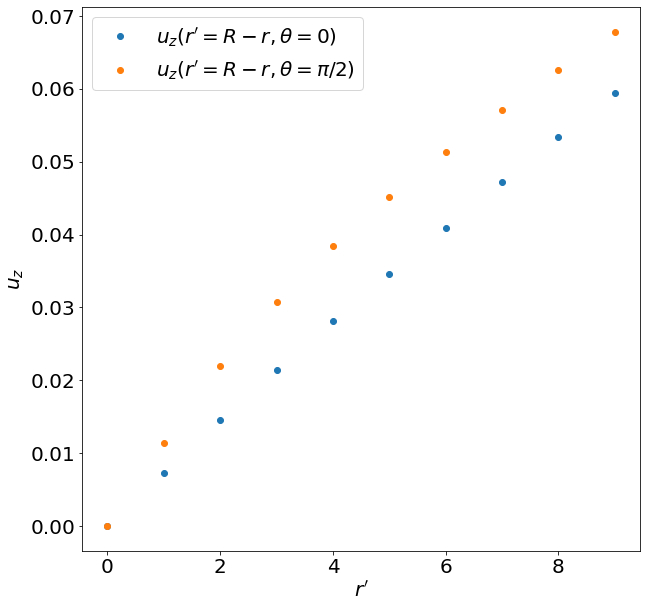}
        \caption{Near-wall velocity profiles along the $\theta=0$ and $\theta=\pi/2$ directions [58].}
    \label{6.10}
\end{figure}

This variation is more prominent up to the peak of the induced magnetic field intensity, depicted by the alternating poles in Fig. \ref{6.7} (b), showing that these local induced effects are relevant for the change of the velocity profiles. Furthermore, as expected, in both directions near the center, the applied magnetic field is either zero or small, resulting in an essentially axisymmetric and locally parabolic profile, as one can see in Fig. \ref{6.7} (a).

\section{Turbulent Pipe Flow}

\subsection{Simulation Setup}

\hspace{0.5 cm} In order to simulate a turbulent pipe flow with a grid resolution fine enough to resolve up to the viscous length scale, the CM approach was changed to the MRT one, with all equilibrium parts --moments and forcing term-- using the equilibrium distributions expanded up to the sixth order in Hermite polynomials (\ref{BoltzmannEquilibriumDistribution}). The main reason behind this change is related to the computational cost of the CM approach, as it is necessary to invert $(n_xn_yn_z)$ $27\times27$ matrices for each time step, while for the MRT scheme, the inversion procedure is only made once, as the moments which define the transformation matrix doesn't vary in time and space. The details of the MRT are given in Appendix \ref{apendiceb}. The computational grid illustrated in Fig. \ref{gridandcell}, had dimensions $n_x \times n_y \times n_z = 300 \times 300 \times 600 $ with the flow driven in the $z$ direction.

\begin{figure}[h]
    \centering
    \includegraphics[width=1.0\textwidth]{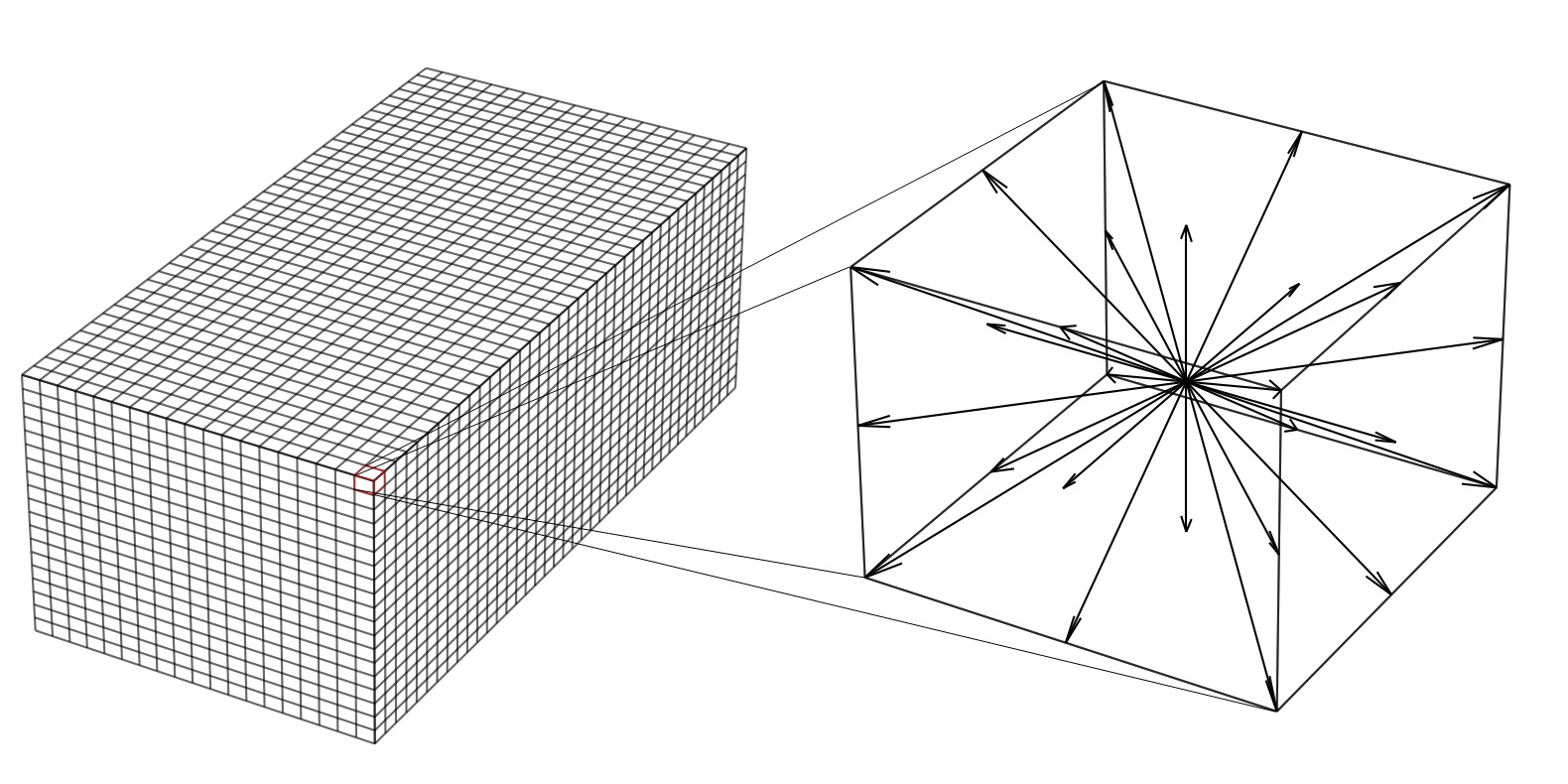}
    \caption{Illustration of computational grid. Each cell is composed by the 27 lattice vectors.}
    \label{gridandcell}
\end{figure}

The pipe radius in lattice units is $R = 149.5$, which results in a streamwise length $L \approx 4.01R $. The flow is driven by a constant body force per unit volume $\rho h$, which, at a fully developed stage should balance the viscous force \cite{Peng2018}, resulting in a friction velocity $u^*$ of
\be
    u^* = \sqrt{\frac{hR}{2}} \ .
\ee 

The viscosity and friction velocity in lattice units are, respectively, $\nu = 0.0032$ and $u^* = 0.00388$, and the friction Reynolds number $Re_{\tau} = u^*R/\nu \approx 181.3$. The viscous length scale is $y^* = \nu/u^*$, which gives the grid spacing in wall units of $\delta_x^+ = \delta_y^+ = 1.212$. All observables with the superscript $+$ from now on, are normalized by the viscous length scale $y^*$ or friction velocity $u^*$.

A No-slip boundary condition is applied through Eqs.~(\ref{fbc1}) and (\ref{fbc2}) and a periodic boundary condition is used in the streamwise direction. The large eddy turnover time in lattice units is $T^{LET} =R/u^* = 38531$ and the Reynolds number based on the bulk velocity and on the pipe's diameter is $Re \approx 5329.3 $.

Previous works have already shown that for $Re_{\tau} = 180$ the smallest length scale in turbulent pipe flow is $\eta^+ = 1.5$ \cite{Peng2018}, which is greater than the aforementioned grid spacing in the pipe's cross-section. The pipe length used, $L\approx 4.01R$, is compared with different datasets also with periodic boundary conditions and with $L\approx (10R, 12.11R)$ \cite{Loulou1997,Peng2018}. Although it does not meet the suggested length to avoid the effects of periodic boundary conditions on turbulent statistics \cite{Chin2010}, the current pipe length was inspired in recent experimental findings which suggests that structures in turbulent pipe flow are correlated up to $L \approx 4R$ for $Re = 24400$ \cite{JackelPRF2023}.

The flow was initialized with the mean profile
\be
    U(\delta^+) = \begin{cases} \delta^+, & \mbox{if } \delta^+ \leq 10.8 \ , \\ \frac{1}{0.4}ln(\delta^+) + 5.0 \ , & \mbox{if } \delta^+ > 10.8 \ , \end{cases}
\ee

\noindent where $\delta = R - r$ is the pipe's wall distance. To trigger turbulence, in addition to the constant body force, a non-uniform and divergence-free force was added during the first three $T^{LET}$, which is given by
\bea
    && F_r = -h\kappa A_0 \frac{R}{r}\frac{k_z l }{L}sin\left( \frac{2\pi t}{T} \right) \left \{ 1-cos\left [ \frac{2 \pi (R - r - l_0)}{l} \right ] \right \}cos \left ( k_z \frac{2 \pi z}{L} \right ) cos(k_{\theta}\theta) \ , \nonumber \\
    && F_{\theta} = h(1 - \kappa) A_0 \frac{k_z}{k_{\theta}}\frac{2\pi R}{L}sin\left( \frac{2\pi t}{T} \right) sin\left [ \frac{2 \pi (R - r - l_0)}{l} \right ] cos \left ( k_z \frac{2 \pi z}{L} \right ) sin(k_{\theta}\theta) \ , \nonumber \\
    && F_z = -hA_0 \frac{R}{r}sin\left( \frac{2\pi t}{T} \right) sin\left [ \frac{2 \pi (R - r - l_0)}{l} \right ] sin \left ( k_z \frac{2 \pi z}{L} \right ) cos(k_{\theta}\theta) \ , 
    \label{TurbForce}
\eea 

\noindent where $k_z = 3$ and $k_{\theta} = 2$ are the streamwise and azimuthal wavenumbers of the perturbation force, $T \approx 0.052T^{LET}$ and $A_0 = 50$ are the forcing period and amplitude, respectively, and $\kappa = 0.5$ is the weighting parameter that distributes the force in radial and azimuthal directions. The force to trigger turbulence was applied only in the region $R - l_0 - l \leq r \leq R - l_0$ with $l_0 = 0.2R$ and $l = 0.4R$. The numerical simulation was performed for $30T^{LET}$ with statistics being done with the last $10T^{LET}$, taking the instantaneous vector fields at every $100$ iteration.

\subsection{Instantaneous Observations}

\hspace{0.5 cm}Instantaneous observations of the turbulent flow are shown in Fig. \ref{snapshots}. As can be seen, the presence of several vortices is displayed seemingly randomly through the vector field and also by the streamwise vorticity. It is worth mentioning how the vortices usually appear in regions close to a change of sign in the streamwise velocity fluctuation, in consonance with the well-known picture of ejections and sweeps related to the $Q_2$ and $Q_4$ quadrant analysis \cite{Wallace2016}.

\begin{figure}[h]
    \centering
    \includegraphics[width=1.0\textwidth]{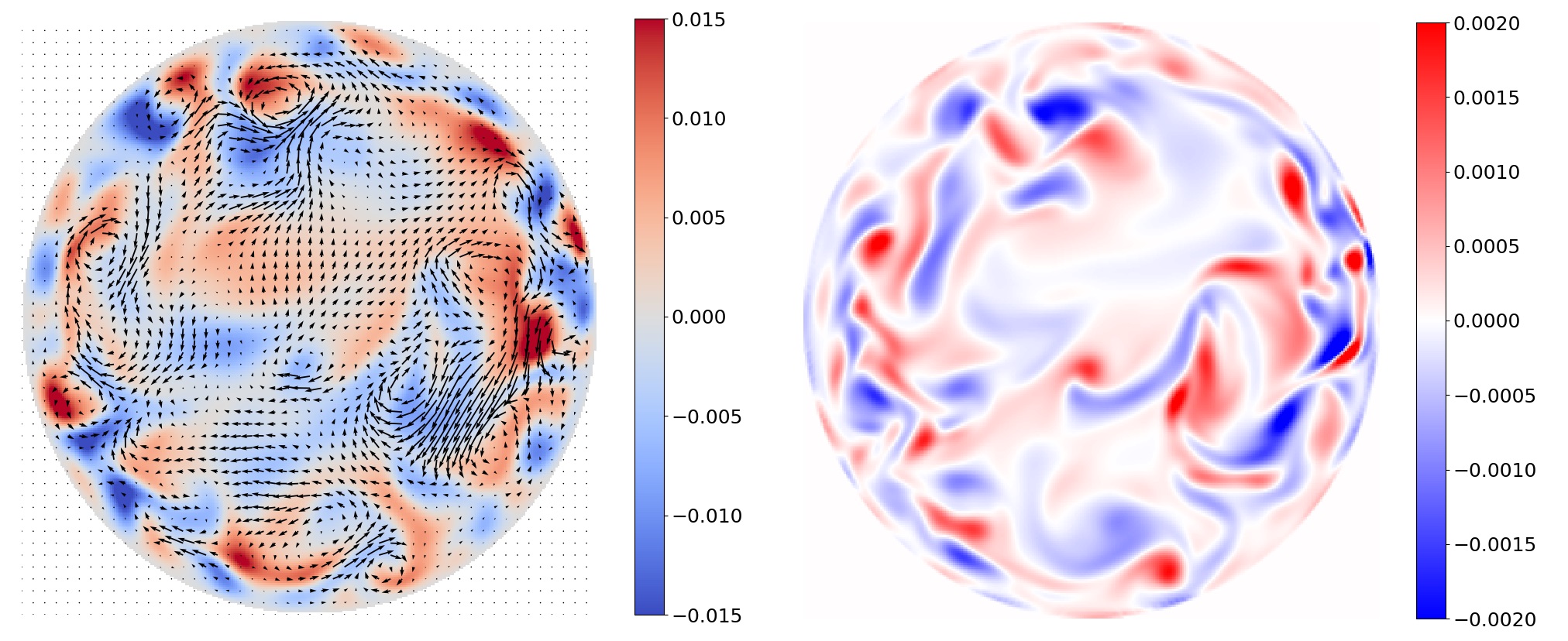}
    \caption{Snapshot observations from turbulent pipe flow simulated through the LBM with $Re_{\tau} \approx 181.3$. Left:In-plane vector field and density plot of streamwise velocity fluctuation in lattice units. Right: Streamwise vorticiy in lattice units.}
    \label{snapshots}
\end{figure}

The instantaneous streamwise velocity component $u_z$ is shown in Fig. \ref{streamwiseuz}, where several near-wall structures can be seen, such as the shape of Kelvin-Helmholtz and Rayleigh-Taylor like instabilities. Also, the thickness of the structures seems to agree relatively well with other simulation techniques at $Re_{\tau} = 180$ \cite{Yao2023}, such as the Pseudo-Spectral based solver Openpipeflow \cite{Willis2017}.

\begin{figure}[!htb]
    \centering
    \includegraphics[width=0.6\textwidth]{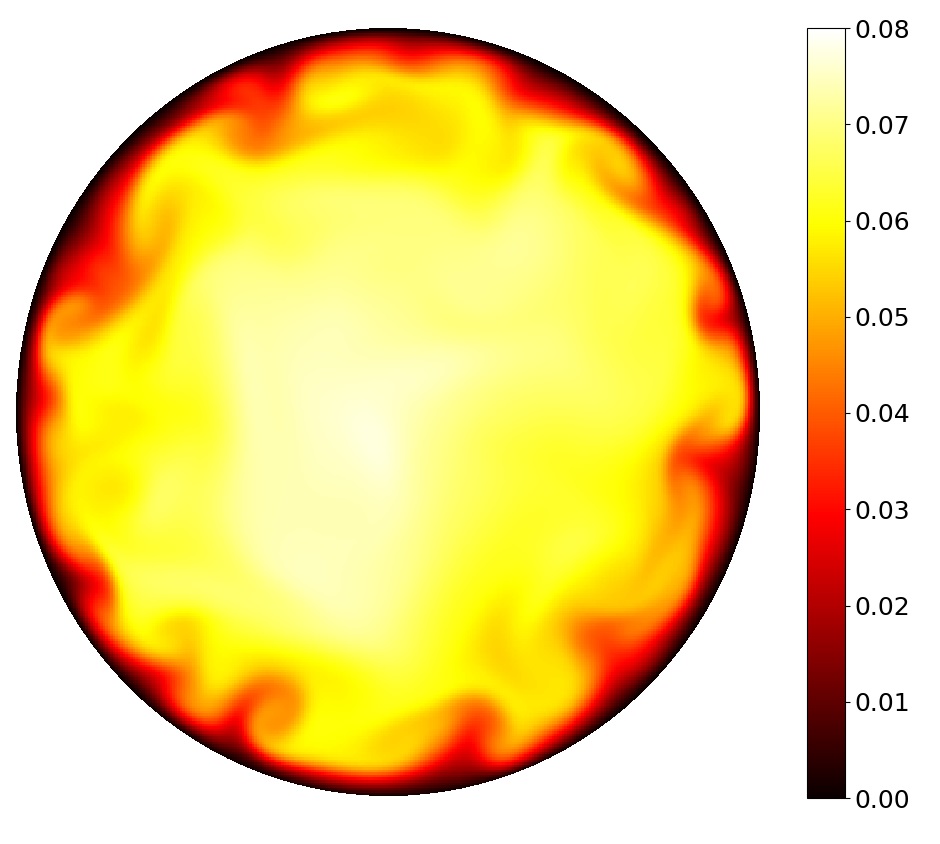}
    \caption{Snapshot of streamwise velocity component $u_z$ in lattice units.}
    \label{streamwiseuz}
\end{figure}
\begin{figure}[!htb]
    \centering
    \includegraphics[width=0.6\textwidth]{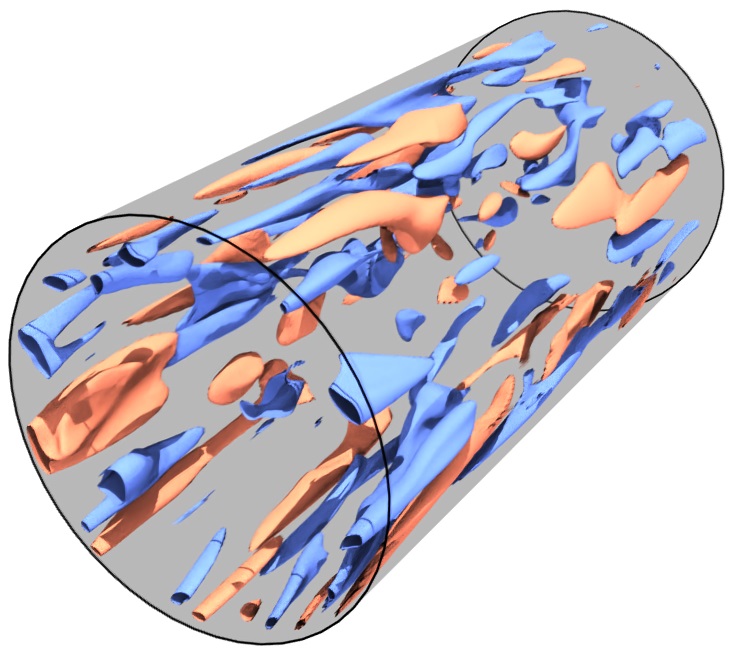}
    \caption{Isosurfaces of streamwise velocity fluctuations. Red/blue isosurfaces are related to high/Low speed streaks, where the velocity fluctuation is $10\%$ above/below the mean velocity profile.}
    \label{highandlowspeedstreaks}
\end{figure}

The high/low-speed streaks can be seen by the red/blue isosurfaces of the streamwise velocity fluctuation $\delta u_z$ in Fig. \ref{highandlowspeedstreaks}. Different structures could be targeted by different thresholds on the isosurface selection, with a decreasing number of structures seen as the value is increased, since they would be related to more extreme events.

\subsection{Turbulent Statistics}

\hspace{0.5 cm}In order to validate the numerical simulations, the turbulent statistics will be compared with a pseudo-spectral approach \cite{Loulou1997}, a LBM approach based on a second-order expansion of the Maxwell Boltzmann equilibrium distribution --which will be referred to as LBM-$\mathcal{O}(2)$-- \cite{Peng2018}, and, whenever available, with experimental results using Laser Doppler Anemometry (LDA) \cite{Tahitu_thesis}, which is known for its accuracy for near-wall measurements.
\begin{table}[h!]
\centering
\begin{tabular}{|c c c c c|} 
 \hline
 \hline
  & Method & $Re_{\tau}$ & $L/R$ & $\Delta T/T^{LET}$ \\ [0.5ex] 
 \hline
 Peng et al. & LBM-$\mathcal{O}(2)$ & 180.0 & 12.12 & 60.1 \\ 
 \hline
 Loulou et al. & PS & 190.0 & 10.00 & 5.8 \\
 \hline
 Tahitu & LDA & 181.2 & 437.81 & 265.0 \\
 \hline
 Present & LBM-$\mathcal{O}(6)$ & 181.3 & 4.01 & 10.0 \\
 \hline
 \hline
\end{tabular}
\caption{Relevant parameters for turbulent pipe flow' statistical properties.}
\label{table:3}
\end{table}

The relevant parameters for all sets to be used are shown in table \ref{table:3}, where $\Delta T$ is the time window used for the statistical averages and LBM-$\mathcal{O}(6)$ is referred to the scheme 

\begin{figure}[h!]
\centering
\begin{minipage}{.5\textwidth}
  \centering
  \includegraphics[width=1.00\linewidth]{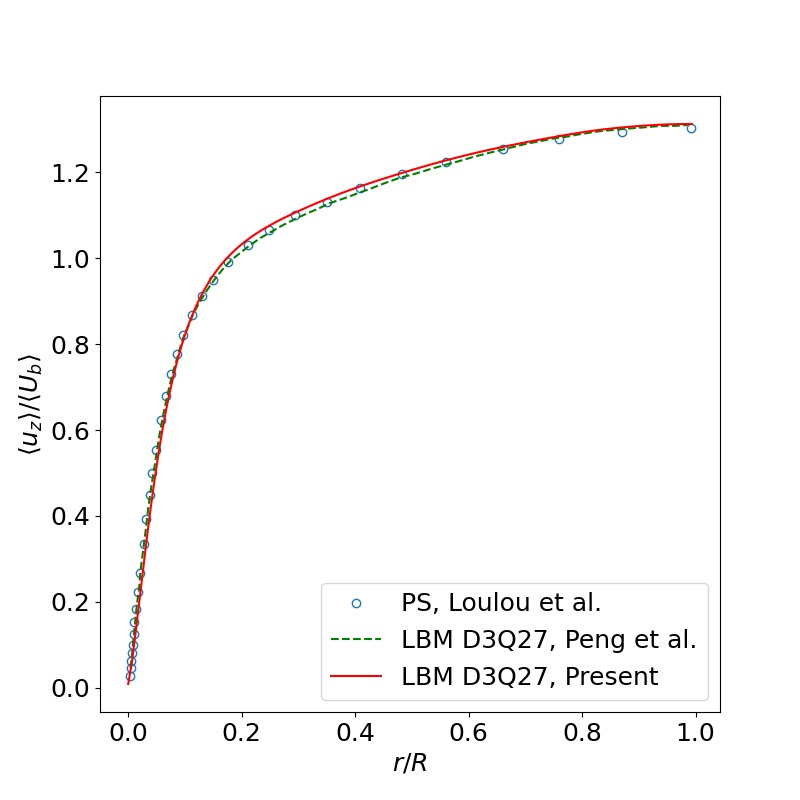}
  \label{meanLBM}
\end{minipage}%
\begin{minipage}{.5\textwidth}
  \centering
  \includegraphics[width=1.00\linewidth]{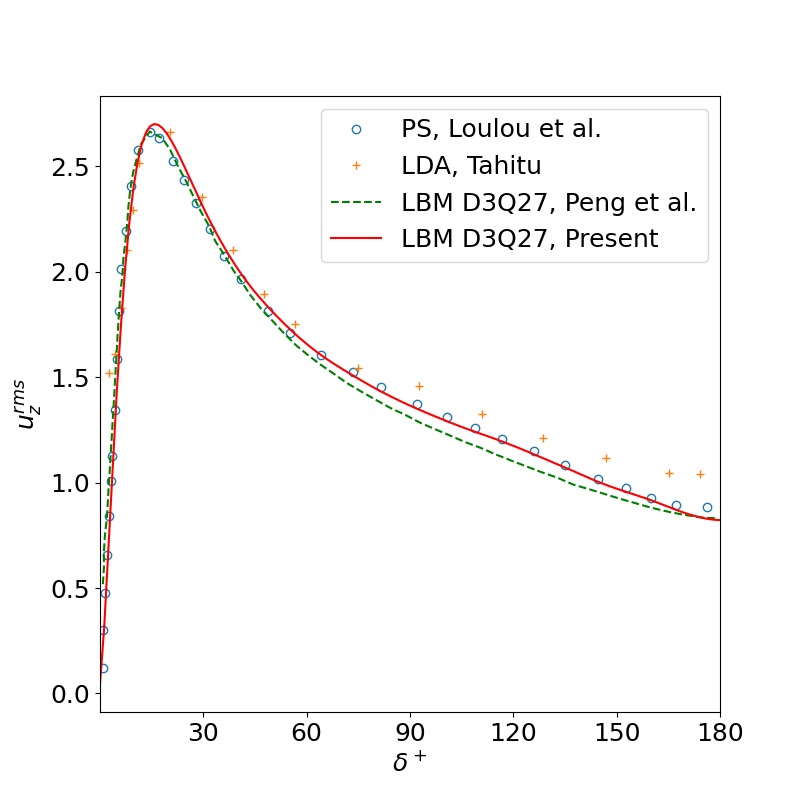}
  \label{urmsLBM}
\end{minipage}
\caption{Left: Mean streamwise velocity profile normalized by the bulk velocity. Right: Root-mean-square streamwise velocity fluctuation in wall units.}
\label{MeanUrms}
\end{figure}

\noindent used in this dissertation with the equilibrium distribution expanded up to the sixth order in Hermite polynomials. The rms of the streamwise velocity fluctuation and high-order statistical moments are computed through Eqs.~(\ref{urmsformula}), and (\ref{statisticalmomentsn}). The mean streamwise velocity profile normalized by the bulk velocity and the rms streamwise velocity fluctuation in wall units are shown in Fig. \ref{MeanUrms}. To verify the accuracy of the statistical properties of the present numerical study, the skewness and the flatness of the streamwise velocity component are shown in Fig. \ref{S3S4LBM}, as high-order statistical moments require a well converged and large enough statistical ensemble.

\begin{figure}[h!]
\centering
\begin{minipage}{.5\textwidth}
  \centering
  \includegraphics[width=1.00\linewidth]{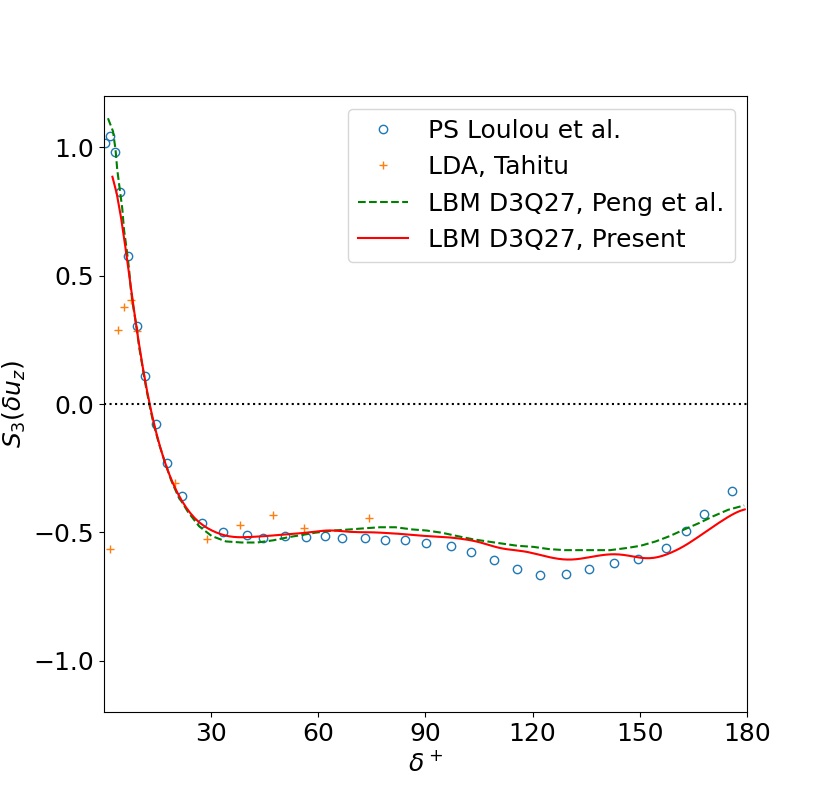}
  \label{skewnessLBM}
\end{minipage}%
\begin{minipage}{.5\textwidth}
  \centering
  \includegraphics[width=1.00\linewidth]{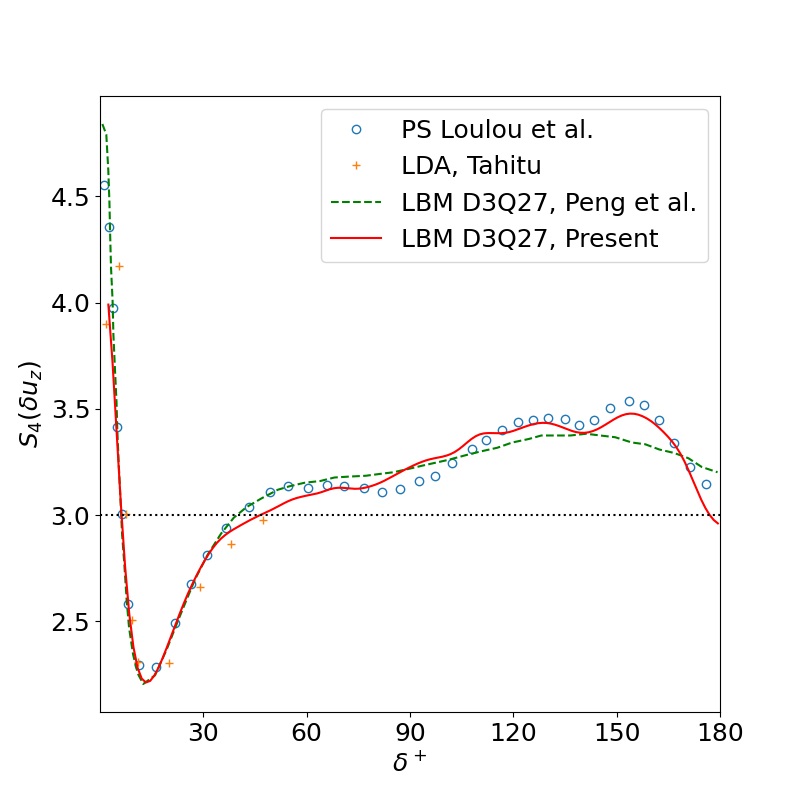}
  \label{flatnessLBM}
\end{minipage}
\caption{High order statistical moments of the streamwise velocity component. Left: Skewness. Right: Flatness.}
\label{S3S4LBM}
\end{figure}

As one can see, all observables agree really well with the literature, even though the used pipe length is smaller than the PS simulations, the time window used was almost twice the size in large eddy turnover times. The comparisons with LDA which used a very large time window for the statistical analysis, as one can see in table \ref{table:3}, is also in very good agreement, with the peak positions and plateau of all observables being well represented by the numerical simulation.

\subsection{CS Identification and Dynamics of Mode Transition}

\hspace{0.5 cm}The identification of CS followed the same approach used for the statistical analysis of the experimental data presented in the previous chapter. The dimensionality reduction based on the dominant azimuthal wave number of Eq.~(\ref{newcorrelation}) was applied to the simulated turbulent data. The turbulent data set of 10$T^{LET}$ was analyzed at every 20 time steps.

\begin{figure}[h]
    \centering
    \includegraphics[width=0.60\linewidth]{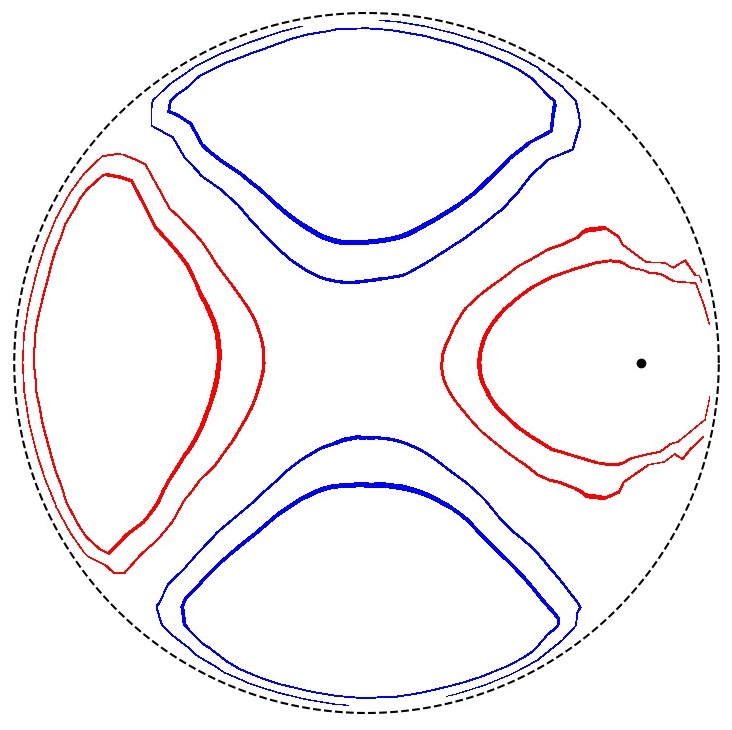}
    \caption{Velocity-velocity correlation contours conditioned to azimuthal wave number $k_{\theta} = 2$. Red countors are related to $R_{uu} = 0.05$ and $0.1$, while blue has the same absolute value with opposite sign.}
    \label{uucorrelationLBM}
\end{figure}

Figure \ref{uucorrelationLBM} shows the contour plot of the velocity-velocity correlation function conditioned to the azimuthal wave number $k_{\theta} = 2$. Remarkably, approximately 10 different azimuthal wave numbers were found in the produced numerical data. Modes higher than $k=10$ were also found, but their statistical weight was negligible, in close agreement to the experimental findings \cite{JackelPOF2023, JackelICHMT2023 ,JackelPRF2023}. A comparison between the probability distribution of dominant azimuthal wave numbers found here and the one experimentally observed is shown in Fig. \ref{modedistributionLBM}.

\begin{figure}[h]
    \centering
    \includegraphics[width=0.80\linewidth]{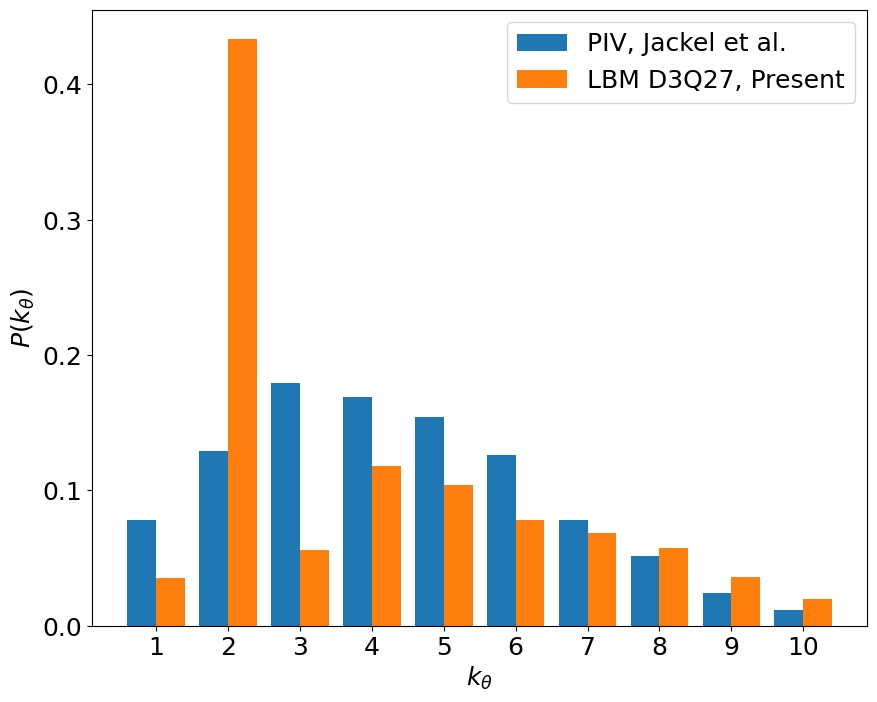}
    \caption{Probability distribution of dominant azimuthal wave numbers.}
    \label{modedistributionLBM}
\end{figure}

The peak in the azimuthal mode $k_{\theta} = 2$ seems to be related to the \emph{principle of permanence of large eddies} \cite{Frisch} and to a non-trivial memory effect regarding the force that was used to trigger turbulence (\ref{TurbForce}), as the force was only active in the first $3T^{LET}$ and the dimensionality reduction was only applied in the range $20 T^{LET}\leq t \leq 30 T^{LET}$. At the same time, the mode distributions, except for the peaked forced mode, resemble the cascade effects in three-dimensional turbulence, as expected, with the energy going from large to small scales.

The time dependence of the identified modes during the 10 analyzed turnover times is shown in Fig. \ref{modetransitions}. As one can see, the dynamics between the modes seem to be approximately random for a large time interval, however, the intriguing phenomenon is the persistence of mode $k_{\theta} = 2$ (the forced mode) intermittently. To illustrate how the modes are finely resolved in time, a zoom-in of 10x is also displayed.

\begin{figure}[h]
    \centering
    \includegraphics[width=0.90\linewidth]{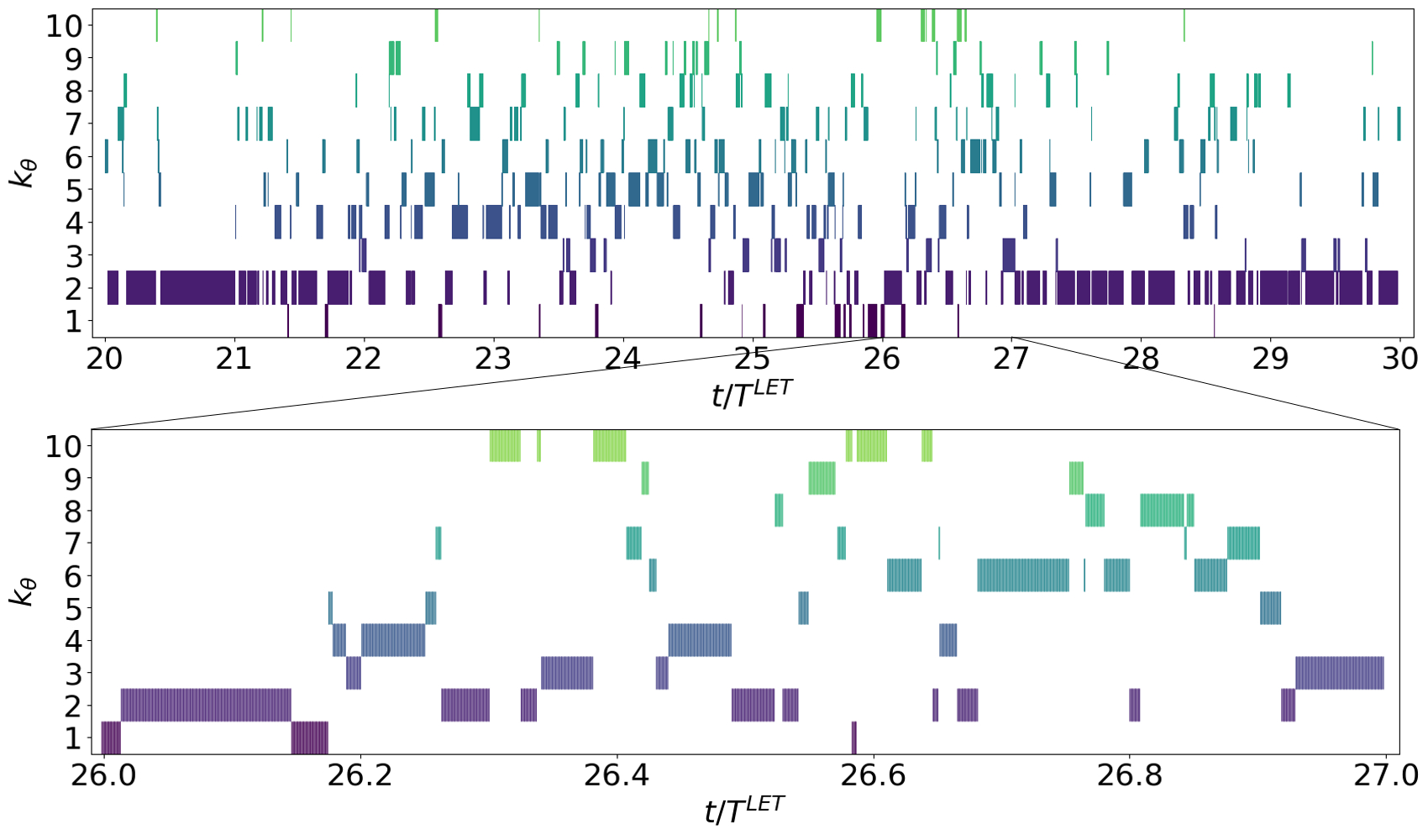}
    \caption{Time dependence of the identified dominant modes.}
    \label{modetransitions}
\end{figure}

A stochastic series based on the identified modes to investigate their transitions was also set, as in Eq.~(\ref{stochasticseries}). In this case, it is given by
\be
    \mathcal{S}^{LBM} \equiv  \{ k^*(t_0), k^*(t_0 + \Delta), \ ... , k^*(30T^{LET}) \  \} \ ,
    \label{stochasticseriesLBM}
\ee

\noindent where $t_0 = 20T^{LET}$ and $\Delta = 20 \delta t$. The transition matrix among the identified modes was found to be
\be
    T = 
    \begin{bmatrix}
    0.975 & 0.000 & 0.001 & 0.000 & 0.002 & 0.001 & 0.001 & 0.003 & 0.000 & 0.011\\
    0.004 & 0.992 & 0.006 & 0.004 & 0.006 & 0.008 & 0.008 & 0.009 & 0.004 & 0.011\\
    0.003 & 0.001 & 0.977 & 0.002 & 0.001 & 0.003 & 0.002 & 0.001 & 0.000 & 0.005\\
    0.000 & 0.002 & 0.007 & 0.980 & 0.002 & 0.003 & 0.002 & 0.002 & 0.006 & 0.005\\
    0.003 & 0.001 & 0.004 & 0.002 & 0.978 & 0.003 & 0.005 & 0.005 & 0.004 & 0.003\\
    0.001 & 0.001 & 0.001 & 0.003 & 0.005 & 0.973 & 0.003 & 0.006 & 0.004 & 0.003\\
    0.004 & 0.001 & 0.004 & 0.002 & 0.005 & 0.002 & 0.969 & 0.003 & 0.004 & 0.005\\
    0.006 & 0.001 & 0.000 & 0.002 & 0.002 & 0.002 & 0.005 & 0.969 & 0.004 & 0.003\\
    0.000 & 0.000 & 0.000 & 0.002 & 0.001 & 0.003 & 0.004 & 0.003 & 0.970 & 0.000\\
    0.003 & 0.000 & 0.001 & 0.001 & 0.001 & 0.002 & 0.002 & 0.000 & 0.001 & 0.953
    \end{bmatrix} \ , \
    \label{transitionmatrixlbm}
\ee 

\noindent which, as one can see, is diagonally dominant, with $k_{\theta} = 2$ being the mode with a higher probability of persistence. The CK equation was also tested by performing a decimation of several orders on the stochastic series (\ref{stochasticseriesLBM}). These results are shown in Fig. \ref{CKLBM}. In contrast with the experimental case, a qualitatively good agreement was found between the comparison of the absolute values of the eigenvalues from the original series with their respective counterparts from the decimated series.

\begin{figure}[h]
    \centering
    \includegraphics[width=0.80\linewidth]{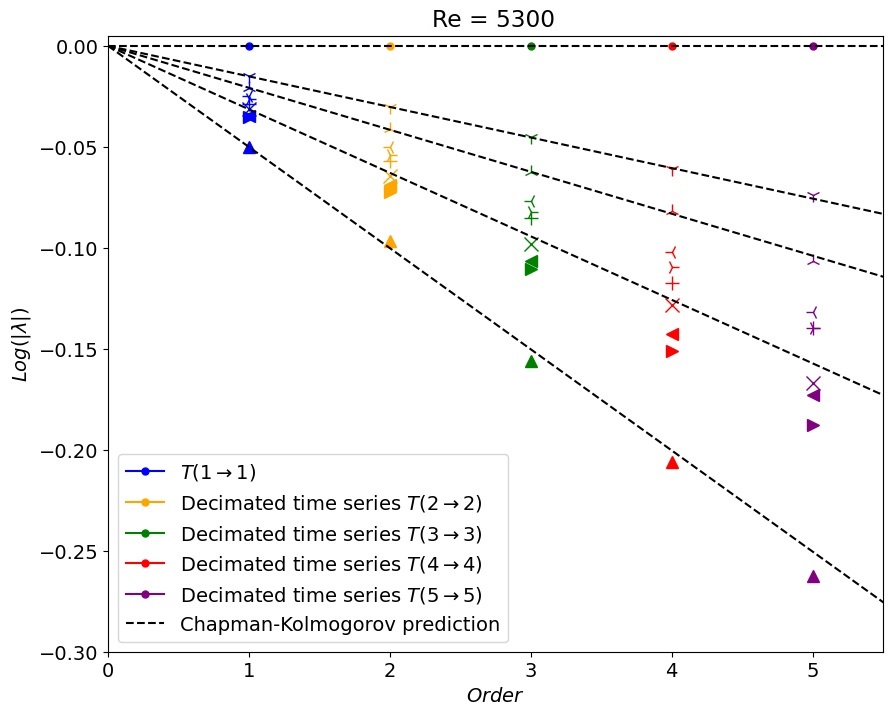}
    \caption{Comparison between eigenvalues of the original transition probability matrix (Order = $1$) with eigenvalues for the transition matrices of the decimated series (Order up to $5$).}
    \label{CKLBM}
\end{figure}

This was observed up to a fifth-order decimation, which reduces the amount of analyzed turbulent fields by a factor of $1/100$, resulting in approximately 3850 snapshots, which might not be large enough, so the effects of low statistics may play a role. Also, the deviation from the CK prediction could be related to a transition to the scenario observed in the experiments, where the dynamics among the transitions display a non-Markovian behavior. It is worth mentioning that the temporal resolution obtained by the LBM is of the order of $10^2$ finer than the experimental case obtained with a laser of 15 Hz. In this case, the agreement with the CK prediction could be associated with the fact that the transition matrix is diagonally dominant, so the modes, in general, don't have a considerably high memory about the transition from different modes.

\begin{figure}[!htb]
    \centering
    \includegraphics[width=0.75\linewidth]{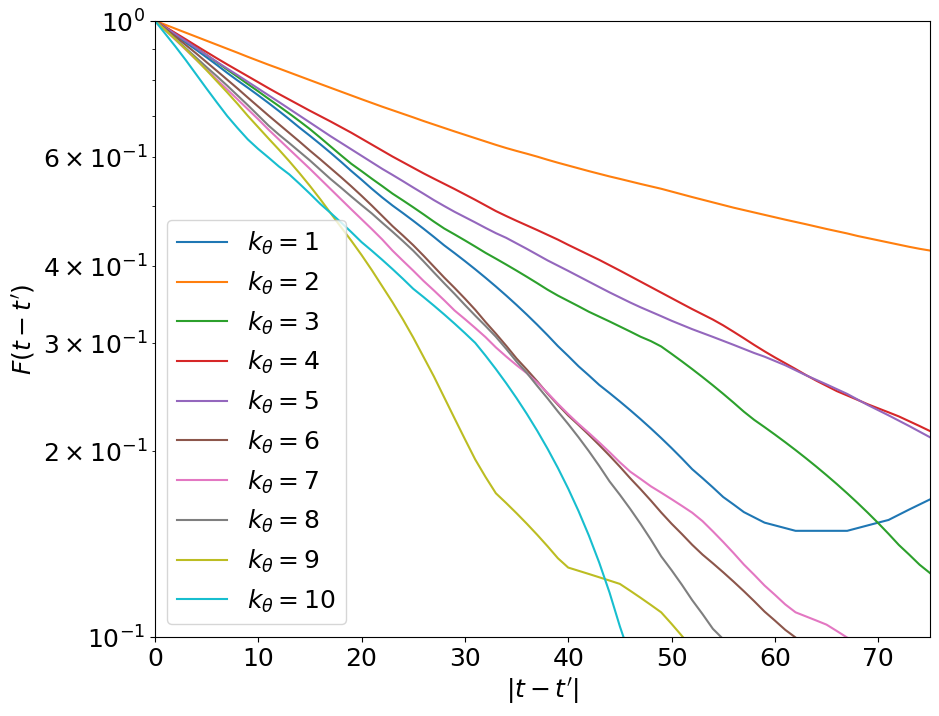}
    \caption{Semi-log plot of self mode correlation for all the identified modes.}
    \label{FcorrelationLBM}
\end{figure}

\begin{figure}[!htb]
    \centering
    \includegraphics[width=0.75\linewidth]{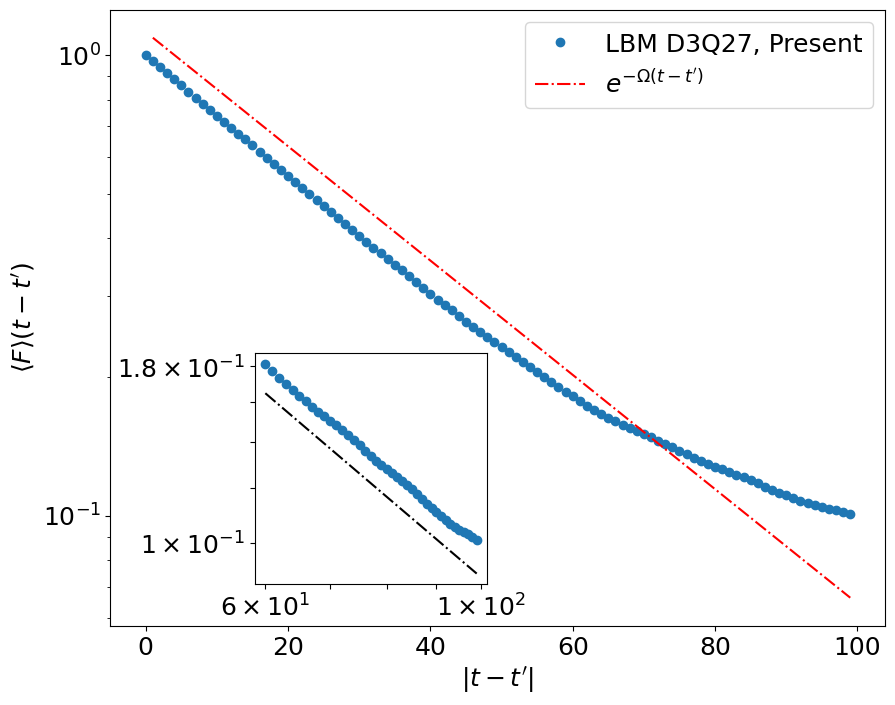}
    \caption{Semi-log plot of average self correlation and straight line of exponential decay for comparison. Inset: behavior of the region $\vert t - t'\vert \geq 60$ in a log-log plot with a straight line of algebric decay with exponent $-1.2$ for comparison.}
    \label{FmeancorrelationLBM}
\end{figure}

To investigate quantitatively the observed agreement between the CK prediction with the decimated stochastic time series, the self mode correlation defined in Eq.~(\ref{Fcorrelationexp}) was analyzed, first individually, for each mode, and afterwards mode averaged. Figure \ref{FcorrelationLBM} shows that, in fact, the correlation function for all modes decays approximately exponentially, which is observed as a straight line in the semi-log plot. Also, as one can see, mode $k_{\theta} = 2$ is the most self correlated one, which agrees with the observed probability distribution. The average self correlation is displayed in Fig. \ref{FmeancorrelationLBM} together with a straight line representing an exponential decay with $\Omega = 1/35$. As it can be seen, the agreement is evident and holds until $\vert t - t'\vert \approx 60$. Larger time separations are shown in the inset, in a log-log plot, which was observed to decay algebraically $\propto \vert t - t'\vert^{-\alpha}$, as indicated by the black straight line, with $\alpha = 1.2$. This fact agrees with previous experimental findings of non-Markovian behavior in turbulent pipe flow with $\alpha \approx 1.0$ for $Re = 24400$, recorded with a smaller frequency \cite{JackelPRF2023}.

The longest and mean streamwise length for all modes are presented in table \ref{table:4} by means of Taylor frozen hypothesis.
\begin{table}[h!]
\centering
\begin{tabular}{|c c c c c c c c c c c|} 
 \hline
 \hline
 $k_{\theta}$ & 1 & 2 & 3 & 4 & 5 & 6 & 7 & 8 & 9 & 10 \\ [0.5ex] 
 \hline
 $\Delta S_{max}/R$ & 0.98 & 8.20 & 1.33 & 1.73 & 1.50 & 1.01 & 1.01 & 0.70 & 0.50 & 0.49 \\ 
 \hline
 $\langle\Delta S \rangle/R$ & 0.27 & 0.87 & 0.32 & 0.36 & 0.33 & 0.27 & 0.21 & 0.23 & 0.22 & 0.15 \\ 
 \hline
 \hline
\end{tabular}
\caption{Longest example and average size by Taylor frozen hypothesis.}
\label{table:4}
\end{table}

As one can see, on average, all modes are within the simulated pipe's length. Although, their longest example might by related to the known large scales of motion (LSMs) and very-LSMs --which are typically longer than 3 pipe radii \cite{Balakumar2007}.

\end{chapter}

\begin{chapter}{Conclusion}
\label{conclusao}

\hspace{0.5 cm} In this dissertation, two correlated lines were followed. In the first, related to the statistical analysis of experimental data, done in Chap. \ref{cap5}, we described the identification of CSs in pipe flows and investigated their mode transitions as a stochastic process. A low-level Markovian model between the microstates was developed, with its macrostates showing a qualitatively good agreement with our empirical observations. The length statistics for individual modes with and without an external magnetic field were also investigated. In the second line, in Chap. \ref{cap6}, Lattice Boltzmann simulations were performed for MHD flows and turbulent pipe flow. A new MRT collision model for the induction equation and boundary conditions was shown for the MHD part. In the turbulent pipe flow, a nontrivial memory effect was observed related to the force that triggered the turbulent state. Also, the stochastic mode transition analysis revealed a Markovian behavior with a transition to the observed non-Markovian behavior in the experimental data for large-time separations.

The dimensionality reduction approach to identify the CSs shown in Chap. \ref{cap5} have shown to be a ``low-cost" methodology that brings important information about the turbulent flow. By definition, it brings the information of the mode which has the largest contribution to the turbulent streamwise kinetic energy in the shell around the reference point. It was also observed that the mode distribution doesn't change considerably in a broad range of Reynolds numbers, which are qualitatively well described by a Poisson distribution. Physically, the modes capture the streamwise low/high-speed streaks, which are related to cross sectional motions of a pair of counter-rotation vortices, that might be related to quasi-streamwise and hairpin-like structures.

One of the great advantages of this CS identification approach lies in collapsing some of the important phenomena of a given snapshot to a single parameter, which allows the study of mode transitions as a stochastic process. The stochastic mode transition analysis with $Re = 24415$ has shown several important information, such as its non-Markovian behavior --displaying a considerably high memory effect-- and the size of correlated packets of turbulent structures. Also, the Markovian model for microstates was shown to give a physically reasonable explanation of the intriguing power law behavior of the studied correlation functions. 

The magnetic MRT collision model for the induction equation here developed, in Chap. \ref{cap6}, allows one to simulate the QS approximation within the LBM. A distance-dependent Dirichlet boundary condition for non-cartesian boundaries was also set. The validations were done extensively with known cases of the literature, such as the 3D Orszag-Tang vortex problem and Gold's analytical solution on MHD laminar pipe flows. Our method was then applied to a more complex situation with a non-uniform external magnetic field with a 6-fold symmetry. To highlight the advantages of the new collision operator, a comparison with the SRT model for the induction equation is shown, which diverges for a few iterations, while the MRT can simulate up to the stationary regime for the same physical parameters.

The turbulent pipe flow simulation was performed by the LBM and validated extensively against experimental and different numerical methodologies. The turbulent data is resolved in space up to the viscous length scale and the instantaneous observations reveal the seemingly random distribution of CSs. Our pipe length followed the suggestion of the size of correlated structures in experimental turbulent pipe flow, as determined in Ref. \cite{JackelPRF2023}, and the statistical moments of streamwise velocity fluctuation matched well previous results in the literature. The CS detection was performed in the same fashion as in the experimental case, which revealed a non-trivial correlation with the force used to trigger the turbulent state. Also, the mode distribution agrees with the experimental findings regarding the turbulence cascade of energy from large to small scales.

The stochastic mode transitions for the finely time-resolved data resulted in a diagonally dominant transition matrix, which culminated in a good agreement of the CK prediction for the decimated time series. The indicated Markovian behavior was then checked by time self-mode correlation functions, which confirmed the exponential decay for all observed modes in a short time window. The average self-mode correlation was plotted for a larger time separation, which revealed a transition from the exponential decay to an algebraically one, recovering the observed experimental behavior of non-Markovianity. The streamwise length of the modes by Taylor's frozen hypothesis showed that, on average, they are all within the length of our pipe simulations, but their longest example might be related to the known VLSM.

This dissertation opens the path for many directions of further research. From the experimental point of view, it would be interesting to investigate the CS detection and stochastic mode transitions for MHD flows with a stronger effect of the magnetic field --increasing the fluid conductivity, the magnetic field intensity, or both.  From the numerical point of view, it would be interesting to investigate larger time series, so the decimation could go up to a higher order, 20 or beyond should be enough to see the same effects of the experimental case with the studied acquisition frequency. New numerical simulations forcing all modes evenly or with the statistical distribution of the experimental findings could give a closer result to the real experiment. Also, related to the investigation of particle deposition, it would be interesting to investigate Lagrangian particle simulations and how they are correlated with CSs of different azimuthal wave numbers.

\end{chapter}



\newpage
\phantomsection



\appendix
\begin{chapter}{Validation of the Extension to the Vector Valued Distribution Boundary Conditions}
\label{apendicea}

\hspace{0.5 cm} 
The bra vector valued distribution $\langle g_{\alpha} \vert$ and the weight coefficient ket $\vert \omega \rangle$ are given by
\bea 
    && \langle g_{\alpha} \vert = (g_{0,\alpha},g_{1,\alpha},...,g_{6,\alpha}) \ , \\
    && \vert \omega \rangle = (\omega_0,\omega_1,...,\omega_6)^T \ .
\eea

\noindent where the latter satisfies
\bea
    &&\langle 1 \vert \omega \rangle = 1 \ , \\
    &&\langle \boldsymbol{\xi} \vert \omega \rangle = 0 \ ,\\
    &&\langle \xi_{\alpha}\xi_{\beta} \vert \omega \rangle = \varepsilon_D \delta_{\alpha \beta} \ .
\eea

The Dirichlet Boundary condition its proposed for the vector valued distributions, similarly to \cite{Li2013} as
\be
    g_{\overline{i},\alpha}(\boldsymbol{x}_f,t+1) = C_1 \tilde{g}_{i,\alpha}(\boldsymbol{x_f},t) + C_2 \tilde{g}_{i,\alpha}(\boldsymbol{x_{ff}},t) + C_3 \tilde{g}_{\overline{i},\alpha}(\boldsymbol{x_f},t) + C_4 \varepsilon_{D}B^d_{\alpha} \ ,
\ee 

\noindent where
\be
    \boldsymbol{x}_f = \boldsymbol{x}_w + \Delta \boldsymbol{\xi}_i \epsilon, \\
    \boldsymbol{x}_{ff} = \boldsymbol{x}_w - (1+\Delta)\boldsymbol{\xi}_i \epsilon \ .
\ee 

Here, to reconstruct the boundary scheme, an asymptotic analysis is used in function of the dimensionless and small parameter
\be 
    \epsilon = \delta x/L,
\ee

\noindent where $\delta x$ is the grid spacing and $L$ the characteristic length scale. Here after $\delta x = \delta t = 1$ for simplicity.

The asymptotic analysis different from the Chapman-Enskog expansion \cite{Junk2005}, relates finite discrete-velocity models with a diffusive scaling and it consist on the expansions of the distributions and macroscopic variables as 
\bea 
    g_{i,\alpha} = g^{(0)}_{i,\alpha} + \epsilon g^{(1)}_{i,\alpha} + \epsilon^2 g^{(2)}_{i,\alpha} + \mathcal{O}(\epsilon^3) \ , \\
    B_{\alpha} = B^{(0)}_{\alpha} + \epsilon B^{(1)}_{\alpha} + \epsilon^2 B^{(2)}_{\alpha} + \mathcal{O}(\epsilon^3) \ .
\eea

The MRT collision operator is written as
\be
    Lg_{i,\alpha} = \mathbf{M}^{-1}\mathbf{S}\mathbf{M}(\mathbf{Q}^{(0)} + \epsilon\mathbf{Q}_{\alpha}^{(1)})g_{i,\alpha} \ ,
\ee 

\noindent where
\be
    \mathbf{Q}^{(0)} = \vert \omega \rangle \langle 1 \vert - \mathcal{I}_{7 \times 7} , \ \  \mathbf{Q}_{\alpha}^{(1)} = (B_{\alpha}v_{\beta} - v_{\alpha}B_{\beta}) \vert \xi_{\beta} \omega \rangle \langle 1 \vert \ .
\ee

Taylor expanding the Dirichlet boundary condition up to second order in $\epsilon$, one has that the equations for the $\mathcal{O}(\epsilon^0)$ and $\mathcal{O}(\epsilon^1)$ are given respectively by
\bea
    && (1-C_3)g^{(0)}_{\overline{i},\alpha} - (C_1 + C_2)g^{(0)}_{i,\alpha} = C_4 \varepsilon_D B^d_{\alpha}, \\
    && (1-C_3)g^{(1)}_{\overline{i},\alpha} - (C_1 + C_2)g^{(1)}_{i,\alpha} + [\Delta(1-C_3) - C_3]c_{\overline{i},\beta}\frac{\partial g^{(0)}_{\overline{i},\alpha}}{\partial x_{\beta}} + \nonumber \\
    && [(\Delta -1)(C_1 + C_2) + C_2]c_{i,\beta}\frac{\partial g^{(0)}_{i,\alpha}}{\partial x_{\beta}} = 0. 
\eea 

The zero and first order equation in powers of $\epsilon$ that comes from the Taylor expansion of the above equations are
\bea
    &&\mathcal{O}(\epsilon^0) : 0 = \mathbf{M}^{-1}\mathbf{S}\mathbf{M}\mathbf{Q}^{(0)} \vert g^{(0)}_{\alpha} \rangle \ , \\
    &&\mathcal{O}(\epsilon^1) : \frac{\partial \vert \xi_{\beta} g^{(0)}_{\alpha} \rangle}{\partial x_{\beta}} = \mathbf{M}^{-1}\mathbf{S} \mathbf{M} \Big (\mathbf{Q}^{(0)} \vert g^{(1)}_{\alpha} \rangle + \mathbf{Q}^{(1)}_{\alpha} \vert g^{(0)}_{\alpha} \rangle \Big ) \ .
\eea 

The solutions to the above equations for $\vert g^{(0)}_{\alpha} \rangle$ and $\vert g^{(1)}_{\alpha} \rangle$ are given by
\be
    \vert g^{(0)}_{\alpha} \rangle = \vert \omega \rangle B^{(0)}_{\alpha}
    \label{g0eq}
\ee 

\noindent and
\be
    \vert g^{(1)}_{\alpha} \rangle = \vert \omega \rangle B^{(1)}_{\alpha} + (B_{\alpha} v_{\beta} - v_{\alpha} B_{\beta})\vert \xi_{\beta} \omega \rangle B^{(0)}_{\alpha} - \mathbf{M}^{-1}\mathbf{S}^{-1}\mathbf{M} \vert \xi_{\beta} \omega \rangle \frac{\partial B^{(0)}_{\alpha}}{\partial x_{\beta}}.
    \label{g1eq}
\ee 

Equation ~\eqref{g0eq} straightforwardly gives that
\bea 
    &&\langle \xi_{\beta} \vert g^{(0)}_{\alpha} \rangle = 0 \ , \\
    \label{annalogousYoshida1}
    &&\langle \xi_{\beta}\xi_{\gamma} \vert g^{(0)}_{\alpha} \rangle = \varepsilon_D B^{(0)}_{\alpha}\delta_{\beta \gamma} \ .
    \label{annalogousYoshida2}
\eea 

From the Eq.~\eqref{g1eq}, we have
\be
    \langle \xi_{\beta}\xi_{\gamma} \vert g^{(1)}_{\alpha} \rangle = \langle \xi_{\beta}\xi_{\gamma} \vert \omega \rangle B^{(1)}_{\alpha} + (B_{\alpha} v_{\delta} - v_{\alpha} B_{\delta})  \langle \xi_{\beta}\xi_{\gamma} \vert \xi_{\delta} \omega \rangle B^{(0)}_{\alpha} -  \langle \xi_{\beta}\xi_{\gamma} \vert \mathbf{M}^{-1}\mathbf{S}^{-1}\mathbf{M} \vert \xi_{\delta} \omega \rangle \frac{\partial B^{(0)}_{\alpha}}{\partial x_{\delta}},
    \label{bracket_g1}
\ee

\noindent with
\bea 
    &&\langle \xi_{\beta}\xi_{\gamma} \vert \omega \rangle B^{(1)}_{\alpha} = \varepsilon_D B^{(1)}_{\alpha} \delta_{\beta \gamma} \ , \\
    &&\langle \xi_{\beta}\xi_{\gamma} \vert \xi_{\delta} \omega \rangle v_{\delta} = \langle \xi_{\beta}\xi_{\gamma} \vert \xi_{\delta} \omega \rangle B_{\delta} = 0 \ \mbox{, and} \\
    &&\langle \xi_{\beta}\xi_{\gamma} \vert \mathbf{M}^{-1}\mathbf{S}^{-1}\mathbf{M} \vert \xi_{\delta} \omega \rangle \frac{\partial B^{(0)}_{\alpha}}{\partial x_{\delta}} = 0 \ .
\eea

Equation ~\eqref{bracket_g1} then, reads
\be
    \langle \xi_{\beta}\xi_{\gamma} \vert g^{(1)}_{\alpha} \rangle = \varepsilon_D B^{(1)}_{\alpha}\delta_{\beta \gamma}.
    \label{annalogousYoshida3}
\ee 

Using equations \eqref{annalogousYoshida1}, \eqref{annalogousYoshida2} and \eqref{annalogousYoshida3}, we have that
\bea 
    &&g^{(0)}_{\overline{i}, \alpha} - g^{(0)}_{i, \alpha} = 0 \ , \\
    &&g^{(0)}_{\overline{i}, \alpha} + g^{(0)}_{i, \alpha} = \varepsilon_D B^{(0)}_{\alpha} \ , \\
    &&g^{(1)}_{\overline{i}, \alpha} + g^{(1)}_{i, \alpha} = \varepsilon_D B^{(1)}_{\alpha} \ .
\eea

The same relations for the link distribution found on \cite{Li2013}, leading then for the same coefficient relations
\be
  \left\{\begin{array}{@{}l@{}}
    C_2 = -\frac{2\Delta C_1 + 1}{2\Delta + 1} \\
    C_3 = \frac{C_1 + 2\Delta}{2\Delta + 1}\\
    C_4 = \frac{1 - C_1}{2\Delta + 1}\\
    C_1 \neq 1.
  \end{array}\right.
\ee

Then, the boundary condition for the vector valued distributions may be written as a simple extension from the one found on \cite{Li2013} for each of the components of the distributions $g_{i,\alpha}$, given by Eq.~\eqref{boundarycondition_gmag}.

\end{chapter}

\begin{chapter}{MRT Model for Hydrodynamic Distributions}
\label{apendiceb}
\setcounter{MaxMatrixCols}{27}

\hspace{0.5 cm} The previous CM \cite{Rosis2019} constructs its transformation matrix (\ref{Ts}) depending on the central-moments defined in (\ref{cm}) as $\mbf{T} = \mbf{T}(\overline{\mbf{c}_i} = \mbf{c}_i - \mbf{u})$. The MRT model can be derived from it, using the same organization of moments, by setting $\mbf{u} = 0$, as
\be 
    \mbf{M} = \mbf{T}(\mbf{c}_i) \ .
    \label{Mrtmatrix}
\ee 

In this case, Eq.~(\ref{k*}) is replaced by
\be
    \vert m^* \rangle = (\mbf{I} - \mbf{\Lambda})\vert m \rangle + \mbf{\Lambda} \vert m^{eq} \rangle + \Bigg ( \mbf{I} - \frac{\mbf{\Lambda}}{2} \Bigg )\vert R^{MRT} \rangle \ , \ \label{m*}
\ee

\noindent where $\vert m \rangle = \mbf{M}\vert f \rangle$, $\vert m^{eq} \rangle = \mbf{M}\vert f^{eq} \rangle$ is the MRT equilibrium moments, which, using the Boltzmann equilibrium distribution expanded up to the 6-th order in Hermite polynomials (\ref{BoltzmannEquilibriumDistribution}), will be given by
\bea 
    && m_0^{eq} = \rho \ , \ \ m_1^{eq} = \rho u_x \ , \ \ m_2^{eq} = \rho u_y \ , \ \ m_3^{eq} = \rho u_z \ , \ \ m_4^{eq} = \rho u_xu_y \ ,  \nonumber \\
    && m_5^{eq} = \rho u_xu_z \ , \ \ m_6^{eq} = \rho u_yu_z \ , \ \ m_7^{eq} = \rho (u_x^2 - u_y^2) \ , \ \ m_8^{eq} = \rho (u_x^2 - u_z^2) \ , \nonumber \\
    && m_9^{eq} = \rho (u_x^2 + u_y^2 + u_z^2 + 1) \ , \ \ m_{10}^{eq} = \rho c_s^2 u_x (3u_y^2 + 3u_z^2 + 2) \ , \nonumber \\
    && m_{11}^{eq} = \rho c_s^2 u_y (3u_x^2 + 3u_z^2 + 2) \ , \ \  m_{12}^{eq} = \rho c_s^2 u_z (3u_x^2 + 3u_y^2 + 2) \ , \nonumber \\
    && m_{13}^{eq} = \rho u_x (u_y^2 - u_z^2) \ , \ \ m_{14}^{eq} = \rho u_y (u_x^2 - u_z^2) \ , \ \ m_{15}^{eq} = \rho u_z (u_x^2 - u_y^2) \ , \nonumber \\
    && m_{16}^{eq} = \rho u_xu_yu_z \ , \ \ m_{17}^{eq} = \rho c_s^2 \left [ 3(u_x^2u_y^2 + u_x^2u_z^2 + u_y^2u_z^2) + 2(u_x^2 + u_y^2 + u_z^2) + 1 \right] \ , \nonumber \\
    && m_{18}^{eq} = \rho c_s^4 \left [ 9(u_x^2u_y^2 + u_x^2u_z^2 - u_y^2u_z^2) + 6u_x^2 + 1 \right] \ , m_{19}^{eq} = \rho c_s^2 (u_y^2 - u_z^2)( 3u_x^2 + 1 ) \ , \nonumber \\
    && m_{20}^{eq} = \rho c_s^2u_yu_z(3u_x^2 + 1 ) \ , \ \ m_{21}^{eq} = \rho c_s^2u_xu_z(3u_y^2 + 1 ) \ , \ \ m_{22}^{eq} = \rho c_s^2u_xu_y(3u_z^2 + 1 ) \ , \nonumber 
\eea 

\bea 
    && m_{23}^{eq} = \rho c_s^4u_x(3u_y^2 + 1)(3u_z^2 + 1 ) \ , \ \ m_{24}^{eq} = \rho c_s^4u_y(3u_x^2 + 1)(3u_z^2 + 1 ) \ , \nonumber \\
    && m_{25}^{eq} = \rho c_s^4u_z(3u_x^2 + 1)(3u_y^2 + 1 ) \ , \ \ m_{26}^{eq} = \rho c_s^6(3u_x^2 + 1)(3u_y^2 + 1)(3u_z^2 + 1 ) \ . 
\eea 

In the absense of a forcing term, the post-collision moments are given by $m_i^{*} = m_i^{eq}$ for $i = (0,...,3,9,...26)$ and
\bea 
    && m_4^{*} = (1 - \omega)m_4 + \omega\rho u_xu_y \ , \ \ m_5^{*} = (1 - \omega)m_5 + \omega\rho u_xu_z \ , \ \ m_6^{*} = (1 - \omega)m_6 + \omega\rho u_yu_z \ ,  \nonumber \\
    && m_7^{*} = (1 - \omega)m_7 + \omega\rho (u_x^2 - u_y^2) \ , \ \ m_8^{*} = (1 - \omega)m_8 + \omega\rho (u_x^2 - u_z^2) \ ,
\eea 

\noindent with
\bea
&& m_4 = \sum_i f_i c_{ix}c_{iy} \ , \ \  m_5 = \sum_i f_i c_{ix}c_{iz} \ , \ \  m_6 = \sum_i f_i c_{iy}c_{iz} \ , \ \nonumber \\
&& m_7 = \sum_i f_i (c_{ix}^2 - c_{iy}^2) \ , \ \  m_8 = \sum_i f_i (c_{ix}^2 - c_{iz}^2) \ . 
\eea

The forcing term $\vert R^{MRT} \rangle$ in Eq.~(\ref{m*}) can be obtained by the matrix multiplication of Eq.~(\ref{Mrtmatrix}) by Eq.~(\ref{Forcing_term}), which, also using the Boltzmann equilibrium distribution expanded up to the 6-th order in Hermite polynomials, will be given by

\bea 
    && R_0^{MRT} = 0 \ , \ \ R_1^{MRT} = \rho F_x \ , \ \ R_2^{MRT} = \rho F_y \ , \ \ R_3^{MRT} = \rho F_z \ , \ \ R_4^{MRT} = \rho (F_yu_x + F_xu_y) \ ,  \nonumber \\
    && R_5^{MRT} = \rho (F_zu_x + F_xu_z) \ , \ \ R_6^{MRT} = \rho (F_zu_y + F_yu_z) \ , \ \ R_7^{MRT} = 2\rho (F_xu_x - F_yu_y) \ ,  \nonumber \\
    && R_8^{MRT} = 2\rho (F_xu_x - F_zu_z) \ , \ \ R_9^{MRT} = 2\rho (\mbf{F}\cdot \mbf{u}) \ , \nonumber \\
    && R_{10}^{MRT} = c_s^2\rho \left[6 u_x (F_yu_y + F_zu_z) + F_x(2 + 3u_y^2 + 3u_z^2) \right ] \ , \nonumber \\
    && R_{11}^{MRT} = c_s^2\rho \left[6 u_y (F_xu_x + F_zu_z) + F_y(2 + 3u_x^2 + 3u_z^2) \right ] \ , \nonumber \\
    && R_{12}^{MRT} = c_s^2\rho \left[6 u_z (F_xu_x + F_yu_y) + F_z(2 + 3u_x^2 + 3u_y^2) \right ] \ , \nonumber \\
    && R_{13}^{MRT} = \rho \left[2 u_x (F_yu_y - F_zu_z) + F_x(u_y^2 - u_z^2) \right ] \ , \nonumber \\
    && R_{14}^{MRT} = \rho \left[2 u_y (F_xu_x - F_zu_z) + F_y(u_x^2 - u_z^2) \right ] \ , \nonumber \\
    && R_{15}^{MRT} = \rho \left[2 u_z (F_xu_x - F_yu_y) + F_z(u_x^2 - u_y^2) \right ] \ , \nonumber 
\eea 

\bea 
    && R_{16}^{MRT} = \rho (F_xu_yu_z + F_yu_xu_z + F_zu_xu_y ) \ , \nonumber \\
    && R_{17}^{MRT} = 2c_s^2\rho \left [ F_xu_x( 2 + 3u_y^2 + 3u_z^2) + F_yu_y( 2 + 3u_x^2 + 3u_z^2) + F_zu_z( 2 + 3u_x^2 + 3u_y^2)  \right ] \ , \nonumber \\
    && R_{18}^{MRT} = 2c_s^2\rho \left [ F_xu_x( 2 + 3u_y^2 + 3u_z^2) + 3F_yu_y( u_x^2 - u_z^2) + 3F_zu_z(u_x^2 - u_y^2)  \right ] \ , \nonumber \\
    && R_{19}^{MRT} = 2c_s^2\rho \left [ 3F_xu_x( u_y^2 - u_z^2) + (F_yu_y - F_zu_z)(1 + 3u_x^2)  \right ] \ , \nonumber \\
    && R_{20}^{MRT} = c_s^2\rho \left [ (F_zu_y + F_yu_z)(1 + 3u_x^2) + 6F_xu_xu_yu_z  \right ] \ , \nonumber \\
    && R_{21}^{MRT} = c_s^2\rho \left [ (F_zu_x + F_xu_z)(1 + 3u_y^2) + 6F_yu_xu_yu_z  \right ] \ , \nonumber \\
    && R_{22}^{MRT} = c_s^2\rho \left [ (F_yu_x + F_xu_y)(1 + 3u_z^2) + 6F_zu_xu_yu_z  \right ] \ , \nonumber \\
    && R_{23}^{MRT} = c_s^4\rho \{ F_x (1 + 3u_y^2)(1 + 3u_z^2) + 6u_x\left [ F_zu_z(1+3u_y^2) + F_yu_y(1+3u_z^2) \right]  \} \ , \nonumber \\
    && R_{24}^{MRT} = c_s^4\rho \{ F_y (1 + 3u_x^2)(1 + 3u_z^2) + 6u_y\left [ F_zu_z(1+3u_x^2) + F_xu_x(1+3u_z^2) \right]  \} \ , \nonumber \\
    && R_{25}^{MRT} = c_s^4\rho \{ F_z (1 + 3u_x^2)(1 + 3u_y^2) + 6u_z\left [ F_yu_y(1+3u_x^2) + F_xu_x(1+3u_y^2) \right]  \} \ , \nonumber \\
    &&R_{26}^{MRT} = 2c_s^4\rho \{ F_xu_x (1 + 3u_y^2)(1 + 3u_z^2) + \nonumber \\ 
    && \hspace{1.7 cm} (1 + 3u_x^2)\left [ F_zu_z(1+3u_y^2) + F_yu_y(1+3u_z^2) \right]  \} \ .
\eea 

After the computation of the post-collision moments by Eq.~(\ref{m*}), the post-collision populations are obtained straightforwardly by
\be
    \vert f^*\rangle = \mbf{M}^{-1}\vert m^*\rangle \ ,
\ee 

\noindent and streamed by Eq.~(\ref{stream}).

\end{chapter}


\end{document}